\documentclass[aps,prd,showpacs,superscriptaddress,preprintnumbers,notitlepage]{revtex4-2}
\usepackage{graphicx,float,wrapfig,subfigure}
\usepackage{amsfonts,amsmath,amssymb,amstext}
\usepackage{latexsym}
\usepackage{bm}
\usepackage{color}
\usepackage[normalem]{ulem}
\usepackage{MnSymbol}
\usepackage[colorlinks=true,linktocpage=true,linkcolor=blue,citecolor=blue]{hyperref}

\newcommand{\tr}{ \mbox{tr}}
\definecolor{red}{rgb}{0.7,0,0}
\definecolor{green}{rgb}{0,0.5,0}

\begin{document}
 
 \title{The fermion self-energy and damping rate in a hot magnetized plasma}
 
 \author{Ritesh Ghosh}
 \email{Ritesh.Ghosh@asu.edu}
 \affiliation{College of Integrative Sciences and Arts, Arizona State University, Mesa, Arizona 85212, USA}

 \author{Igor A. Shovkovy}
 \email{igor.shovkovy@asu.edu}
 \affiliation{College of Integrative Sciences and Arts, Arizona State University, Mesa, Arizona 85212, USA}
 \affiliation{Department of Physics, Arizona State University, Tempe, Arizona 85287, USA}
 
 \date{April 17, 2024}
 
 \begin{abstract}
We derive a general expression for the fermion self-energy in a hot magnetized plasma by using the Landau-level representation. In the one-loop approximation, the Dirac structure of the self-energy is characterized by five different functions that depend on the Landau-level index $n$ and the longitudinal momentum $p_z$. We derive general expressions for all five functions and obtain closed-form expressions for their imaginary parts. The latter receive contributions from three types of on-shell processes, which are interpreted in terms of Landau-level transitions, accompanied by a single photon (gluon) emission or absorption. By making use of the imaginary parts of the self-energy functions, we also derive the Landau-level dependent fermion damping rates $\Gamma_{n}(p_z)$ and study them numerically in a wide range of model parameters. We also demonstrate that the two-spin degeneracy of the Landau levels is lifted by the one-loop self-energy corrections. While the spin splitting of the damping rates is small, it may be important for some spin and chiral effects. We argue that the general method and the numerical results for the rates can have interesting applications in heavy-ion physics, astrophysics, and cosmology, where strongly magnetized QED or QCD plasmas are ubiquitous.
 \end{abstract} 
 
 \maketitle

\section{Introduction}
\label{Introduction}

The influence of magnetic fields on relativistic matter has been a topic of continued investigations and interest for decades. Strong magnetic fields appear and play an important role in cosmology~\cite{Vachaspati:1991nm,Widrow:2002ud}, and astrophysics~\cite{Raffelt:1996wa,Harding:2006qn}, and heavy-ion collisions \cite{Liao:2014ava,Kharzeev:2015znc,Huang:2020xyr,Shovkovy:2021yyw}. They can affect physics of magnetars~\cite{Kaspi:2017fwg}, supernovae~\cite{Hardy:2000gg}, and gamma ray bursts~\cite{Granot:2015xba}. Theoretical estimates show that extremely strong magnetic fields up to $|eB|\simeq m_\pi^2$ are produced in high-energy noncentral heavy-ion collisions \cite{Skokov:2009qp,Voronyuk:2011jd,Deng:2012pc,Bloczynski:2012en,Guo:2019mgh}. Of course, the strength and temporal evolution of these fields can be affected by many factors, including the collision energy, the impact parameter, and the electrical conductivity of the plasma \cite{McLerran:2013hla,Tuchin:2013apa,Gursoy:2014aka,Zhong:2014cda,Tuchin:2015oka,Li:2016tel}. Even in condensed matter physics, strong magnetic fields can trigger some relativisticlike phenomena when topological features of the band structure give rise to low-energy quasiparticles described by Dirac and Weyl equations \cite{Gorbar:2021ebc}.  

The groundwork for understanding relativistic systems in the presence of a magnetic field was laid by Heisenberg and Euler \cite{Heisenberg:1936nmg} and later by Schwinger \cite{Schwinger:1951nm}. Many field-theoretical studies have been done over the years since. The key developments and foundations can be found in many books and reviews, e.g., see Refs.~\cite{Dittrich:1985yb,Andersen:2014xxa,Miransky:2015ava}. Despite broad theoretical knowledge gained, surprisingly few quantitive results are known about the Green functions and radiative corrections for relativistic plasmas in background magnetic fields beyond the two extremes of the lowest Landau level approximation and the weak-field limit \cite{Bandyopadhyay:2016fyd,Das:2019nzv,Ghosh:2019kmf}. For some of the recent developments, see Refs.~\cite{Hattori:2012je,Hattori:2012ny,Tuchin:2013bda,Karbstein:2013ufa,Ishikawa:2013fxa,Sadooghi:2016jyf,Ayala:2019akk,Ayala:2020wzl,Hattori:2020htm,Ghosh:2018xhh,Ghosh:2020xwp,Wang:2020dsr,Wang:2021ebh,Chaudhuri:2021skc,Das:2021fma,Wang:2022jxx}.

In a uniform magnetic field, the usual transverse momenta are not good quantum numbers for charged particles. Instead, their eigenstates are given by the Landau-level orbitals. This fact has profound implications on the field theory formalism. The most natural form of the fermion propagator is given in the Landau-level representation \cite{Miransky:2015ava}. The inherent complexity of such a representation makes the evaluation of Feynman diagrams difficult even at the lowest one-loop order. 

The main objective of this study is a rigorous derivation of the fermion self-energy in a strongly magnetized hot relativistic plasma. In particular, the emphasis will be made on the proper treatment of the self-energy in the Landau-level representation. We will follow the approach developed previously in the context of the quantum Hall effect in graphene \cite{Gorbar:2011kc,Shovkovy:2015kja}. Similar methodology was also utilized in the studies of chiral asymmetry in magnetized QED at nonzero density \cite{Gorbar:2013uga,Gorbar:2013upa}. Here we will focus on the fermion self-energy in the Landau-level representation and investigate in detail its imaginary part. Such an imaginary part defines the fermion damping rate in the plasma. It is also a critical input in determining the particle mean free path and some transport properties. We will derive explicit expressions for different components of the self-energy and discuss their interpretation in terms of underlying physical processes. We will also study the quantitative dependence of the fermion damping rate on the Landau level index and the longitudinal momentum. 
 
Several attempts at studying the fermion self-energy in strongly magnetized vacuum can be found in the literature \cite{Tsai:1974df,Jancovici:1969exc,Gepraegs:1994hy,Gusynin:1998nh,Machet:2015swa}. Most notably, the authors of Refs.~\cite{Das:2017vfh,Ayala:2021lor,Chaudhuri:2023djv} had the most of progress in recent years, where they calculated the Fourier transform of the transitionary invariant part of the self-energy but stopped short of projecting the results onto the Landau levels. As we argue here, the latter procedure is necessary in order to extract observable features of the self-energy.
 
The paper is organized as follows. We start from the definition of the fermion self-energy in coordinate space in Sec.~\ref{sec:self-energy}. After removing  the Schwinger phase and performing a Fourier transform on the translation invariant part of the self-energy, we derive a relation that resembles but is not the usual momentum space representation. To extract physics information, the corresponding result is mapped onto the Landau levels in Sec.~\ref{analytical}. The numerical results for the imaginary parts of the functions, defining the Dirac structure of the self-energy, are presented in Sec.~\ref{parameters-numerics}. By utilizing the imaginary part of the self-energy, we derive the fermion damping rate and study its dependence on the Landau-level index $n$ and the longitudinal momentum $p_z$ in Sec.~\ref{damping_rate}. Note that we use two different methods in Subsecs.~\ref{sec:Damping1} and \ref{sec:Damping2}, but they give the same spin-averaged expression for the damping rate. However, the use of the poles of the full propagator in Subsec.~\ref{sec:Damping2} reveals that the rates for the two spin states of each Landau level are slightly different. Finally, we summarize our main results and conclusions in Sec.~\ref{Summary}. Several technical derivations and auxiliary results are given in the appendices at the end of the paper.

\section{Fermion self-energy in magnetized plasma}
\label{sec:self-energy}

To keep our analysis as simple as possible, we consider a hot magnetized QED-like plasma with a single fermion flavor of mass $\bar{m}_{0}$ and charge $q$. With minor adjustments, accounting for a different coupling constant and the number of gauge bosons, the one-loop expression for the self-energy will be also valid for the QCD plasma. Without loss of generality, we will assume that the background magnetic field $\bm{B}$ points in the $+z$ direction. 

At the leading order in coupling, the coordinate space representation of the fermion self-energy is given by 
\begin{equation}
\Sigma(u,u^{\prime})=-4i\pi\alpha\gamma^\mu S(u,u^{\prime}) \gamma^\nu D_{\mu\nu}(u-u^{\prime}),
\label{self-energy}
\end{equation}
where $\alpha= q^2/(4\pi)$ in the coupling constant, $S(u,u^{\prime}) $ is the free fermion propagator, and $D_{\mu\nu}(u-u^{\prime})$ is the photon (gauge-field) propagator. Note that, by definition, $\Sigma(u,u^{\prime}) =i\left[S^{-1}(u,u^{\prime})- G^{-1}(u,u^{\prime}) \right]$, where $G^{-1}(u,u^{\prime})$ is the inverse of the full fermion propagator (at the leading one-loop order). In the case of the QCD plasma, one would need to replace the coupling constant $\alpha$ with $\alpha_s C_F $, where $ \alpha_s = g_s^2/(4\pi)$ and $C_F=(N_c^2-1)/(2N_c)$.

Because of the broken translation symmetry, the free fermion propagator $S(u,u^{\prime})$ and, in turn, the self-energy $\Sigma(u,u^{\prime})$ depend on spacetime coordinates $u=(t,x,y,z)$ and $u^{\prime}=(t^{\prime},x^{\prime},y^{\prime},z^{\prime})$ as follows \cite{Schwinger:1951nm}:
\begin{eqnarray}
 S(u,u^{\prime}) &=& e^{i\Phi(u_\perp,u^{\prime}_\perp)} \bar{S}(u-u^{\prime}) , 
 \label{S-coordinate-space}\\
 \Sigma(u,u^{\prime}) &=& e^{i\Phi(u_\perp,u^{\prime}_\perp)} \bar{\Sigma}(u-u^{\prime}) ,
 \label{Sigma-coordinate-space}
\end{eqnarray}
where $\Phi(u_\perp,u^{\prime}_\perp)$ is the famous Schwinger phase. Note that the translation-invariant parts $\bar{S}(u-u^{\prime})$ and $\bar{\Sigma}(u-u^{\prime})$ depend on the difference $u-u^{\prime}$ only. Assuming the Landau gauge for the background field, i.e., $\bm{A}=(0,Bx,0)$, the explicit form of the Schwinger phase is given by $\Phi(u_\perp,u^\prime_\perp)=\frac{qB}{2}(x+x^\prime)(y-y^\prime)$, where $q$ is the fermion charge.

For reference, we derive an explicit form of the fermion propagator in a background magnetic field in Appendix~\ref{app:propagator}. We make sure to emphasize its coordinate space dependence and the Landau-level structure. Using the same approach, we also obtain the inverse fermion propagator in Appendix~\ref{app:inverse-propagator}. Note that both propagator and its inverse (and, by extension, the self-energy) have exactly the same Schwinger phase. It is consistent with the structure of Eq.~(\ref{self-energy}) and the spacetime dependence in Eqs.~(\ref{S-coordinate-space}) and (\ref{Sigma-coordinate-space}).

After removing the Schwinger phase and performing the Fourier transform on both sides of Eq.~(\ref{self-energy}), we arrive at the following expression for the self-energy function: 
\begin{equation}
\bar{\Sigma} (p_{\parallel},\bm{p}_{\perp}) =- 4 i \pi \alpha \int\frac{d^2k_{\parallel}d^2\bm{k}_{\perp}}{(2\pi)^4}
 \gamma^{\mu}\,\bar{S}(k_{\parallel},\bm{k}_{\perp})\gamma^{\nu} D_{\mu\nu}(p-k) ,
 \label{self-energy-LO}
\end{equation}
where $p_{\parallel}^\mu =(p_0,p_z)$ and $\bm{p}_{\perp}^\mu =(p_x,p_y)$. Interestingly, this expression coincides with the usual definition of the self-energy in a theory with unbroken translation symmetry. Clearly, it is not the case, and the vector-like variable $\bm{p}_{\perp}$ cannot be interpreted as a conserved transverse momentum in a magnetized plasma. (In contrast, the two components of $p_{\parallel}$, i.e., the energy $p_0$ and the longitudinal momentum $p_z$, are conserved quantities in a uniform magnetic field.) Despite the appearance, the functions $\bar{S}(k_{\parallel},\bm{k}_{\perp})$ and $\bar{\Sigma} (p_{\parallel},\bm{p}_{\perp})$ are not the momentum-space representations of the fermion propagator and the self-energy, respectively. Yet, they encode all information about the propagator and self-energy. 

\begin{figure}[t]
\includegraphics[width=0.25\textwidth]{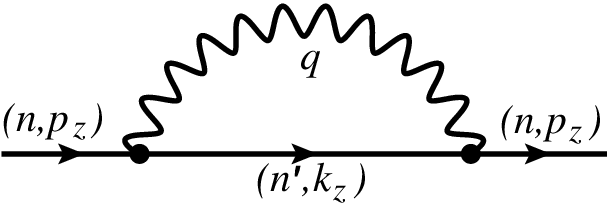}
\caption{The leading order Feynman diagram for the fermion self-energy in the Landau-level representation.}
\label{fig.SelfEnergy}
\end{figure}

The main advantage of the representation in Eq.~(\ref{self-energy-LO}) is its simplicity. To extract its observables effects, however, we will need to render it in the Landau-level basis. The corresponding projection will be discussed and implemented in Sec.~\ref{sec:self-energy-LL-rep}. While technically nontrivial, its outcome is obvious in the diagrammatic form shown in Fig.~\ref{fig.SelfEnergy}.

At this point, we will proceed with the calculation of the self-energy in Eq.~(\ref{self-energy-LO}). In the derivation, we will use the following Feynman gauge for the free gauge-field propagator:
\begin{equation}
D_{\mu\nu}(p-k) = -i \frac{g_{\mu\nu}}{(p-k)^2},
 \label{photon-propagator}
\end{equation}
where $g_{\mu\nu} =\mbox{diag}(1,-1,-1,-1)$ is the Minkowski metric. We note that a more refined analysis of a hot magnetized plasma may require using the hard-thermal \cite{Braaten:1989mz} and hard-magnetic loop \cite{Miransky:2015ava} resummations. The corresponding refinements are beyond the scope of the present exploratory study but should be undertaken in the future. 

It is instructive to emphasize that the Feynman gauge for the gauge-field propagator is convenient but not the most general. In fact, it is well known that the fermion self-energy depends on a gauge choice. In this study, however, we will be concerned primarily with the imaginary (dissipative) part of the self-energy and the fermion damping rate. For these purposes, the simplest Feynman gauge should be sufficient \cite{Rebhan:1992ak,Nakkagawa:1992ew}.
 
By substituting the free fermion propagator, whose explicit form  is given in Appendix~\ref{app:propagator}, and the photon propagator in Eq.~(\ref{photon-propagator}) into the expression for the self-energy in Eq.~(\ref{self-energy-LO}), we obtain 
 \begin{equation}
\bar{\Sigma} (p_{\parallel},\bm{p}_{\perp}) 
  =- 4i \pi \alpha \sum_{n^{\prime}=0}^\infty  \int \frac{d^2k_{\parallel}d^2\bm{k}_{\perp}}{(2\pi)^4} e^{-k_\perp^2 l^2}\gamma^\mu \frac{(-1)^{n^{\prime}} D^{(0)}_{n^{\prime}}(k_{\parallel},\bm{k}_{\perp}) }{k_\parallel^2-\bar{m}_{0}^2-2n^{\prime} |qB|}\gamma_\mu \frac{1}{q_\parallel^2-q_{\perp}^2}.
  \label{Self-energy-1}
 \end{equation}
Here $q_\parallel=p_\parallel-k_\parallel$, $\bm{q}_{\perp}=\bm{p}_{\perp}-\bm{k}_{\perp}$, and 
 \begin{equation}
 D^{(0)}_{n^{\prime}}(k_{\parallel},\bm{k}_{\perp})  = 2\left[(k_\parallel \cdot \gamma_\parallel) +\bar{m}_{0}\right]\left[\mathcal{P}_{+}L_{n^{\prime}}\left(2 k_{\perp}^2 \ell^{2}\right) -\mathcal{P}_{-}L_{n^{\prime}-1}\left(2 k_{\perp}^2 \ell^{2}\right)\right]
 + 4 (\bm{k}_{\perp}\cdot\bm{\gamma}_{\perp}) L_{n^{\prime}-1}^1\left(2 k_{\perp}^2 \ell^{2}\right) ,
 \label{Dn0}  
\end{equation}
where $\mathcal{P}_{\pm}=\left(1 \pm s_{\perp}i\gamma^1\gamma^2\right)/2$ are spin projectors, $\ell = 1/\sqrt{|qB|}$ is the magnetic length, $s_{\perp}=\mbox{sign}(qB)$, and $L_{n}^\alpha(z)$ are the generalized Laguerre polynomials \cite{Gradshteyn:1943cpj}. We assume that, by definition, $L_{-1}^\alpha(z) =0$.

To account for a nonzero temperature $T$, we use Matsubara's formalism. In particular, we replace the fermion energies $p_0$ and $k_0$ with $i\omega_{n_p}\equiv i \pi T (2n_p+1)$ and $i\omega_{n_k}\equiv i \pi T (2n_k+1)$, respectively, and replace the integral over $k_0$ with the Matsubara sum, i.e., 
\begin{equation}
\int\frac{dk_0}{2\pi} F\left(p_0, k_0\right) \to  i T\sum_{n_k=-\infty}^{\infty} F\left( i\omega_{n_p}, i\omega_{n_k}\right) .
\end{equation}
Then, the self-energy (\ref{Self-energy-1}) becomes
 \begin{equation}
\bar{\Sigma} (i\omega_{n_p},p_{z},\bm{p}_{\perp}) 
  = 4\pi \alpha T \sum_{n^{\prime}=0}^\infty \sum_{n_k=-\infty}^{\infty} \int \frac{dk_{z}d^2\bm{k}_{\perp}}{(2\pi)^3}   \frac{(-1)^{n^{\prime}} e^{-k_\perp^2 l^2} \tilde{D}^{(0)}_{n^{\prime}}(i\omega_{n_k},k_{z},\bm{k}_{\perp})}{\left( \omega_{n_k}^2+E_{n^{\prime},k_z}^2 \right)\left[(\omega_{n_p}-\omega_{n_k})^2+E_{q}^2\right]} ,
  \label{Self-energy-2}
 \end{equation}
where we used the shorthand notations for the Landau-level energies $E_{n^{\prime},k_z}\equiv\sqrt{2n^{\prime} |qB|+\bar{m}_{0}^2+k_z^2}$ and the gauge boson energy $E_{q}\equiv \sqrt{\bm{q}_{\perp}^2+q_z^2}$, and introduced the following new function:
\begin{eqnarray}
\tilde{D}^{(0)}_{n^{\prime}}(i\omega_{n_k},k_{z},\bm{k}_{\perp}) &\equiv & \gamma^\mu  D^{(0)}_{n^{\prime}}(i\omega_{n_k},k_{z},\bm{k}_{\perp})\gamma_\mu = 4\bar{m}_{0}\left[L_{n^{\prime}}\left(2 k_{\perp}^2 \ell^{2}\right) -L_{n^{\prime}-1}\left(2 k_{\perp}^2 \ell^{2}\right)\right]
\nonumber\\
&-&4\left(i\omega_{n_k}\gamma^0-k_{z}\gamma^3\right)\left[\mathcal{P}_{-}L_{n^{\prime}}\left(2 k_{\perp}^2 \ell^{2}\right) -\mathcal{P}_{+}L_{n^{\prime}-1}\left(2 k_{\perp}^2 \ell^{2}\right)\right]
-8 (\bm{k}_{\perp}\cdot\bm{\gamma}_{\perp}) L_{n^{\prime}-1}^1\left(2 k_{\perp}^2 \ell^{2}\right).
 \end{eqnarray}
 After performing the Matsubara sum, we obtain
\begin{eqnarray}
\bar{\Sigma} (p_{\parallel},\bm{p}_{\perp}) 
  &=& 4\pi \alpha \sum_{n^{\prime}=0}^\infty  \sum_{s_1=\pm} \sum_{s_2=\pm} (-1)^{n^{\prime}} \int \frac{dk_{z}d^2\bm{k}_{\perp}}{(2\pi)^3}  e^{-k_\perp^2 l^2}  \frac{s_1 s_2 \left[ 1-n_F(s_1 E_{n^{\prime},k_z})+n_B(s_2 E_{q}) \right] }{E_{n^{\prime},k_z} E_{q} 
  \left(p_0 -s_1 E_{n^{\prime},k_z} -s_2 E_{q} +i\epsilon \right) }\nonumber\\
  &&\times \bigg\{
  \left(s_1 E_{n^{\prime},k_z} \gamma^0-k_{z}\gamma^3 \right)\left[\mathcal{P}_{-}L_{n^{\prime}}\left(2 k_{\perp}^2 \ell^{2}\right) -\mathcal{P}_{+}L_{n^{\prime}-1}\left(2 k_{\perp}^2 \ell^{2}\right)\right]
 \nonumber\\ 
&&    -\bar{m}_{0}\left[L_{n^{\prime}}\left(2 k_{\perp}^2 \ell^{2}\right) -L_{n^{\prime}-1}\left(2 k_{\perp}^2 \ell^{2}\right)\right]
+2  (\bm{k}_{\perp}\cdot\bm{\gamma}_{\perp})L_{n^{\prime}-1}^1\left(2 k_{\perp}^2 \ell^{2}\right) 
  \bigg\} ,
  \label{Self-energy-3}
 \end{eqnarray}
 where we used the standard Fermi-Dirac and Bose-Einstein distribution functions, $n_F(E)=\left(e^{E/T}+1\right)^{-1}$ and $n_B(E)=\left(e^{E/T}-1\right)^{-1}$, respectively. In the derivation, we used the following result for the Matsubara sum:
\begin{eqnarray}
T\sum_{n_k=-\infty}^{\infty} \frac{i\omega_{n_k} A+ B}{\left( \omega_{n_k}^2+E_{a}^2 \right)\left[(\omega_{n_p}-\omega_{n_k})^2+E_{b}^2\right]}
&=& -\frac{1}{4}\sum_{s_1,s_2=\pm} \frac{\left(s_1 E_{a}  A+ B\right)\left[ 1-n_F(s_1 E_{a})+n_B(s_2 E_{b}) \right]}{s_1 s_2E_{a} E_{b}\left( i\omega_{n_p}-s_1 E_{a}-s_2 E_{b} \right) }.
\end{eqnarray}
To separate the real and imaginary parts of the self-energy, we perform the analytical continuation $i\omega_{n_p}\to p_0+i\epsilon$ and use the Sokhotski formula,
\begin{equation}
\frac{1}{p_0 -s_1 E_{n^{\prime},k_z} -s_2 E_{q} +i\epsilon} = \mathcal{P}\frac{1}{p_0 -s_1 E_{n^{\prime},k_z} -s_2 E_{q} +i\epsilon } -i\pi \delta\left(p_0 -s_1 E_{n^{\prime},k_z} -s_2 E_{q}\right).
\label{eq:Sokhotski}
\end{equation}

In the rest, we will concentrate on the imaginary (absorptive) part. The corresponding expression can be simplifies by taking into account that 
\begin{eqnarray}
\delta(p_0-s_1E_{n^{\prime},k_z}-s_2E_q) &=& \sum_{s^\prime=\pm} \frac{E_{n^{\prime},k_z} E_q\delta(k_z-k_z^{s^\prime})}{|(k_z^{s^\prime}-p_z)s_1E_{n^{\prime},k_z} +k_z^{s^\prime}s_2 E_q|} 
= \sum_{s^\prime=\pm}\frac{2 E_{n^{\prime},k_z} E_q\delta(k_z-k_z^{s^\prime})}{\sqrt{\left[q_\perp^2-(q_\perp^{-})^2\right]\left[q_\perp^2-(q_\perp^{+})^2\right]}} ,
\end{eqnarray}
where $q_\perp^\pm=|\sqrt{2n^{\prime}|qB|+\bar{m}_{0}^2}\pm \sqrt{p_0^2-p_z^2}|$ and the explicit expressions for the two solutions $k_z^{\pm}$ to the energy-conservation condition read
\begin{eqnarray}
 k_z^{\pm}&=&\frac{p_z}{2}\left(1+\frac{2n^{\prime} |qB|+\bar{m}_{0}^2-q_\perp^2}{p_0^2-p_z^2}\pm \frac{p_0}{p_z(p_0^2-p_z^2)}\sqrt{\left[q_\perp^2-(q_\perp^{-})^2\right]\left[q_\perp^2-(q_\perp^{+})^2\right]} \right).  
\end{eqnarray}
Note that, for the fermions on the mass shell, we should set $p_0^2-p_z^2=2n |qB|+\bar{m}_{0}^2$, and the two thresholds will become $q_{\perp}^{\pm}= \left|\sqrt{2n^{\prime}|qB|+\bar{m}_{0}^2}\pm \sqrt{2n |qB|+\bar{m}_{0}^2}\right|$.

By substituting the solutions of the energy-conservation condition ($k_z=k_z^{\pm}$), we derive the following two expressions for the particle energies:
\begin{eqnarray}
E_{n^{\prime},k_z} \Big|_{k_z\to k_z^{\pm}}&=& 
\frac{s_1p_0}{2}\left(1+\frac{2n^{\prime} |qB|+\bar{m}_{0}^2-q_\perp^2}{p_0^2-p_z^2} 
\pm  \frac{p_z}{p_0(p_0^2-p_z^2)}\sqrt{\left[q_\perp^2-(q_\perp^{-})^2\right]\left[q_\perp^2-(q_\perp^{+})^2\right]} \right),
\label{Enkz} \\
E_q\Big|_{k_z\to k_z^{\pm}}&=& 
\frac{s_2p_0}{2}\left(1-\frac{2n^{\prime} |qB|+\bar{m}_{0}^2-q_\perp^2}{p_0^2-p_z^2}
\mp  \frac{p_z}{p_0(p_0^2-p_z^2)}\sqrt{\left[q_\perp^2-(q_\perp^{-})^2\right]\left[q_\perp^2-(q_\perp^{+})^2\right]} \right).
\label{Eqkz}
\end{eqnarray}
Without loss of generality, below we will concentrate on the case of Landau-level states with positive-energies, $p_0>0$. On the mass shell, they will be given by the Landau-level energies, $p_0=\sqrt{2n |qB|+\bar{m}_{0}^2+p_z^2}$. 
If needed, the self-energy results for the Landau-level states with negative energies could be obtained by using the charge-conjugation symmetry. 

\begin{figure}[t]
  \subfigure[]{\includegraphics[width=0.22\textwidth]{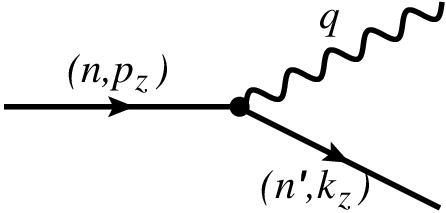}}
  \hspace{0.12\textwidth}
  \subfigure[]{\includegraphics[width=0.22\textwidth]{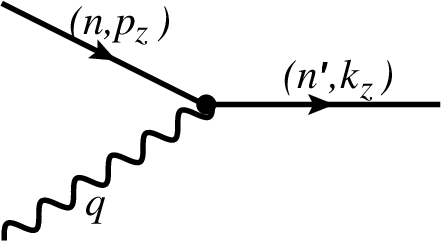}}
  \hspace{0.12\textwidth}
  \subfigure[]{\includegraphics[width=0.22\textwidth]{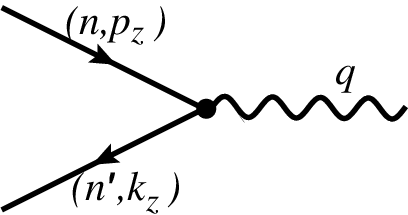}}
\caption{Feynman diagrams for the three processes contributing to the fermion damping in the $n$th Landau-level state: 
(a) quantum transition to a lower Landau level with emission of a photon $\psi_{n}\to \psi_{n^{\prime}}+\gamma$ ($n >n^{\prime}$), 
(b) quantum transition to a higher Landau level with absorption of a photon $\psi_{n}+\gamma\to\psi_{n^{\prime}}$ ($n<n^{\prime}$),
(c) particle-antiparticle annihilation $\psi_{n}+\bar{\psi}_{n^{\prime}}\to\gamma$.}
\label{fig.FeynmanDiagrams}
\end{figure}

By analyzing the solutions for energy-conservation relation $p_0=s_1E_{n^{\prime},k_z}+s_2E_q$ with the assumption $p_0>0$, we identify the following three kinematic cases:
\begin{eqnarray}
s_1>0, ~ s_2>0: &\quad& 0<q_\perp<q_\perp^{-},
\label{proc-1}\\
s_1>0, ~ s_2<0: &\quad& 0<q_\perp<q_\perp^{-},
\label{proc-2}\\
s_1<0, ~ s_2>0: &\quad& q_\perp^{+}<q_\perp< \infty. 
\label{proc-3}
\end{eqnarray}
The first one describes a transition to a lower energy particle state with emission of a photon ($\psi_{n}\to \psi_{n^\prime} +\gamma$ with $n>n^\prime$), see Fig.~\ref{fig.FeynmanDiagrams}a. The second one describes a transition to a higher energy particle state with absorption of a photon ($\psi_{n}+\gamma\to \psi_{n^\prime} $ with $n<n^\prime$), see Fig.~\ref{fig.FeynmanDiagrams}b. Finally, the third one describes a transition to an antiparticle state with emission of a photon (i.e., annihilation process $\psi_{n}+\bar{\psi}_{n^\prime}\to \gamma$ for any $n$ and $n^\prime$), see Fig.~\ref{fig.FeynmanDiagrams}c. 

It is instructive to emphasize that the three processes in Fig.~\ref{fig.FeynmanDiagrams} contribute to the fermion damping rate only when the background magnetic field is nonzero. Without magnetic field, these processes of order $\alpha$ are forbidden by the energy-momentum conservation. Instead, the fermion damping is dominated by diagrams of order $\alpha^2$ such as two-to-two scattering and annihilation processes (i.e., $\psi_{k}+\gamma \to \psi_{k^{\prime}} +\gamma$ and $\psi_{k}+\bar{\psi}_{k^{\prime}}\to\gamma+\gamma$). Turning the argument around, this also implies that contributions from higher-order processes will compete with those in Fig.~\ref{fig.FeynmanDiagrams} when the magnetic field is sufficiently weak. 

The final expression for the imaginary part reads
\begin{eqnarray}
\mbox{Im}\left[\bar{\Sigma} (p_{\parallel},\bm{p}_{\perp}) \right]
  &=& - 4 \pi \alpha \sum_{n^{\prime}=0}^\infty  \sum_{\{s\}} (-1)^{n^{\prime}} \int \frac{d^2\bm{k}_{\perp}}{(2\pi)^2}  e^{-k_\perp^2 l^2}  \frac{ 1-n_F(s_1 E_{n^{\prime},k_z^{s^\prime}})+n_B(s_2 E_{q}) }{s_1 s_2 \sqrt{\left[q_\perp^2-(q_\perp^{-})^2\right]\left[q_\perp^2-(q_\perp^{+})^2\right]}  }\nonumber\\
  &&\times \bigg\{
  \left(s_1 E_{n^{\prime},k_z^{s^\prime}} \gamma^0-k_{z}^{s^\prime}\gamma^3 \right)\left[\mathcal{P}_{-}L_{n^{\prime}}\left(2 k_{\perp}^2 \ell^{2}\right) -\mathcal{P}_{+}L_{n^{\prime}-1}\left(2 k_{\perp}^2 \ell^{2}\right)\right]
 \nonumber\\ 
&& -\bar{m}_{0}\left[L_{n^{\prime}}\left(2 k_{\perp}^2 \ell^{2}\right) -L_{n^{\prime}-1}\left(2 k_{\perp}^2 \ell^{2}\right)\right]
+2  (\bm{k}_{\perp}\cdot\bm{\gamma}_{\perp})L_{n^{\prime}-1}^1\left(2 k_{\perp}^2 \ell^{2}\right) 
  \bigg\} ,
  \label{Self-energy-4}
\end{eqnarray}
where the shorthand notation $\sum_{\{s\}}$ represents the sum over three signs, i.e., $s_1,s_2,s^\prime=\pm1$. 
This expression for the Fourier transform of the translation invariant part of the self-energy, as defined in Eq.~(\ref{Sigma-coordinate-space}), does not reveal explicitly the Landau-level structure. Indeed, as we show in Appendix~\ref{app:self-energy-LL}, its Landau-level representation (\ref{app:self-energy-2}) should be given as an expansion in Laguerre polynomials $L_{n}^\alpha\left(2 p_{\perp}^2 \ell^{2}\right) $, where $p_{\perp}$ is the Fourier variable for the external line. In the next section, we will use the properties of the Laguerre polynomials to render the self-energy in such a Landau-level form.

\section{Fermion self-energy in Landau-level representation}
\label{sec:self-energy-LL-rep}

\subsection{Analytical expressions for the self-energy functions}
\label{analytical}

By making use of explicit wave functions for the Landau-level orbitals in Appendix~\ref{app:self-energy-LL}, we find  that the self-energy must take the following general form:
\begin{eqnarray}
 \bar{\Sigma} (p_{\parallel},\bm{p}_{\perp})  &=&- 2  e^{-p_{\perp}^2 \ell^2}  \sum\limits_{n=0}^{\infty} (-1)^n
 \left[ \delta v_{\parallel,n} (p_\parallel\cdot\gamma_\parallel)
 +i\gamma^1\gamma^2 (p_\parallel\cdot\gamma_\parallel) \tilde{v}_{n}
 -\delta m_{n}  -i\gamma^1\gamma^2\tilde{m}_{n}
 \right] \left[\mathcal{P}_{+} L_n(2p_{\perp}^2 \ell^2)- \mathcal{P}_{-} L_{n-1}(2p_{\perp}^2 \ell^2) \right] \nonumber\\
 &-& 4  e^{-p_{\perp}^2 \ell^2}  \sum\limits_{n=0}^{\infty} (-1)^n \delta v_{\perp,n} (\bm{\gamma}_\perp\cdot \bm{p}_{\perp}) L^1_{n-1}(2p_{\perp}^2 \ell^2) .
 \label{inversefull-2}
\end{eqnarray}
As is easy to verify, it contains all the same Dirac matrices as the main expression for the one-loop self-energy $ \bar{\Sigma} (p_{\parallel},\bm{p}_{\perp}) $ in Eq.~(\ref{Self-energy-3}), or its imaginary part in Eq.~(\ref{Self-energy-4}). Of course, it is not accidental since we made an educated choice for a general form of the full propagator in Appendices~\ref{app:propagator} and ~\ref{app:inverse-propagator}. In this connection, we should mention that, if higher-order calculations would reveal the need for additional Dirac structures (allowed by symmetries), they could be easily incorporated into a general ansatz for the full propagator.
 
The physical meaning of $\delta v_{\parallel,n}$, $\delta v_{\perp,n}$, and $\delta m_{n}$ is clear. They measure the one-loop corrections to the maximum spin-averaged particle speed (in the directions parallel and perpendicular to the magnetic field) and the corrections to the particle mass in the $n$th Landau level. As for the other two functions, i.e., $\tilde{v}_{n}$ and $\tilde{m}_{n}$, they determine the splitting of the parallel velocities and masses of the two spin states.

The functional form of the self-energy dependence on $p_{\perp}$, obtained in the previous section, does not seem to match the Landau-level representation in Eq.~(\ref{inversefull-2}), which is an expansion in the Laguerre polynomials $L_{n}^\alpha\left(2 p_{\perp}^2 \ell^{2}\right)$. However, by making use of the orthogonality property of the Laguerre polynomials, i.e.,
\begin{equation}
 \int_{0}^{\infty} e^{-x} x^{\alpha} L_{n}^{\alpha}(x)L_{n^{\prime}}^{\alpha}(x) dx = \delta_{nn^{\prime}} \frac{\Gamma(n+\alpha+1)}{n!} ,
\end{equation}
it is straightforward to render the result in the form of such an expansion.

In particular, after separating different Dirac structures, we can match the self-energy functions
$\delta v_{\parallel,n}$, $\delta v_{\perp,n}$, $\delta m_{n}$, $\tilde{v}_{n}$, and $\tilde{m}_{n}$ in the Landau-level representation to the projection of function $ \bar{\Sigma} (p_{\parallel},\bm{p}_{\perp}) $ onto the Landau-level orbitals. The corresponding relations read:
\begin{eqnarray}
  \delta v_{\parallel,n}^{+} &\equiv&  \delta v_{\parallel,n} +s_\perp \tilde{v}_{n} = -\frac{(-1)^{n} \ell^2}{2\pi  p_\parallel^2 }
  \int d^2 \bm{p}_\perp e^{-p_{\perp}^2 \ell^2}  \tr\left[(p_\parallel\cdot\gamma_\parallel)  \mathcal{P}_{+} \Sigma(p_{\parallel},\bm{p}_{\perp})
  \right]  L_{n}(2p_{\perp}^2 \ell^2)   ,
  \label{Laguerre-projection-1}   \\
  \delta v_{\parallel,n}^{-} &\equiv&  \delta v_{\parallel,n} - s_\perp \tilde{v}_{n} = \frac{(-1)^{n} \ell^2}{2\pi p_\parallel^2 }
  \int d^2 \bm{p}_\perp e^{-p_{\perp}^2 \ell^2} 
\tr\left[(p_\parallel\cdot\gamma_\parallel)  \mathcal{P}_{-} \Sigma(p_{\parallel},\bm{p}_{\perp})
  \right]  L_{n-1}(2p_{\perp}^2 \ell^2) ,
  \label{Laguerre-projection-2} \\
   \delta m_{n}^{+} &\equiv&  \delta m_{n} +s_\perp \tilde{m}_{n} =  \frac{(-1)^{n} \ell^2}{2\pi}
  \int d^2 \bm{p}_\perp e^{-p_{\perp}^2 \ell^2}  \tr\left[\mathcal{P}_{+} \Sigma(p_{\parallel},\bm{p}_{\perp})
  \right]  L_{n}(2p_{\perp}^2 \ell^2) ,
  \label{Laguerre-projection-3} \\
 \delta m_{n}^{-} &\equiv& \delta m_{n}  -s_\perp \tilde{m}_{n} = - \frac{(-1)^{n} \ell^2}{2\pi}
  \int d^2 \bm{p}_\perp e^{-p_{\perp}^2 \ell^2} 
\tr\left[ \mathcal{P}_{-} \Sigma(p_{\parallel},\bm{p}_{\perp})
  \right]  L_{n-1}(2p_{\perp}^2 \ell^2)  ,
  \label{Laguerre-projection-4} \\ 
 \delta v_{\perp,n}  &=&  \frac{(-1)^{n} \ell^4}{4 \pi n}
  \int d^2 \bm{p}_\perp e^{-p_{\perp}^2 \ell^2} 
\tr\left[ (\bm{\gamma}_\perp\cdot \bm{p}_{\perp}) \Sigma(p_{\parallel},\bm{p}_{\perp})
  \right]  L_{n-1}^{1}(2p_{\perp}^2 \ell^2) .
  \label{Laguerre-projection-5} 
\end{eqnarray}
Note that the self-energy component functions $\delta v_{\parallel,n}^{\pm}$ and $ \delta m_{n}^{\pm}$ have a simple meaning. They describe corrections to the velocity and mass parameters for the two spin states in the $n$th Landau level. Only two of such functions, namely Eqs.~(\ref{Laguerre-projection-1}) and (\ref{Laguerre-projection-3}), are defined for all the Landau levels, $n\geq0$. The other three are defined only for the higher Landau levels with $n\geq 1$. This is related to the unique property of the lowest Landau level, which has only one spin polarization (i.e., pointing along the field direction if the fermions carry a positive charge, or opposite to the field if they carry a negative charge). As a result, the self-energy in the lowest Landau level ($n=0$) is fully characterized by the longitudinal velocity (or the wave-function renormalization) $v_{\parallel,n}^{+}= \delta v_{\parallel,n} +s_\perp  \tilde{v}_{n}$ and the mass renormalization $ \delta m_{n}^{+} =\delta m_{n} + s_\perp \tilde{m}_{n}$.

By substituting the expression for the one-loop result (\ref{Self-energy-3}) into the above definitions (\ref{Laguerre-projection-1}) through (\ref{Laguerre-projection-5}), we will have all Dirac components of the self-energy in the Landau-level representation. Here we will concentrate on the imaginary parts of the self-energy functions by using the result in Eq.~(\ref{Self-energy-4}). The corresponding results read 
\begin{eqnarray}
 \mbox{Im}\left[ \delta v_{\parallel,n}^{+}  \right] &=&\frac{\alpha}{p_\parallel^2}  \sum_{n^{\prime}=0}^\infty  \sum_{\{s\}}  
\int q_\perp d q_\perp 
\mathcal{I}_{0}^{n,n^{\prime}-1}\left( \frac{q_\perp^2 \ell^2}{2} \right) 
  \frac{\left(s_1 E_{n^{\prime},k_z^{s^\prime}} p_0-k_{z}^{s^\prime}p_z \right)\left[ 1-n_F(s_1 E_{n^{\prime},k_z^{s^\prime}})+n_B(s_2 E_{q}) \right] }{s_1 s_2  \sqrt{\left[q_\perp^2-(q_\perp^{-})^2\right]\left[q_\perp^2-(q_\perp^{+})^2\right]}  } ,
  \label{pars-1}   \\
 \mbox{Im}\left[ \delta v_{\parallel,n}^{-} \right] &=& \frac{\alpha}{p_\parallel^2}  \sum_{n^{\prime}=0}^\infty  \sum_{\{s\}}  
\int q_\perp d q_\perp
\mathcal{I}_{0}^{n-1,n^{\prime}}\left( \frac{q_\perp^2 \ell^2}{2} \right) 
  \frac{\left(s_1 E_{n^{\prime},k_z^{s^\prime}} p_0-k_{z}^{s^\prime}p_z \right)\left[ 1-n_F(s_1 E_{n^{\prime},k_z^{s^\prime}})+n_B(s_2 E_{q}) \right] }{s_1 s_2   \sqrt{\left[q_\perp^2-(q_\perp^{-})^2\right]\left[q_\perp^2-(q_\perp^{+})^2\right]}  }  ,
  \label{pars-2} \\
 \mbox{Im}\left[  \delta m_{n}^{+}  \right] &=& \alpha \bar{m}_{0} \sum_{n^{\prime}=0}^\infty  \sum_{\{s\}}  
\int q_\perp d q_\perp \left[
\mathcal{I}_{0}^{n,n^{\prime}}\left( \frac{q_\perp^2 \ell^2}{2} \right)  
+\mathcal{I}_{0}^{n,n^{\prime}-1}\left( \frac{q_\perp^2 \ell^2}{2} \right)  
\right]
  \frac{ 1-n_F(s_1 E_{n^{\prime},k_z^{s^\prime}})+n_B(s_2 E_{q}) }{s_1 s_2  \sqrt{\left[q_\perp^2-(q_\perp^{-})^2\right]\left[q_\perp^2-(q_\perp^{+})^2\right]}  }   ,
  \label{pars-3} \\ 
   \mbox{Im}\left[  \delta m_{n}^{-}  \right] &=& \alpha \bar{m}_{0} 
   \sum_{n^{\prime}=0}^\infty  \sum_{\{s\}}  
\int q_\perp d q_\perp\left[
\mathcal{I}_{0}^{n-1,n^{\prime}}\left( \frac{q_\perp^2 \ell^2}{2} \right)  
+\mathcal{I}_{0}^{n-1,n^{\prime}-1}\left( \frac{q_\perp^2 \ell^2}{2} \right)  
\right]  \frac{ 1-n_F(s_1 E_{n^{\prime},k_z^{s^\prime}})+n_B(s_2 E_{q}) }{s_1 s_2  \sqrt{\left[q_\perp^2-(q_\perp^{-})^2\right]\left[q_\perp^2-(q_\perp^{+})^2\right]}  }   ,
  \label{pars-4} \\ 
 \mbox{Im}\left[  \delta v_{\perp,n}  \right] &=&  \frac{\alpha}{2 n}
  \sum_{n^{\prime}=0}^\infty  \sum_{\{s\}}  
\int q_\perp d q_\perp
\mathcal{I}_{2}^{n-1,n^{\prime}-1}\left( \frac{q_\perp^2 \ell^2}{2} \right) 
  \frac{ 1-n_F(s_1 E_{n^{\prime},k_z^{s^\prime}})+n_B(s_2 E_{q}) }{s_1 s_2  \sqrt{\left[q_\perp^2-(q_\perp^{-})^2\right]\left[q_\perp^2-(q_\perp^{+})^2\right]}  }   .
  \label{pars-5} 
\end{eqnarray}
Here we introduced two unitless kernel functions that depend on  $q_\perp^2 \ell^2/2$. They are defined in Appendix~\ref{app:kernels}. There we also prove that the kernels reduce to functions $\mathcal{I}_{0}^{n,n^\prime}(\xi)$ and $\mathcal{I}_{2}^{n,n^\prime}(\xi)$ introduced previously in Ref.~\cite{Wang:2021ebh}. The explicit expressions for these functions are given in Eqs.~(\ref{I0f-LL-form1}) -- (\ref{I2f-LL-form1}) of our Appendix~\ref{app:kernels}.

As expected, all parameters are Landau-level dependent functions of the longitudinal momentum $p_z$.
We can further simplify the integrand in Eqs.~(\ref{pars-1}) and (\ref{pars-2}) by taking into account the following relation:
\begin{eqnarray}
s_1 E_{n^{\prime},k_z^{s^\prime}} p_0-k_{z}^{s^\prime}p_z &=& \frac{1}{2}\left(p_{\parallel}^2+2n^{\prime} |qB|+\bar{m}_{0}^2-q_\perp^2\right) \Big|_{\rm m. s.}
=\left(n+n^{\prime}\right) |qB|+\bar{m}_{0}^2-\frac{q_\perp^2}{2}.
\label{Ep-relation}
\end{eqnarray} 
In the last expression, we used the mass-shell condition to express the parallel components of the fermion momentum in terms of the Landau-level index: $p_{\parallel}^2 = 2n|qB|+\bar{m}_{0}^2$. For numerical calculations later, it will help that the result is independent of the signs $s_1$ and $s^\prime$.

We should note that, despite the appearance, the combination of the Fermi-Dirac and Bose-Einstein distribution functions, $1-n_F(s_1 E_{n^{\prime},k_z^{s^\prime}})+n_B(s_2 E_{q})$, is also independent of the signs $s_1$ and $s_2$. Indeed, it is obvious after taking into account the energy expressions in Eqs.~(\ref{Enkz}) and (\ref{Eqkz}), which contain the overall factors of $s_1$ and $s_2$, respectively.

The only dependence of the integrands in Eqs.~(\ref{pars-1}) through (\ref{pars-5}) on the signs $s_1$ and $s_2$ comes from the overall factor $s_1 s_2$. It is instructive to recall that different sign choices determine the process types contributing to the imaginary part, see Eqs.~(\ref{proc-1}) -- (\ref{proc-3}). Therefore, up to overall sign $s_1 s_2$, the integrands are formally the same for all processes. The contributions of quantum transitions of fermions to lower Landau-level states (accompanied by photon emission) come with a plus sign. The contributions of transitions to higher Landau-level states (accompanied by  photon absorption) and the annihilation processes (accompanied by photon emission) come with a minus sign. While the integrands are formally the same for all three processes (up to a sign), the range of integration over $q_\perp$ differs. Namely, it is $0<q_\perp<q_\perp^{-}$ for transitions to lower/higher Landau-level states and $q_\perp^{+}<q_\perp< \infty$ for the annihilation processes.

By using the five functions in Eqs.~(\ref{pars-1}) through (\ref{pars-4}), we can obtain the spin-average Landau-level dependent values of the parallel velocity and mass, i.e.,
\begin{eqnarray}
\mbox{Im}\left[ \delta v_{\parallel,n} \right]  &=& \frac{1}{2} \mbox{Im}\left[ \delta v_{\parallel,n}^{+} +\delta v_{\parallel,n}^{-} \right]  ,\\
\mbox{Im}\left[ \delta m_{n}  \right]  &=& \frac{1}{2} \mbox{Im}\left[\delta m_{n}^{+} +\delta m_{n}^{-} \right] .
\end{eqnarray}
as well as the corresponding spin-splitting functions, i.e.,
\begin{eqnarray}
\mbox{Im}\left[ \tilde{v}_{n}  \right]  &=& \frac{s_\perp}{2} \mbox{Im}\left[ \delta v_{\parallel,n}^{+} -\delta v_{\parallel,n}^{-} \right] ,\\
\mbox{Im}\left[ \tilde{m}_{n}  \right]  &=& \frac{s_\perp}{2} \mbox{Im}\left[\delta m_{n}^{+} -\delta m_{n}^{-} \right] .
\end{eqnarray}
As expected, all of these parameters, as well as $ \mbox{Im}\left[  \delta v_{\perp,n}  \right] $, are Landau-level dependent functions of the longitudinal momentum $p_z$.

\subsection{Self-energy in QCD plasma}
\label{parameters-numerics}

To demonstrate the proof of concept, here we study numerically the imaginary part of the self-energy functions in a hot magnetized QCD plasma. Keeping in mind their potential applications to heavy-ion physics, we will assume that the plasma temperature $T$ is of the order of $200~\mbox{MeV}$ to $400~\mbox{MeV}$ and the magnetic field is of the order of $|qB|\sim m_\pi^2$, where $m_\pi=135~\mbox{MeV}$. 

Because of different electric charges of the up and down quarks ($q_u=+2e/3$ and $q_d=-e/3$), the effect of a background magnetic field on their self-energies differs. Nevertheless, their dependence on $|qB|$ will remain essentially the same. (Strictly speaking, the roles of spin-up and spin-down states in the lowest Landau level will be interchanged because their charges have opposite signs.) Instead of considering the cases of up- and down-quarks separately, we will consider several fixed values of $|qB|$. This will suffice to demonstrate the qualitative effects of the magnetic field on the quark self-energy in the QCD plasma. We will also assume that the quark mass is the same for both flavors, i.e., $\bar{m}_{0}=5~\mbox{MeV}$. 

In the case of QCD plasma, the expressions for the self-energy functions have the same form as in Eqs.~(\ref{pars-1}) -- (\ref{pars-5}), but the coupling constant $\alpha$ should be replaced with $\alpha_s C_F$, where $ \alpha_s = g^2/(4\pi)$ and $C_F=(N_c^2-1)/(2N_c)=4/3$. To get an order of magnitude estimate, we will assume that the strong coupling is $\alpha_s \simeq 1/2$. In this case, $\alpha_s C_F = 2/3$. This choice is sufficient to get order of magnitude estimates. One could try to improve the approximation, for example, by incorporating the running of the coupling constant at the scale of temperature or the momentum transfer. For the purposes of the current proof-of-concept study, however, it is unnecessary. In any case, the overall benefit from this and other improvements is likely very limited. Because of the strong coupling in QCD, the quantitative validity of the one-loop correction will remain questionable. Thus, our numerical result should be interpreted with great caution and, at best, viewed as reasonable estimates rather than true quantitative results. 

When calculating the self-energy functions defined in Eqs.~(\ref{pars-1}) through (\ref{pars-5}), one needs to add up contributions from all three processes, sum over Landau level index $n^\prime$, and integrate over the transverse momentum $q_\perp$ in the appropriate kinematic range, see Eqs.~(\ref{proc-1}) -- (\ref{proc-3}). We will limit the analysis to the first $50$ Landau levels (i.e., $n\leq n_{\rm max}=50$). In this case, to achieve a good numerical precision in calculations, we include all transitions to Landau levels with the indices up to $n_{\rm max}^\prime = 100$. 

\begin{figure}
	\begin{center}
		{\includegraphics[width=0.31\textwidth]{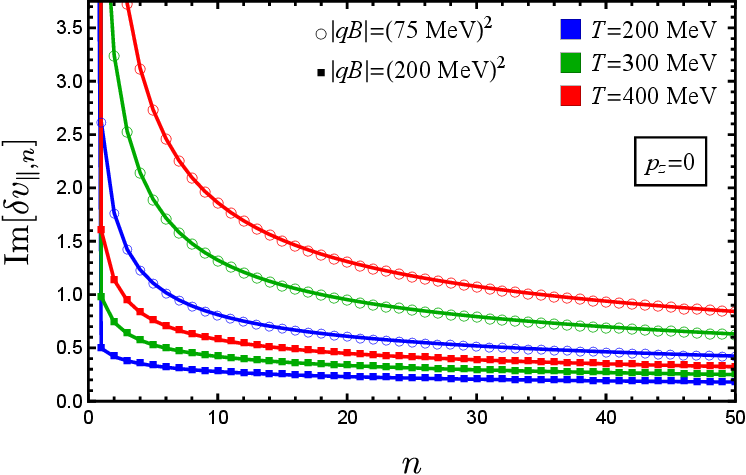}}
		\hspace{0.02\textwidth}
		{\includegraphics[width=0.31\textwidth]{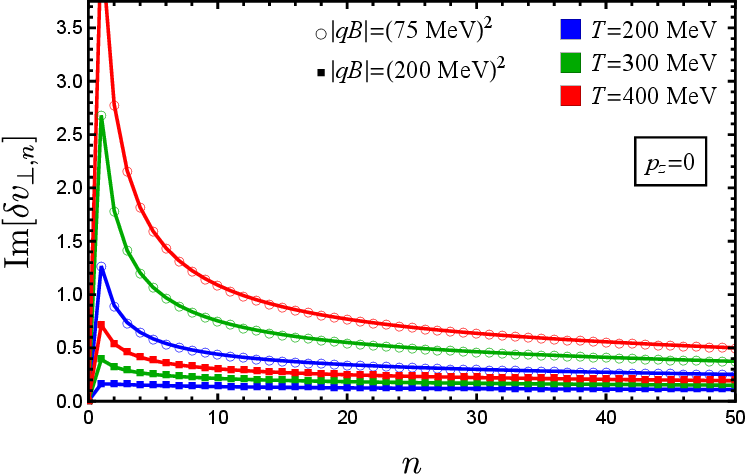}}
		\hspace{0.02\textwidth}
		{\includegraphics[width=0.31\textwidth]{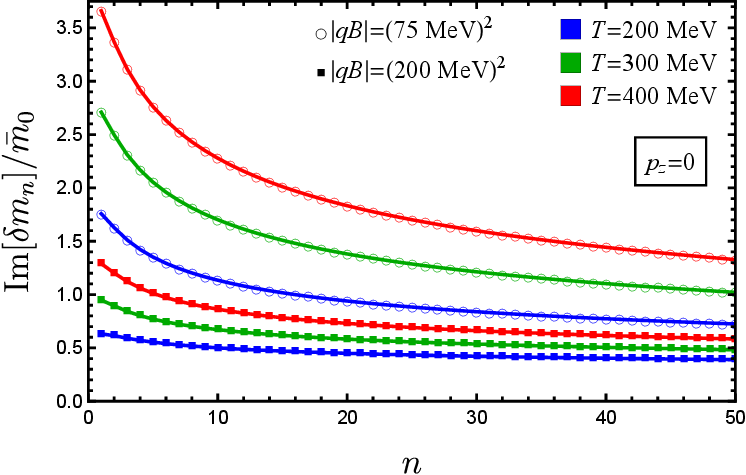}} \\[3mm]
		{\includegraphics[width=0.31\textwidth]{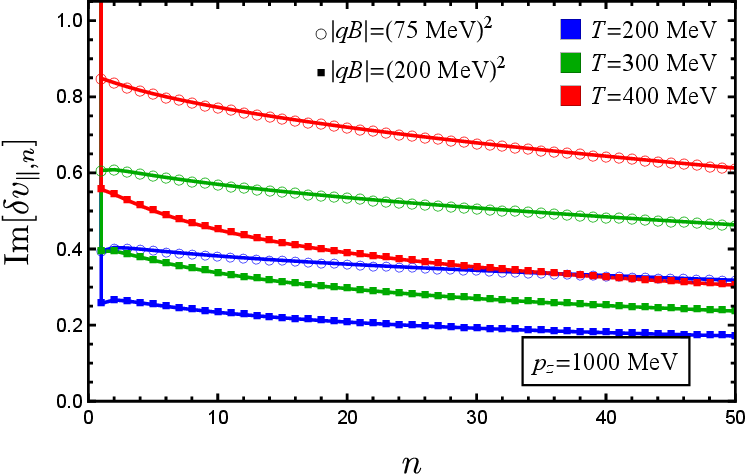}}
		\hspace{0.02\textwidth}
		{\includegraphics[width=0.31\textwidth]{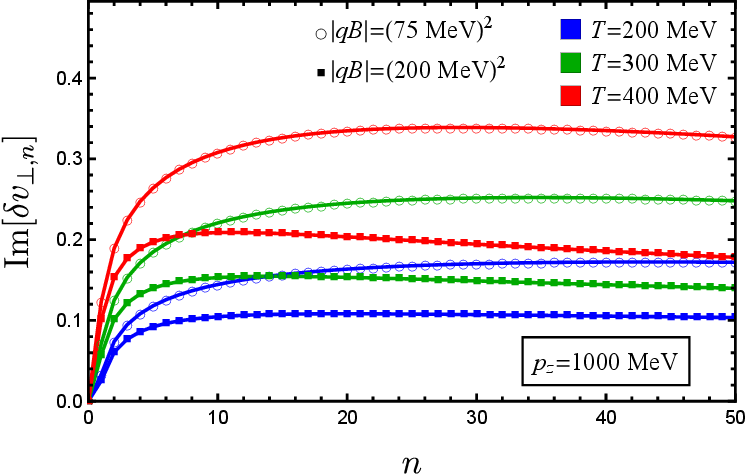}}
		\hspace{0.02\textwidth}
		{\includegraphics[width=0.31\textwidth]{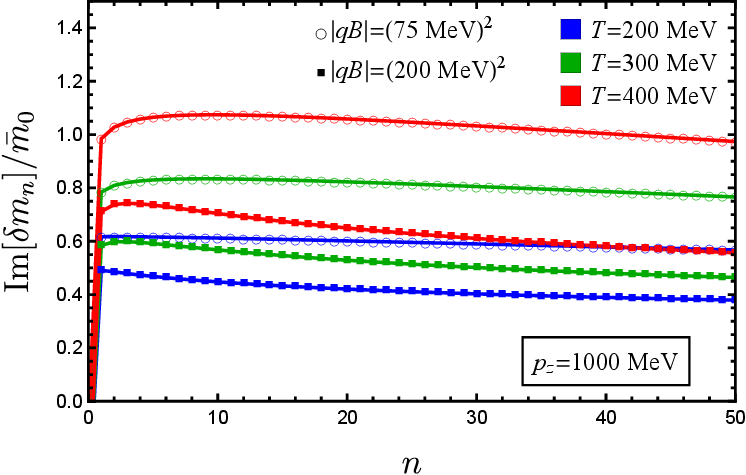}}
		\caption{The dependence of the self-energy functions $\mbox{Im}[\delta v_{\parallel,n}]$,  $\mbox{Im}[\delta v_{\perp,n}]$, and $\mbox{Im}[\delta m_{n}]/\bar{m}_{0}$ on  the Landau-level index $n$ for two fixed values of the longitudinal momentum: $p_z=0$ (top panels) and $p_z=1000~\mbox{MeV}$ (bottom panels). Each panel contains results for three different temperatures: $T=200~\mbox{MeV}$ (blue), $T=300~\mbox{MeV}$ (green), and $T=400~\mbox{MeV}$ (red); and two magnetic fields: $|qB|=(75~\mbox{MeV})^2$ (open circles) and $|qB|=(200~\mbox{MeV})^2$ (filled squares).}
		\label{3parameters}
	\end{center}
\end{figure}

The representative results for the imaginary parts of the velocity and the mass as functions of the Landau-level index $n$ are shown in Fig.~\ref{3parameters}. Each panel displays numerical data for three different temperatures, i.e., $T=200~\mbox{MeV}$ (blue lines),  $T=300~\mbox{MeV}$ (green lines),  $T=400~\mbox{MeV}$ (red lines),  and two different magnetic fields, i.e., $|qB|=(75~\mbox{MeV})^2$ (open circles),  $|qB|=(200~\mbox{MeV})^2$ (filled squares). The top three panels show the results for $p_z=0$, while the bottom three panels show the results for $p_z=1000~\mbox{MeV}$.

The multi-panel Fig.~\ref{3parameters} provides only a limited view of the numerical data for two fixed values of the longitudinal momentum. A large set of additional data for a wide range of $p_z$ values is included as Supplementary Data files \cite{DataFiles:2024}. Overall, we find that the imaginary parts of the velocity and mass functions tend to increase with the temperature and decrease with the magnetic field. Beyond these general tendencies, one finds that their dependence on the Landau-level index is nonmonotonous in general and differs at  small and large values of $p_z$. 

Since the imaginary parts of the Landau-level dependent velocity and mass functions have no clear physical meaning by themselves, we will not be discussing them in more detail. We note, however, that they are needed as an input to calculate the fermion damping rate. The latter will be discussed in the next section.

\section{Damping rate}
\label{damping_rate}

In quantum field theory without a background magnetic field, the fermion damping rate is related to the imaginary part of self-energy \cite{Weldon:1983jn}. In some recent studies, e.g., see Refs.~\cite{Hattori:2016cnt,Bandyopadhyay:2021zlm,Bandyopadhyay:2023hiv}, a similar formula was used rather heuristically in the case of magnetized plasmas. It should be noted, however, that no formal justification was given to utilize the Fourier transform of the translation invariant part of the self-energy in such calculations. Since the transverse momenta are not good quantum numbers in the field theory in a magnetic field, the underlying foundation of  Weldon's  arguments in Ref.~\cite{Weldon:1983jn} cannot be transferred to an unphysical representation. Below we provide a more rigorous derivation of the damping rate in terms of the self-energy in Landau-level representation. 

\subsection{Damping rate from the imaginary part of self-energy}
\label{sec:Damping1}

Following the general approach of Ref.~\cite{Weldon:1983jn}, we define the damping rate using the wave functions in coordinate space as follows:
\begin{equation}
\Gamma_n(p_z) =  \frac{1}{2p_0} \int d^4 u^\prime \int d^4 u\mbox{Tr}\left[ \frac{2\pi \ell^2}{V_\perp} \int dp \sum_s \bar{\Psi}_{n,p,s} (u^\prime)  \mbox{Im}\Sigma(u^\prime,u) \Psi_{n,p,s} (u)   \right]. 
\label{damping-rate-def}
\end{equation}
Note that $1/(2\pi  \ell^2)$ is the number of degenerate states per unit transverse area (excluding the spin degeneracy) and $V_\perp$ is the volume (area) of the transverse plane. Thus, $V_\perp/(2\pi  \ell^2)$ is the total number of such degenerate states. 

By making use of the fermion wave functions in a constant magnetic field, discussed in Appendix~\ref{Wave-functions}, we then derive 
\begin{eqnarray}
\Gamma_n(p_z) &=& \frac{1}{p_0} 
 \Bigg\{ \frac{\delta_{n,0}}{2} \left[p_\parallel^2  \mbox{Im}(\delta v_{\parallel,n} +s_\perp \tilde{v}_{n}) -\bar{m}_{0} \mbox{Im}(\delta m_{n} +s_\perp \tilde{m}_{n}) \right]
\nonumber\\
&&+\left(1- \delta_{n,0} \right) \left[ p_\parallel^2  \mbox{Im}(\delta v_{\parallel,n})  -\bar{m}_{0}  \mbox{Im}(\delta m_{n} ) -2n|qB|  \mbox{Im}(\delta v_{\perp,n})   \right]
  \Bigg\},
\label{damping-rate-ave}
\end{eqnarray}
where we used the result in Eq.~(\ref{SumPhi-PsiBar}). In the final expression, one should assume that the fermion is on the mass shell, i.e., $p_0=\sqrt{2n|qB|+\bar{m}_{0}^2+p_z^2}$. 

By definition, the Landau-level dependent fermion damping rate in Eq.~(\ref{damping-rate-ave}) is a spin-averaged quantity. Indeed, in the derivation, we summed up contributions of the spin states indiscriminately. In the presence of a nonzero magnetic field, however, the spin-degeneracy of each Landau level is likely to be lifted. Thus, the damping rates of the corresponding states are expected to be different. As we show in the next subsection, it is indeed the case. Moreover, we will be able to calculate the spin-dependent damping rates from the imaginary part of the one-loop self-energy. 

By substituting the results in Eqs.~(\ref{pars-1}) through (\ref{pars-5}) into the general expression for the rate (\ref{damping-rate-ave}), we derive the following damping rate in the zeroth Landau level:
\begin{eqnarray}
\Gamma_0(p_z) &=& \frac{\alpha |qB| }{4p_0}  
  \sum_{n^{\prime}=0}^\infty  \sum_{\{s\}}  \int   d\xi\,
  \left[n^\prime \mathcal{I}_{0}^{0,n^{\prime}-1}(\xi) 
  - \left(n^\prime +\bar{m}_{0}^2\ell^2 \right) \mathcal{I}_{0}^{0,n^{\prime}}(\xi)  \right]
  \frac{\left[ 1-n_F(s_1 E_{n^{\prime},k_z^{s^\prime}})+n_B(s_2 E_{q}) \right] }{s_1 s_2 \sqrt{ (\xi-\xi^{-})(\xi-\xi^{+})} } ,
\label{Gamma_0_pz}
\end{eqnarray}
where we used the identity $\xi  \mathcal{I}_{0}^{0,n^{\prime}-1}(\xi) = n^{\prime}  \mathcal{I}_{0}^{0,n^{\prime}}(\xi)$.
The expression for the damping rate in the higher Landau levels ($n\geq 1$) reads
\begin{eqnarray}
\Gamma_n(p_z) &=&\frac{\alpha |qB| }{4p_0}  
  \sum_{n^{\prime}=0}^\infty  \sum_{\{s\}}  \int  d\xi\, \left[
\mathcal{I}_{0}^{n,n^{\prime}-1}(\xi)  +\mathcal{I}_{0}^{n-1,n^{\prime}}(\xi)  \right]  \frac{ (n+n^\prime) \left[ 1-n_F(s_1 E_{n^{\prime},k_z^{s^\prime}})+n_B(s_2 E_{q}) \right] }{s_1 s_2 \sqrt{(\xi-\xi^{-})(\xi-\xi^{+}) }  } \nonumber\\
&-&\frac{\alpha}{4p_0}
  \sum_{n^{\prime}=0}^\infty  \sum_{\{s\}}  
\int  d\xi\,  \left[
\mathcal{I}_{0}^{n,n^{\prime}}(\xi)  +\mathcal{I}_{0}^{n-1,n^{\prime}-1}(\xi)  \right]  
  \frac{ \left( n+n^\prime + \bar{m}_{0}^2 \ell^2 \right)
  \left[ 1-n_F(s_1 E_{n^{\prime},k_z^{s^\prime}})+n_B(s_2 E_{q}) \right]  }{s_1 s_2  \sqrt{ (\xi-\xi^{-})(\xi-\xi^{+}) } }  .
\label{Gamma_n_pz}
\end{eqnarray}
Here, we introduced shorthand notations $\xi=q_\perp^2 \ell^2/2$ and $\xi^{\pm}=(q_\perp^{\pm})^2 \ell^2/2$, and  used Eq.~(\ref{Ep-relation}) to simplify the integrands. Also, to express $\mathcal{I}_{2}^{n-1,n^{\prime}-1}(\xi)$ in terms of $\mathcal{I}_{0}^{n,n^{\prime}}(\xi)$, we  used Eq.~(\ref{I2-I0}) from Appendix~\ref{app:kernels}. 

We can rewrite the above expressions for the damping rates in a form valid for all $n\geq 0$ as follows:
\begin{eqnarray}
\Gamma_n(p_z) &=&\frac{\alpha |qB|}{4p_0}  
  \sum_{n^{\prime}=0}^\infty  \sum_{\{s\}}  
\int   d\xi\,   \frac{{\cal M}_{n,n^{\prime}} (\xi) \left[ 1-n_F(s_1 E_{n^{\prime},k_z^{s^\prime}})+n_B(s_2 E_{q}) \right] }{s_1 s_2 \sqrt{(\xi-\xi^{-})(\xi-\xi^{+})} } ,
\label{Gamma_n_pz-short}
\end{eqnarray}
where we introduced the following function:
\begin{equation}
{\cal M}_{n,n^{\prime}}(\xi)  =-  \left(n+n^{\prime}+ \bar{m}_{0}^2\ell^2\right)\left[\mathcal{I}_{0}^{n,n^{\prime}}(\xi)+\mathcal{I}_{0}^{n-1,n^{\prime}-1}(\xi) \right]
+(n+n^{\prime}) \left[\mathcal{I}_{0}^{n,n^{\prime}-1}(\xi)+\mathcal{I}_{0}^{n-1,n^{\prime}}(\xi) \right]. 
\end{equation} 
As one can verify, the damping rate in Eq.~(\ref{Gamma_n_pz-short}) is a positive definite quantity. This is expected since Weldon's method \cite{Weldon:1983jn} should produce a result proportion to the squared amplitudes of the three underlying processes. As we show below, the same expression (\ref{Gamma_n_pz-short}) for the rate (after spin averaging) is obtained also from the poles of the propagator in Subsec.~\ref{sec:Damping2} below.

To further scrutinize the result in Eq.~(\ref{Gamma_n_pz-short}), we note that photon emission in a strongly magnetized plasma must be determined by the same squared amplitudes at the leading order in coupling. By making use of the analytical expression in Ref.~\cite{Wang:2021ebh}, we verified that the photon emission rate is indeed determined by the same function ${\cal M}_{n,n^{\prime}}(\xi)$.

\begin{figure}[t]
  {\includegraphics[width=0.47\textwidth]{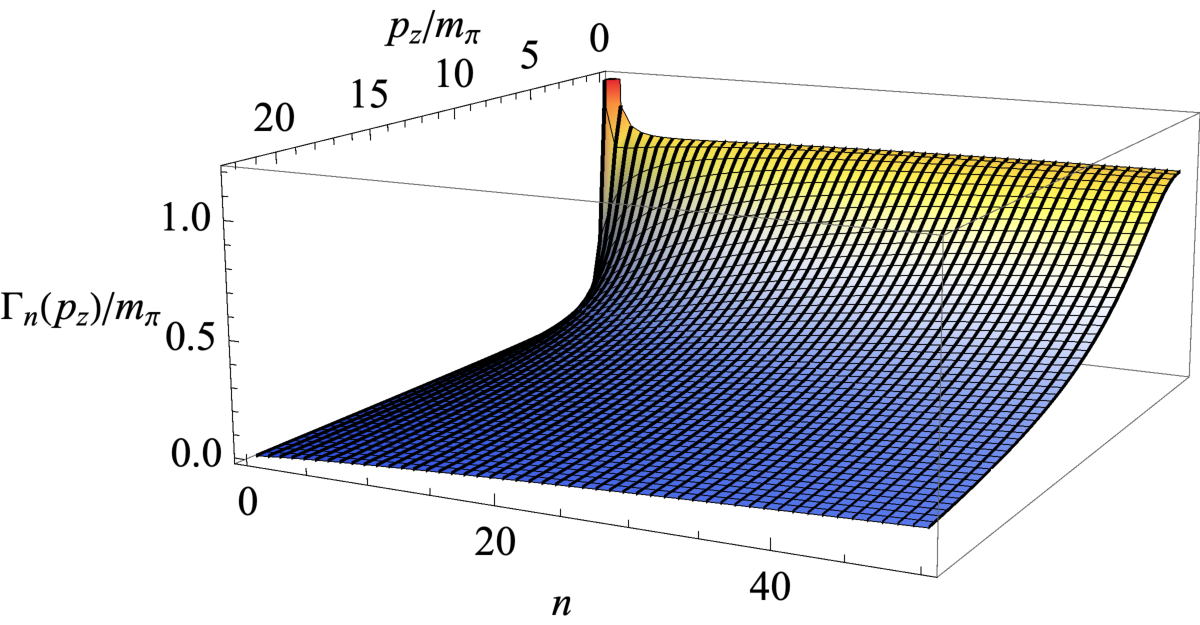}}
  \hspace{0.02\textwidth}
  {\includegraphics[width=0.47\textwidth]{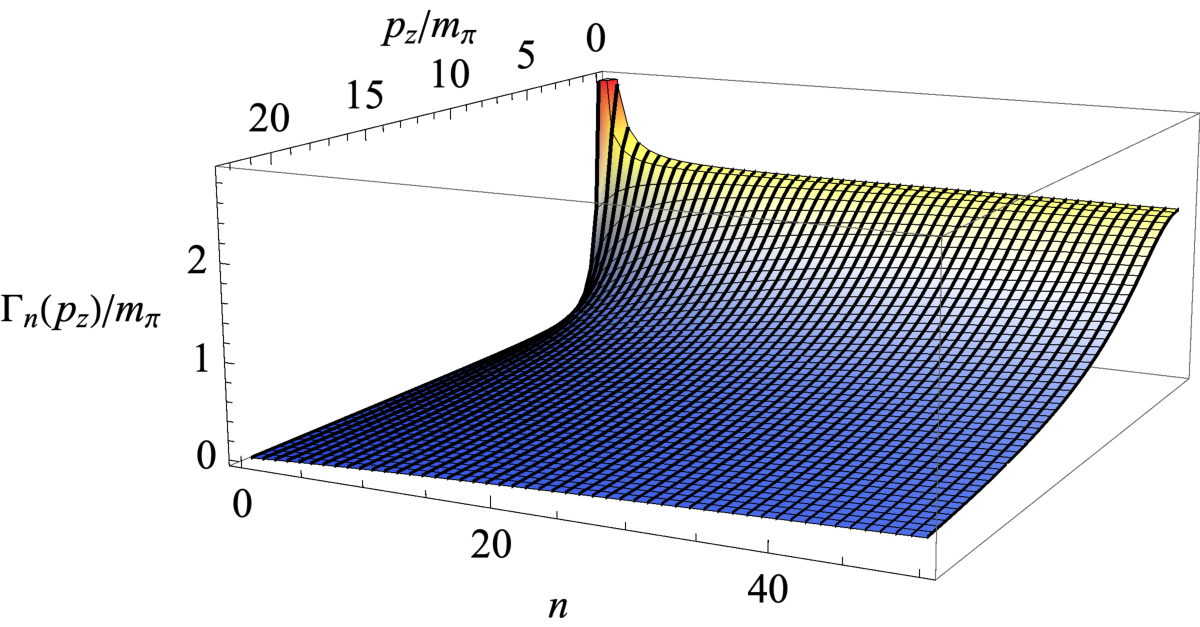}}
  \\[3mm]
  {\includegraphics[width=0.47\textwidth]{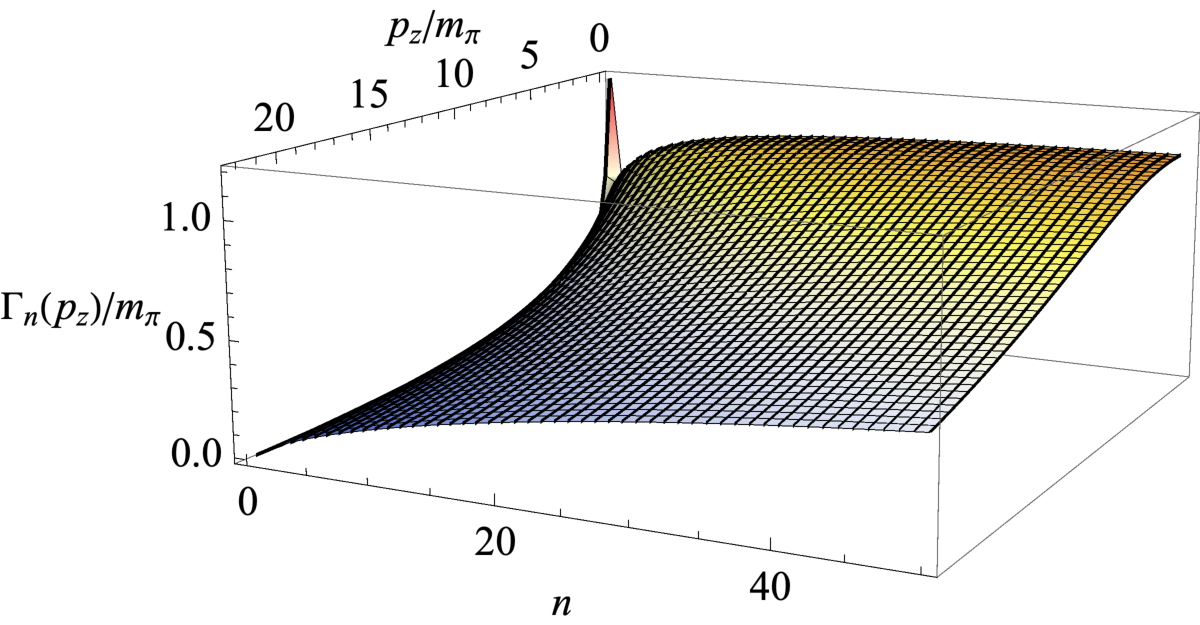}}
  \hspace{0.02\textwidth}
  {\includegraphics[width=0.47\textwidth]{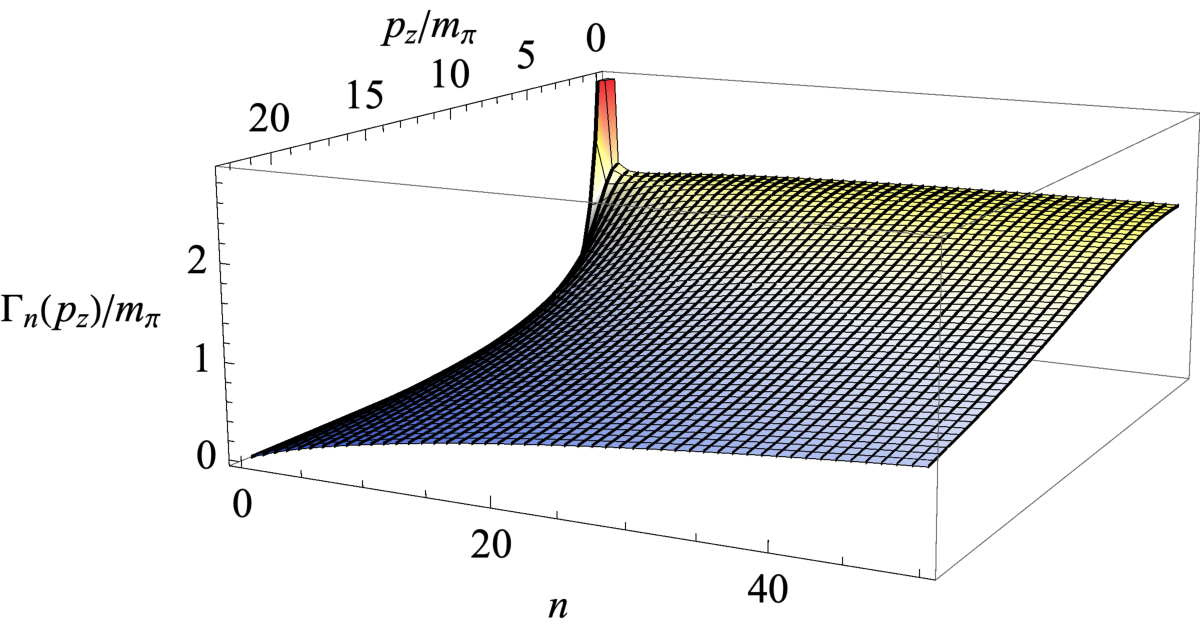}}
\caption{The fermion damping rate as a function of the longitudinal momentum $p_z$ and the Landau-level index $n$. The damping rate is measured in units of the pion mass. Four separate panels display results for two different temperatures, $T=200~\mbox{MeV}$ (left panels) and $T=400~\mbox{MeV}$ (right panels), and two magnetic fields, $|qB|=(75~\mbox{MeV})^2$ (top panels) and $|qB|=(200~\mbox{MeV})^2$ (bottom panels).}
\label{fig.DampingRate-units-of-mPi}
\end{figure}

The numerical results for the fermion damping rate (\ref{Gamma_n_pz-short})  as a function of the Landau-level index $n$ and the longitudinal momentum are shown in Fig.~\ref{fig.DampingRate-units-of-mPi}. Note that the values of the rate and the longitudinal momentum $p_z$ are given in units of the pion mass $m_\pi=135~\mbox{MeV}$. We use the same value of the QCD coupling as in Subsec.~\ref{parameters-numerics}. Four different panels display results for two different temperatures, i.e., $T=200~\mbox{MeV}$ (left panels) and $T=400~\mbox{MeV}$ (right panels), and two different magnetic fields, i.e., $|qB|=(75~\mbox{MeV})^2$ (top panels) and $|qB|=(200~\mbox{MeV})^2$ (bottom panels).

By comparing the compilation of numerical data in the four panels of Fig.~\ref{fig.DampingRate-units-of-mPi}, representing different temperatures and magnetic fields, we see that both temperature and magnetic field have a tendency to increase the damping rates. Such an enhancement is not surprising since both have the tendency to increase the phase space for transitions to other Landau levels. In connection to the magnetic field, in particular, its presence is critical to trigger the three processes responsible for the damping rate at the leading order in coupling. In the absence of the field, the only processes contributing to the fermion damping rate are of the subleading order in coupling. The findings are further reinforced by the numerical data for intermediate values of temperature, $T=300~\mbox{MeV}$, and magnetic field, $|qB|=(125~\mbox{MeV})^2$, which are not shown in the figures but included in the Supplementary Data files \cite{DataFiles:2024}. 

A careful analysis shows that the enhancement factors, resulting from increasing the temperatures and magnetic field, are nonuniform functions of the Landau-level index $n$ and longitudinal momentum $p_z$. For example, the increase of temperature from $T=200~\mbox{MeV}$ to $T=400~\mbox{MeV}$ leads to enhancement factors of the order of $2$ to $4$ in the whole region of $n$ and $p_z$ investigated. The largest increase is seen in the low-lying Landau levels at small longitudinal momenta.

The effect of the magnetic field is also nonuniform across the whole range of $n$ and $p_z$ values. Quantitatively, the increase of the magnetic field from $|qB|=(75~\mbox{MeV})^2$ to $|qB|=(200~\mbox{MeV})^2$ gives the largest enhancement factors of the order of 5 to 6, which occurs at large values of $p_z$ and small $n$. While, in absolute terms, the damping rates are the highest at small values of $p_z$, the increase due to the magnetic field is moderate (of the order of 2 or less). In fact, when both $n$ and $p_z$ are small, we find that the rate can even decrease by a factor of about 2 or less. We should note, however, that this part of the parameter space must be treated with great caution because of a limited validity of the one-loop approximation. 

\begin{figure}[t]
  {\includegraphics[width=0.47\textwidth]{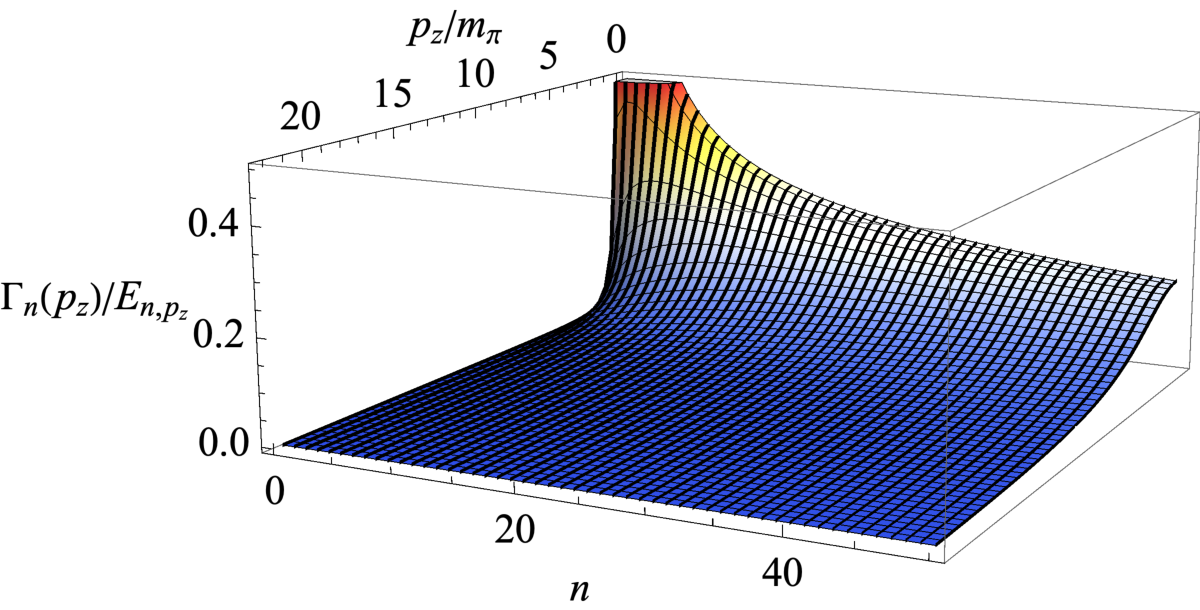}}
  \hspace{0.02\textwidth}
  {\includegraphics[width=0.47\textwidth]{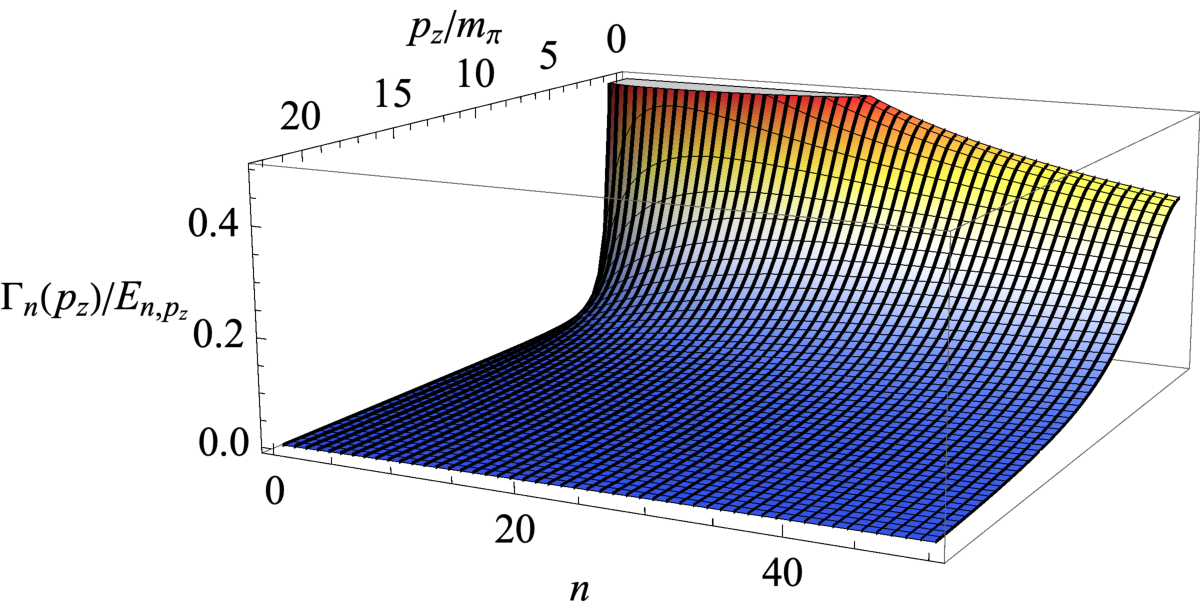}}
  \\[3mm]
  {\includegraphics[width=0.47\textwidth]{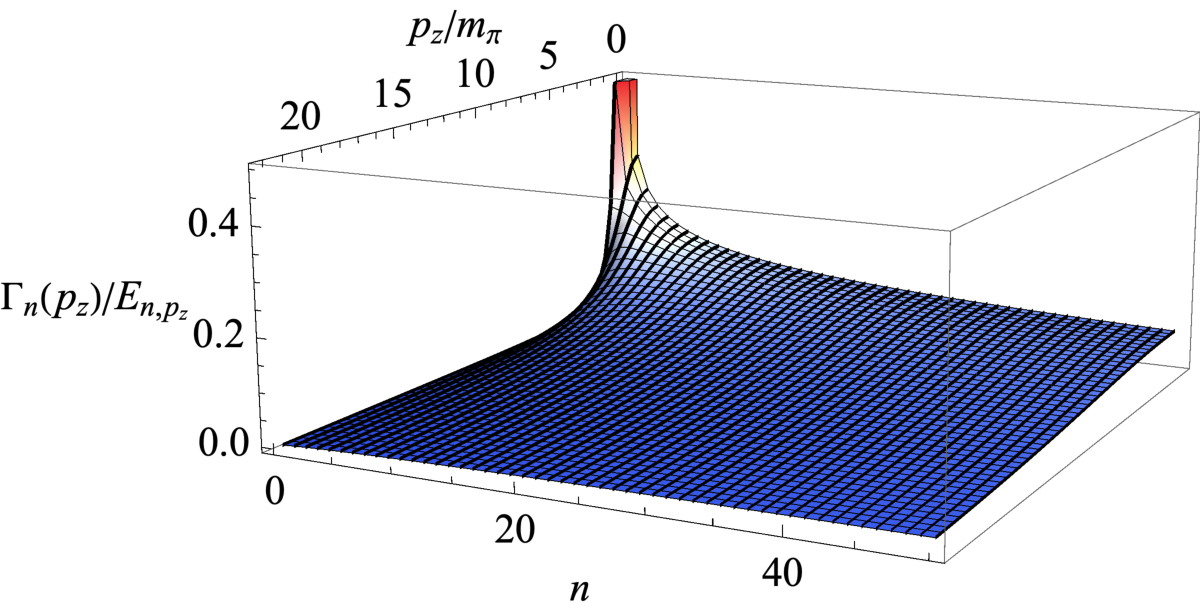}}
  \hspace{0.02\textwidth}
  {\includegraphics[width=0.47\textwidth]{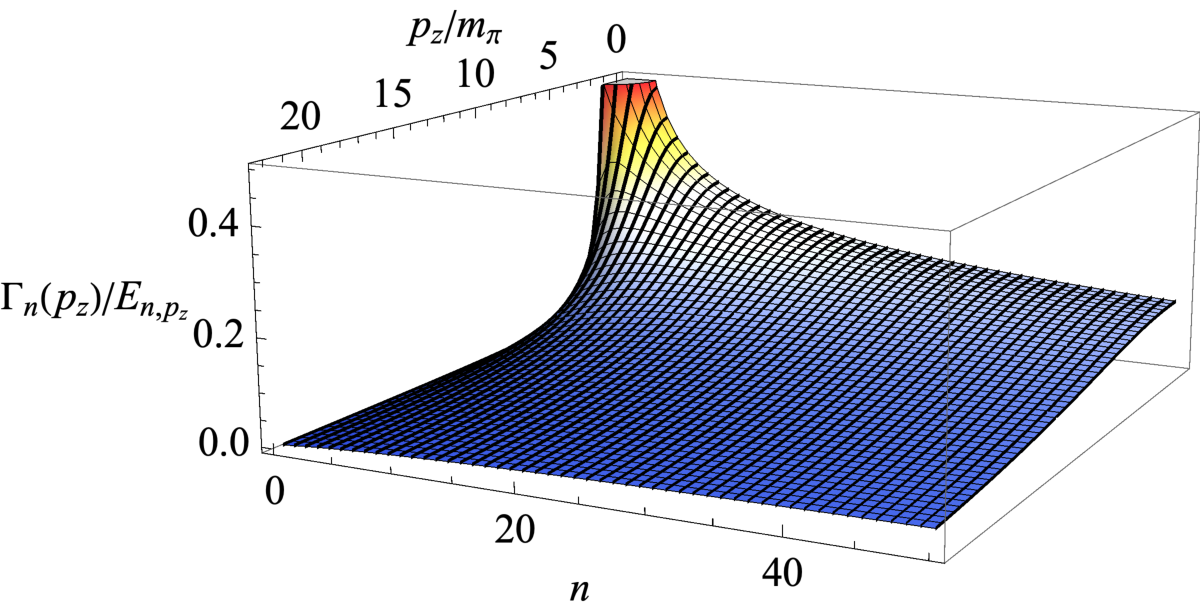}}
\caption{The ratio of the fermion damping rate to its energy as a function of the longitudinal momentum $p_z$ and the Landau-level index $n$. Four separate panels display results for two different temperatures, $T=200~\mbox{MeV}$ (left panels) and $T=400~\mbox{MeV}$ (right panels), and two magnetic fields, $|qB|=(75~\mbox{MeV})^2$ (top panels) and $|qB|=(200~\mbox{MeV})^2$ (bottom panels).}
\label{fig.DampingRate-units-of-Energy}
\end{figure}

Before proceeding further, it is instructive to investigate the ratio of the damping rate and the real part of the fermion energy, $\Gamma_n(p_z)/E_{n,p_z}$. Note that the knowledge of the real part of particle energy at the zeroth order is sufficient for calculating first-order corrections to $\Gamma_n(p_z)/E_{n,p_z}$. The corresponding results are presented in Fig.~\ref{fig.DampingRate-units-of-Energy}. The four panels correspond to the same choices of two temperatures and two magnetic fields. In essence, this is the ratio of the imaginary and real parts of the fermion energy that shows whether the quantum state (with given $n$ and $p_z$) is a well-defined quasiparticle. When the ratio value is comparable to $1$ or larger, the quasiparticle description is inapplicable. Indeed, this is the case when the particle's lifetime $\tau_n  = 1/\Gamma_n$ is comparable to or shorter than the time needed to measure its energy $\Delta t \lesssim 1/E_{n,p_z}$ according to the uncertainty principle. Alternatively, the uncertainty in particle's energy $\Gamma_n$ is larger than the energy $E_{n,p_z}$ itself. 

As we see from Fig.~\ref{fig.DampingRate-units-of-Energy}, the ratio $\Gamma_n(p_z)/E_{n,p_z}$ remains small almost in the whole range of $n$ and $p_z$ values. However, the damping rate becomes very large in the lowest few Landau levels ($n\lesssim 1$) when the longitudinal momentum $p_z$ is sufficiently small ($p_z\lesssim m_\pi$). Formally, these results indicate that the concept of well-defined quasiparticles breaks down for the corresponding lowest Landau level states. We believe this might be a premature conclusion, however. It seems more likely that the validity of the perturbative one-loop calculation breaks down in this case. Because of the high degeneracy of the Landau levels, it is plausible that the one-loop calculation breaks down, especially in the region of small fermion energies. 

As in the absence of a magnetic field, hard thermal loop resummations might be very important in the strongly magnetized QCD plasma \cite{Braaten:1989mz}. Additionally, somewhat similar hard magnetic loop resummation \cite{Miransky:2015ava} may be needed when there is a strong magnetic field. Both are very likely to affect the self-energy at small energies. Therefore, we reiterate that the large damping rates at small $n$ and $p_z$ should be accepted with great caution. Most likely, the corresponding results are outside of the range of validity of the approximations used. Qualitatively, however, it is intriguing to think that the damping rates can be indeed large in the low-lying Landau levels. They could dramatically affect some observables in heavy-ion collisions, e.g., the electrical conductivity of plasma \cite{Hattori:2016cnt,Ghosh:2024fkg} and the heavy-quark energy loss and dissipation rate \cite{Bandyopadhyay:2021zlm,Bandyopadhyay:2023hiv}.

\subsection{Damping rates from the poles of the propagator}
\label{sec:Damping2}

In the previous subsection, we used the definition of the damping rate in terms of the imaginary part of the self-energy by generalizing the general approach of Ref.~\cite{Weldon:1983jn} to the case of quantum field theory in a quantizing magnetic field. Here we consider an alternative definition that follows from the structure of the full propagator, calculated in the one-loop approximation. 

When the full propagator is known, the fermion damping rate can be also determined from the location of its poles in the complex energy plane. At the leading order in coupling, the explicit structure of the fermion propagator is derived in Appendix~\ref{app:propagator}. As expected, the self-energy functions $v_{\parallel,n}$, $m_{n}$, $v_{\perp,n}$, $\tilde v_{n}$, and $\tilde m_{n}$ modify the fermion propagator, see Eqs.~(\ref{app:G-mixed-2}), (\ref{1overM-2nqB}) and (\ref{Un}). Most importantly for our purposes here, one can extract the quasiparticles energies from the location of the poles in the propagator, see Eq.~(\ref{p0_sol}). Assuming that the self-energy corrections are small, the approximate expressions for the (positive) energies can be written as follows:
\begin{eqnarray}
p_0^{(\pm)}&\simeq& \sqrt{2n|qB|+ \bar{m}_{0}^2+p_z^2}\left(1+\frac{ \bar{m}_{0}\delta m_n
- (2n|qB|+ \bar{m}_{0}^2)\delta v_{\parallel,n} + 2n|qB|\delta v_{\perp,n} \pm \sqrt{2n|qB|+ \bar{m}_{0}^2} (\bar{m}_{0}\tilde{v}_n-\tilde{m}_n) }{2n|qB|+ \bar{m}_{0}^2+p_z^2}\right).
\label{p0_expansion}
\end{eqnarray}
Note that there are two different branches of solutions that correspond to two spin states. Recall that the corresponding two states were degenerate in the free propagator. However, already at the leading order in coupling, the degeneracy is lifted by the self-energy corrections $\tilde{v}_n$ and $\tilde{m}_n$. Since we did not calculate explicitly the real parts of the self-energy functions $v_{\parallel,n}$, $m_{n}$, $v_{\perp,n}$, $\tilde v_{n}$, and $\tilde m_{n}$, we cannot quantify the corresponding corrections to the real parts of particle energies. 

Nevertheless, using the imaginary parts of self-energy functions, see Eqs.~(\ref{pars-1}) through (\ref{pars-5}), we can determine leading-order corrections to the imaginary parts of particle energies, i.e., $\mbox{Im}[\delta p_{0,n}^{(\pm)}]$. Since the latter should coincide up an  overall sign with the damping rate, we derive
\begin{eqnarray}
\Gamma_n^{(\pm)} &\simeq& \frac{ (2n|qB|+ \bar{m}_{0}^2)\mbox{Im}[\delta v_{\parallel,n} ]-\bar{m}_{0} \mbox{Im}[\delta m_n]
- 2n|qB|\mbox{Im}[\delta v_{\perp,n}]\mp \sqrt{2n|qB|+ \bar{m}_{0}^2} (\bar{m}_{0}\mbox{Im}[\tilde{v}_n]-\mbox{Im}[\tilde{m}_n])}{\sqrt{2n|qB|+ \bar{m}_{0}^2+p_z^2}}.
\label{damping-rate-pm}
\end{eqnarray} 
As expected, this result demonstrates that the two spin-split Landau-level states have different damping rates. At the same time, it is rewarding to see that the spin-averaged damping rate, $\Gamma_{n}^{\rm (ave)} \equiv (\Gamma_n^{(+)}+\Gamma_n^{(-)})/2$, agrees perfectly with the result obtained by a very different method in the previous subsection, see Eq.~(\ref{damping-rate-ave}).

\begin{figure}
	\begin{center}
		\includegraphics[width=0.47\textwidth]{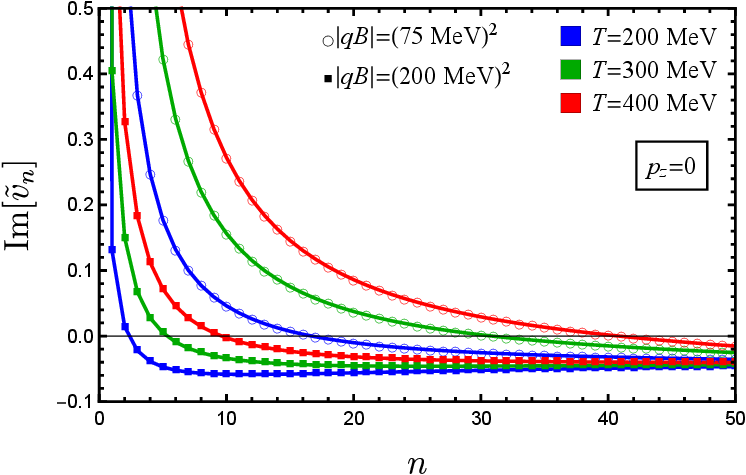}
		\hspace{0.02\textwidth}
		\includegraphics[width=0.47\textwidth]{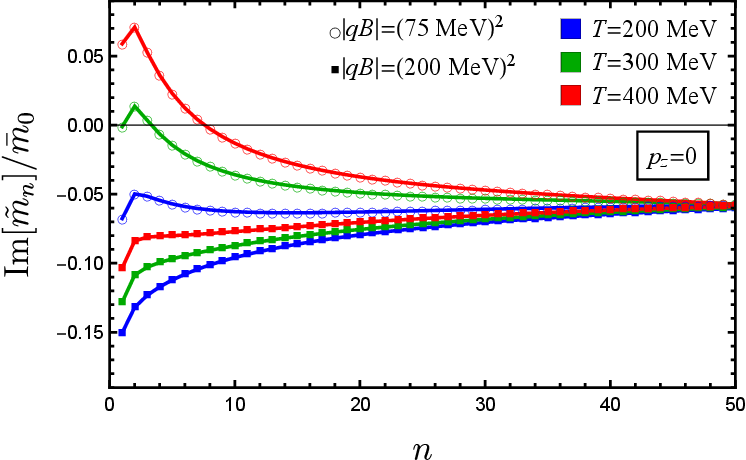}\\[3mm]
		\includegraphics[width=0.47\textwidth]{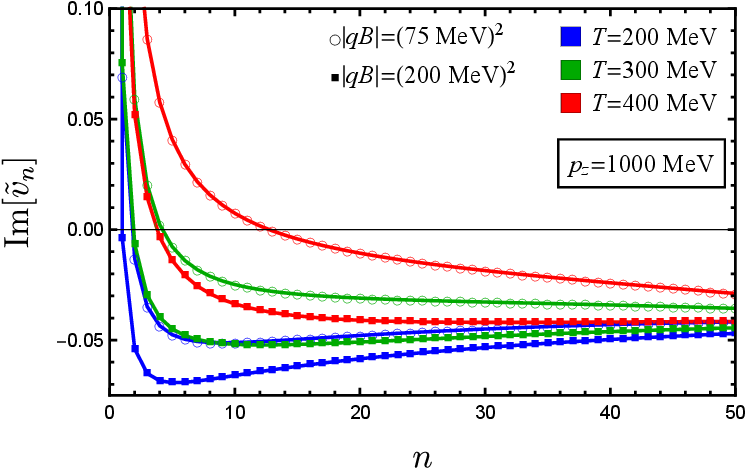}
		\hspace{0.02\textwidth}
		\includegraphics[width=0.47\textwidth]{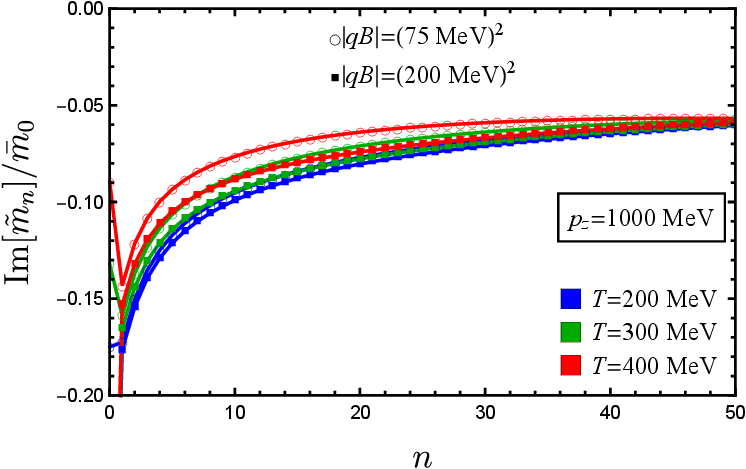}
		\caption{The dependence of the self-energy functions $\mbox{Im}[\tilde v_{n}]$ and $\mbox{Im}[\tilde m_{n}]/\bar{m}_{0}$ on  the Landau-level index $n$ for two fixed values of the longitudinal momentum: $p_z=0$ (top panels) and $p_z=1000~\mbox{MeV}$ (bottom panels). Each panel contains results for three different temperatures: $T=200~\mbox{MeV}$ (blue), $T=300~\mbox{MeV}$ (green), and $T=400~\mbox{MeV}$ (red); and two magnetic fields: $|qB|=(75~\mbox{MeV})^2$ (open circles) and $|qB|=(200~\mbox{MeV})^2$ (filled squares).}
\label{2parameters}
	\end{center}
\end{figure}

It is natural to ask how large the spin splitting effects on the quasiparticle damping rate are. As we see from Eq.~(\ref{damping-rate-pm}), they are determined by the self-energy functions $\mbox{Im}[\tilde v_{n}]$ and $\mbox{Im}[\tilde m_{n}]$. The representative results for both, as functions of the Landau-level index $n$, are shown in Fig.~\ref{2parameters}. Each panel displays numerical data for three different temperatures, i.e., $T=200~\mbox{MeV}$ (blue lines),  $T=300~\mbox{MeV}$ (green lines),  $T=400~\mbox{MeV}$ (red lines),  and two different magnetic fields, i.e., $|qB|=(75~\mbox{MeV})^2$ (open circles),  $|qB|=(200~\mbox{MeV})^2$ (filled squares). The top panels show the results for $p_z=0$, while the bottom panels show the results for $p_z=1000~\mbox{MeV}$. Since the imaginary parts of $\tilde{v}_n$ and $\tilde{m}_n$ themselves have no direct physical meaning, there is no need to display more data here. However, an interested reader could find a large set of additional data for a wide range of $p_z$ values in the Supplementary Data files \cite{DataFiles:2024}. 

\begin{figure}
	\begin{center}
		\includegraphics[width=0.47\textwidth]{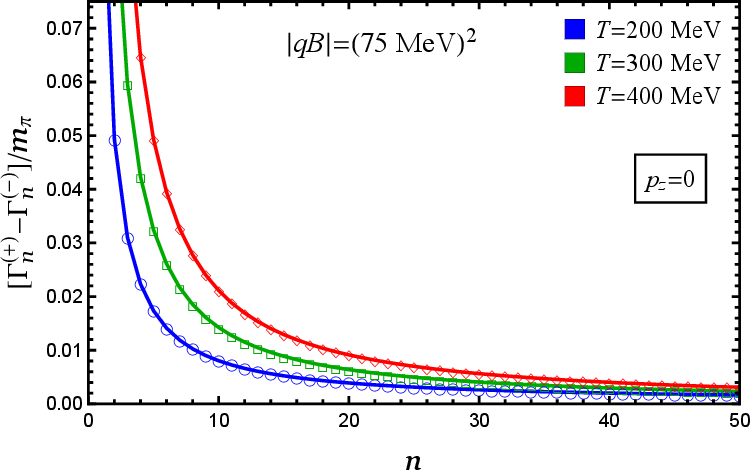}
		\hspace{0.02\textwidth}
		\includegraphics[width=0.47\textwidth]{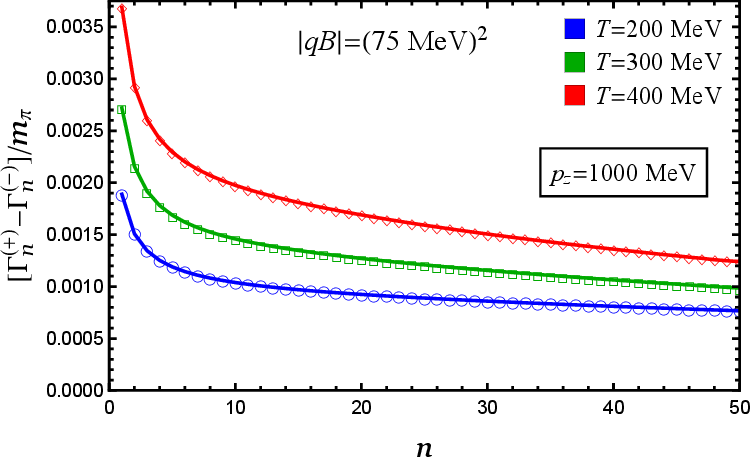}
\caption{The spin-splitting of damping rates as functions of Landau-level index $n$ for two fixed values of the longitudinal momentum: $p_z=0$ (left panel) and $p_z=1000~\mbox{MeV}$ (right panel). The magnetic field is $|qB|=(75~\mbox{MeV})^2$. Each panel contains results for three different temperatures: $T=200~\mbox{MeV}$ (blue), $T=300~\mbox{MeV}$ (green), and $T=400~\mbox{MeV}$ (red). }
\label{fig.DampingRates-pm}
	\end{center}
\end{figure}

Let us now turn to the spin-splitting effects on the damping rates. Two sets of representative results are shown in Fig.~\ref{fig.DampingRates-pm}. We display the difference between the rates of the spin up and down states as functions of the Landau-level index $n$. The two panels display the results for the same (smallest) value of the magnetic field, $|qB|=(75~\mbox{MeV})^2$ but two different longitudinal momenta, $p_z=0$ (left panel) and $p_z=1000~\mbox{MeV}$ (right panel). The data for three different temperatures are represented by different colors. By comparing the magnitude of spin splitting with the average damping rates in Fig.~\ref{fig.DampingRate-units-of-mPi}, we see that the effect of spin splitting is really small. The same is true for other values of the magnetic field. Quantitatively, a typical difference between the rates of the spin up and down states is of the order of a few percent of the average rate or less. However, it may reach up to about $10\%$ in low-lying Landau levels at small longitudinal momenta. In general, we find that the relative spin splitting decreases with increasing of the magnetic field. Therefore, one can argue that, for most purposes, it is sufficient to use the spin-averaged damping rate, $\Gamma_{n}^{\rm (ave)} \equiv (\Gamma_n^{(+)}+\Gamma_n^{(-)})/2$, which was investigated in detail in the previous subsection. This argument can be further reinforced by the observation that systematic uncertainties of the one-loop approximation used in the study are probably larger than the effects of spin splitting. 
 
In conclusion of this section, let us emphasize that the spin splitting is a qualitatively new feature that can play an important role in strongly magnetized plasmas. While the differences between the damping rates for spin-split states in each Landau level remain quantitatively small, they may affect some spin physics phenomena, chiral magnetic or chiral separation effects. In this connection, it should be emphasized that not only the imaginary parts of the Landau-level energies but also their real parts will be spin split. While we did not calculate the latter, such a conclusion is supported by the general expression for the self-energy derived.

\section{Discussion and Summary}
\label{Summary}

In this paper we derived a general expression for the fermion self-energy in a hot and strongly magnetized plasma by using the Landau-level representation. As we show, the leading-order one-loop expression for the self-energy is characterized by three velocity and two mass functions. The velocity functions include a pair of spin-split parallel components and a perpendicular component of the velocity. The other two functions are the masses of the spin-split pair of states in each Landau level. As we demonstrated, all of these five functions have a nontrivial dependence on the Landau-level index $n$ and the longitudinal momentum $p_z$. 

Here we focused primarily on the imaginary (dissipative) part of the fermion self-energy. We derived closed-form expressions for the imaginary parts of all five  functions that define the Dirac structure of the self-energy. At the leading order in coupling, the contributions to the imaginary parts of the velocity and mass functions in the $n$th Landau level come from the following three types of on-shell processes: (i) transitions to other Landau levels with lower indices $n^\prime$  ($\psi_{n}\to \psi_{n^\prime} +\gamma$ with $n>n^\prime$), (ii) transitions to other Landau levels with higher indices $n^\prime$ ($\psi_{n}+\gamma\to \psi_{n^\prime} $ with $n<n^\prime$), and (iii) transitions to Landau-level states with negative energies (i.e., the annihilation process $\psi_{n}+\bar{\psi}_{n^\prime}\to \gamma$ for any $n$ and $n^\prime$). 

We used the imaginary parts of the self-energy functions to derive the Landau-level dependent fermion damping rates $\Gamma_{n}(p_z)$. We employed two different methods to get the corresponding results. On one hand, we obtain the damping rate by utilizing the general approach of Weldon \cite{Weldon:1983jn}. To apply it to the case of hot plasma in a quantizing magnetic field, first we had to modify the method to account for the correct set of quantum numbers characterizing the Landau-level states. As expected, the final result is expressed in terms of the imaginary parts of the spin-averaged velocity and mass functions, see Eq.~(\ref{damping-rate-ave}). 

The second method for extracting the damping rates used the location of the poles in the full propagator. This approach revealed that the two-spin degeneracy of the Landau level states was lifted by radiative corrections. Furthermore, by using the imaginary parts of particle energies, we were able to extract the damping rates for the spin-split states $\Gamma_{n}^{(\pm)}(p_z)$. It is important to note that the spin-averaged rate, $\Gamma_{n}^{\rm (ave)} \equiv (\Gamma_n^{(+)}+\Gamma_n^{(-)})/2$, agrees perfectly with the result obtained by Weldon's method. Since the effect of spin splitting on the rate is not large, one may argue that the use of Weldon's method might be sufficient in most applications.

The analytical expression for the damping rate in Eq.~(\ref{Gamma_n_pz-short}) is remarkable in many ways. It defines a positive definite damping rate as a function of the Landau-level $n$ and the longitudinal momentum $p_z$. We also showed that it is determined by the same amplitudes that appear in photon emission from a magnetized plasma.

To demonstrate the Landau-level dependent description of the self-energy effects, we studied numerically the fermion damping rates in a wide range of model parameters, considering three different temperatures and three different magnetic fields. The choice of model parameters, with temperatures between $200~\mbox{MeV}$ and $400~\mbox{MeV}$ and magnetic fields of the order of $m_\pi^2$, were motivated by potential applications in heavy-ion physics. The main results are summarized in Figs.~\ref{fig.DampingRate-units-of-mPi} and \ref{fig.DampingRate-units-of-Energy}. In absolute terms, the largest values of the rates are found for the low-lying Landau levels and small values of the longitudinal momentum. In fact, in some cases (at small $n$ and $p_z$), the damping rates appear to be formally much larger than the real parts of the particle energies. This suggests that the quasiparticle picture may fail for such quantum states. These extreme cases should be treated with great caution, however, since the one-loop approximation may become particularly bad in those regions of the parameter space. 

Generally, we find that the rates have an overall tendency to grow with increasing both temperature and magnetic field. However, the enhancement is nonuniform in the range of Landau-level indices $n$ and longitudinal momenta $p_z$ explored. The thermal effects are pronounced the most in the region of small values of $n$. The magnetic field enhancement, in contrast, is most prominent at large values of $p_z$. The latter may not be as surprising after one recalls that the magnetic field is essential for allowing the leading-order, one-photon processes (i.e., $\psi_{n}\to \psi_{n^\prime} +\gamma$,  $\psi_{n}+\gamma\to \psi_{n^\prime} $,  and $\psi_{n}+\bar{\psi}_{n^\prime}\to \gamma$) to occur in the first place.

We hope that the results for the fermion damping rates, as well as the general method for calculating the self-energy in the Landau-level representation, can be useful in a wide range of studies of strongly magnetized relativistic plasma. They can be useful in the calculation of transport properties such as the electrical conductivity \cite{Hattori:2016cnt,Ghosh:2024fkg} and the particle loss or dissipation rate \cite{Bandyopadhyay:2021zlm,Bandyopadhyay:2023hiv}. In addition to heavy-ion physics, our self-energy results can be useful in studies of QED plasmas in astrophysics and cosmology.

While this study provides a clear proof of concept for utilizing the Landau-level representation to describe self-energy effects in strongly magnetized relativistic plasmas, there are many theoretical issues left outstanding. The most obvious of them is the calculation of the real part of the self-energy. Unlike the imaginary part, the expression for the real part contains ultraviolet divergences. Therefore, its evaluation requires a careful renormalization procedure, which is complicated by the Landau-level structure of the self-energy. Despite these difficulties, we believe, the problem can be solved by using the general expression for the self-energy derived here as the starting point. We plan to consider this problem in the follow-up studies.

\acknowledgments{This research was funded in part by the U.S. National Science Foundation under Grant No.~PHY-2209470.}

\appendix

\section{Fermion propagator in the Landau-level representation}
\label{app:propagator}

In this Appendix, we derive an explicit form of the fermion propagator in a magnetic field in the Landau-level representation by using the method developed in Ref.~\cite{Miransky:2015ava}. By definition, the corresponding propagator in coordinate space is given by the following matrix element: 
\begin{eqnarray}
  G(u,u')  &=& i  \langle u | \left[  (i\partial_t \gamma^0-\pi^3 \gamma^3)
  -  (\bm{\pi}_\perp \cdot\bm{\gamma}_\perp) -\bar{m}_{0} -\Sigma
  \right]^{-1} | u^\prime\rangle \nonumber\\
  &=& i  \langle u | \left[ v_\parallel (i\partial_t \gamma^0-\pi^3 \gamma^3)
  - v_\perp (\bm{\pi}_\perp \cdot\bm{\gamma}_\perp) 
   + i \gamma^1\gamma^2 \tilde{v}  (i\partial_t \gamma^0-\pi^3 \gamma^3)
  -m  - i\gamma^1\gamma^2 \tilde{m}
  \right]^{-1} | u^\prime\rangle  ,
\end{eqnarray}  
where $\bm{\pi}_\perp = -i \left(\bm{\nabla} -i q \bm{A}\right)$ and the vector potential in the Landau gauge is used, i.e., $\bm{A}=(0,Bx,0)$. Here we took into account all possible Dirac structures of the full propagator at the leading order in coupling. In particular, functions $m$, $v_\parallel$ and $v_\perp$ include radiative corrections to the mass, the parallel and perpendicular components of the velocity, respectively. The two additional functions $ \tilde{v}$ and $\tilde{m}$ capture the effects of spin splitting corrections to the parallel velocity and the mass. A self-consistency check shows that there is no spin splitting correction to $v_\perp$. Note that, strictly speaking, all five are operator-valued functions. When acting on the Landau-level orbitals (see below), they will become functions of the Landau-level index $n$ and the longitudinal momentum $p_\parallel$. For example, $m_n$ will be the mass function in the $n$th Landau level. (Note that we use notation $\bar{m}_{0}$ for the tree-level mass to distinguish it from the mass in the lowest Landau level $m_0$.)

Considering that translation symmetry remains intact in the time and the $z$ direction, it is convenient to switch to the corresponding momentum subspace represented by the longitudinal momentum $p_{\parallel} = (p_0, p_z)$. The resulting propagator in a mixed representation reads
\begin{eqnarray}
  G(p_\parallel, u_\perp,u^\prime_\perp)   &=&\langle u_\perp | \left[ v_\parallel (p_\parallel \cdot \gamma_\parallel)
  - v_\perp (\bm{\pi}_\perp \cdot\bm{\gamma}_\perp) 
  - i \gamma^1\gamma^2 \tilde{v}  (p_\parallel \cdot \gamma_\parallel)
  +m  - i\gamma^1\gamma^2 \tilde{m}
  \right]\nonumber\\ 
  &\times& \Big[\left( v_\parallel^2 -\tilde{v}^2 \right) p_\parallel^2-v_\perp^2 \bm{\pi}^2-m^2  +\tilde{m}^2
 + 2 i \gamma^1\gamma^2(m\tilde{v} - \tilde{m} v_\parallel)(p_\parallel \cdot \gamma_\parallel)
 -i \gamma^1\gamma^2 v_\perp^2 qB  \Big]^{-1} | u^\prime_\perp \rangle.
 \label{app:G-mixed-1}
 \end{eqnarray}
Here we took into account that  $ -(\bm{\pi}_\perp\cdot\bm{\gamma}_\perp)^2=\bm{\pi}_\perp^{2}-qBi\gamma^1\gamma^2$. The  eigenvalues of the operator $\bm{\pi}_\perp^{2}$ are $(2n+1)|qB|$, where $n=0,1,2,\dots$, and the corresponding normalized eigenfunctions are given by the Landau orbitals, i.e.,
\begin{equation}
 \psi_{np}( u_{\perp})\equiv \langle  u_{\perp} | n p \rangle=\frac{1}{\sqrt{2\pi \ell}}\frac{1}{\sqrt{2^nn!\sqrt{\pi}}}
 H_n\left(\frac{x}{\ell}+p\ell\right)e^{-\frac{1}{2\ell^2}(x+p\ell^2)^2} e^{- is_\perp py },
\end{equation}
where $s_{\perp}=\mbox{sign}(qB)$, $\ell=1/\sqrt{|qB|}$ is the magnetic length, and $H_{n}(x)$ are the Hermite polynomials. It is useful to note that
\begin{eqnarray}
\pi_x\psi_{n,p}( u_{\perp}) &=& -i\partial_x \psi_{n,p}( u_{\perp})= \frac{i}{2\ell}\left(\sqrt{2(n+1)}\psi_{n+1,p}( u_{\perp})-\sqrt{2n}\psi_{n-1,p}( u_{\perp})\right),   \label{property1} \\
\pi_y \psi_{n,p}( u_{\perp}) &=& \left(-i\partial_y -qBx \right)\psi_{n,p}( u_{\perp})=-\frac{s_\perp}{2\ell}\left(\sqrt{2(n+1)}\psi_{n+1,p}( u_{\perp})+\sqrt{2n}\psi_{n-1,p}( u_{\perp})\right) ,   \label{property2} \\
 (\bm{\pi}_\perp\cdot\bm{\gamma}_\perp) \psi_{n,p}( u_{\perp})  &=& \frac{i}{2\ell} \sqrt{2(n+1)}\psi_{n+1,p}( u_{\perp})\left( \gamma^1 +i s_\perp \gamma^2 \right)
 -\frac{i}{2\ell} \sqrt{2n}\psi_{n-1,p}( u_{\perp})\left( \gamma^1 -i s_\perp \gamma^2 \right) ,
 \label{property3}
\end{eqnarray}
where we took into account that $H^\prime_n(x) = 2nH_{n-1}(x)$ and $H_{n+1}(x) =2xH_{n}(x) -2n H_{n-1}(x) $.

These wave functions satisfy the condition of completeness
\begin{equation}
  \sum\limits_{n=0}^{\infty}\int\limits_{-\infty}^{\infty}
 dp \psi_{np}( u_{\perp}) \psi^{*}_{np}( u_{\perp}^{\prime})
 =\delta^2( u_{\perp}- u_{\perp}^{\prime}) ,
 \label{completeness}
\end{equation}
which can be written in a compact form as $\sum_{n,p}  \langle u_\perp | n p\rangle \langle p n  | u^\prime\rangle =\langle u_\perp |u^\prime\rangle$. 

By inserting the unit operator $\sum_{n,p} | n p\rangle \langle p n|$ in front of $| u^\prime_\perp \rangle$ on the right hand of Eq.~(\ref{app:G-mixed-1}) and making use of the properties in Eqs.~(\ref{property1}) -- (\ref{property3}), we derive the propagator in the following form:
\begin{equation}
 G(p_{\parallel}, u_{\perp}, u_{\perp}^{\prime})= e^{i\Phi( u_{\perp}, u_{\perp}^{\prime})} 
 \bar{G}(p_{\parallel}, u_{\perp}- u_{\perp}^{\prime}),
\end{equation} 
 where $\Phi( u_{\perp}, u_{\perp}^{\prime})=\frac{qB}{2}(x+x^\prime)(y-y^\prime) $ is the Schwinger phase, and  
\begin{eqnarray}
\bar{G}(p_\parallel, u_{\perp})&=& i \frac{e^{-u_{\perp}^2/(4\ell^2)}}{2\pi \ell^2} \sum_{n=0}^{\infty}
\Bigg\{ \left[v_{\parallel,n} (p_\parallel \cdot \gamma_\parallel)
   - i \gamma^1\gamma^2 \tilde{v}_n  (p_\parallel \cdot \gamma_\parallel)
  +m_n  - i\gamma^1\gamma^2 \tilde{m}_n\right]
\left[L_{n}\left(\frac{u_{\perp}^2}{2\ell^2}\right) \mathcal{P}_{+} + L_{n-1}\left(\frac{u_{\perp}^2}{2\ell^2}\right) \mathcal{P}_{-}\right] 
\nonumber\\
&&-i \frac{v_{\perp,n}}{\ell^2}(\bm{u}_{\perp}\cdot\bm{\gamma}_{\perp})L_{n-1}^{1}\left(\frac{u_{\perp}^2}{2\ell^2}\right)\Bigg\} \frac{1}{\mathcal{M}_{n}-2nv_{\perp,n}^2 |qB|} ,
\label{app:G-mixed-2}
\end{eqnarray} 
where $L_n^{\alpha}(z)$ are the Laguerre polynomials (by definition, $L_{-1}^\alpha(z) \equiv 0$), $\mathcal{P}_{\pm}=\left(1 \pm s_{\perp}i\gamma^1\gamma^2\right)/2$ are spin projectors, and 
\begin{equation}
  \mathcal{M}_{n} = \left( v_{\parallel,n}^2 -\tilde{v}_{n}^2 \right) p_\parallel^2 - m_{n}^2   +\tilde{m}_{n}^2
 + 2 i \gamma^1\gamma^2(m_{n}\tilde{v}_{n} - \tilde{m}_{n} v_{\parallel,n})(p_\parallel \cdot \gamma_\parallel).
\end{equation}   
In derivation, we used the following relations:
\begin{eqnarray}
 \int_{-\infty}^{\infty} dp\, \psi_{np}( u_{\perp}) \psi^{*}_{np}( u_{\perp}^{\prime}) &=&
 \frac{e^{-\zeta/2+i\Phi( u_{\perp}, u_{\perp}^{\prime})}}{2\pi \ell^2}L_n\left(\frac{(u_{\perp}-u_{\perp}^{\prime})^2}{2\ell^2}\right) , \\
 \int_{-\infty}^{\infty} dp\,   \psi_{n+1,p}( u_{\perp}) \psi^{*}_{np}( u_{\perp}^{\prime}) &=&
 \frac{e^{-\zeta/2+i\Phi( u_{\perp}, u_{\perp}^{\prime})}}{2\pi \ell^2 \sqrt{2(n+1)} } \frac{x-x^\prime- i s_\perp \left(y-y^\prime\right)}{\ell} L_{n}^{1}\left(\frac{(u_{\perp}-u_{\perp}^{\prime})^2}{2\ell^2}\right) ,\\
 \int_{-\infty}^{\infty} dp\,  \psi_{n-1,p}( u_{\perp}) \psi^{*}_{np}( u_{\perp}^{\prime}) &=&
 \frac{e^{-\zeta/2+i\Phi( u_{\perp}, u_{\perp}^{\prime})}}{2\pi \ell^2 \sqrt{2n} }\frac{x^\prime-x-i s_\perp \left(y-y^\prime\right)}{\ell}  L_{n-1}^{1}\left(\frac{(u_{\perp}-u_{\perp}^{\prime})^2}{2\ell^2}\right),
 \label{n-level}
\end{eqnarray}
where $\zeta = (u_{\perp}-u_{\perp}^{\prime})^2/(2\ell^2)$. To obtain Eq.~(\ref{n-level}), we used the following table integral 7.377 \cite{Gradshteyn:1943cpj}:
\begin{equation}
 \int_{-\infty}^{\infty}  e^{-x^2}H_{m}(x+y)H_{n}(x+z) dx= 2^n \sqrt{\pi} m! z^{n-m}L_m^{n-m}(-2yz),
\end{equation}
which assumes $m\leq n$.

Note that the last factor in Eq.~(\ref{app:G-mixed-2}) is a matrix. It can be rendered in the following more convenient form:
\begin{eqnarray}
 \frac{1}{\mathcal{M}_{n}-2n v_{\perp,n}^2 |qB|}&=&\frac{1}{U_n}\left[ \left( v_{\parallel,n}^2  -\tilde{v}_{n}^2\right) p_\parallel^2 
 -2n v_{\perp,n}^2 |qB| - m_{n}^2 +\tilde{m}_{n}^2
 - 2 i \gamma^1\gamma^2(m_{n}\tilde{v}_{n} - \tilde{m}_{n} v_{\parallel,n})(p_\parallel \cdot \gamma_\parallel) \right],
 \label{1overM-2nqB}
\end{eqnarray}
where 
\begin{equation}
 U_n =  \left[\left( v_{\parallel,n}^2  - \tilde{v}_{n}^2\right) p_\parallel^2 -2n v_{\perp,n}^2 |qB|  - m_{n}^2 + \tilde{m}_{n}^2\right]^2 -4p_\parallel^2 \left(m_{n}\tilde{v}_{n}   - \tilde{m}_{n} v_{\parallel,n}  \right)^2.
 \label{Un}
\end{equation}
The poles of the full propagator are determined by setting $ U_n=0$. Its solutions determine the modified energies of the Landau-level states, i.e.,
\begin{eqnarray}
 p_0^2&=& p_z^2+ \frac{\left(v_{\parallel,n}^2 - \tilde{v}_{n}^2 \right)\left(2  n v_{\perp,n}^2 |qB| +m_{n}^2-\tilde{m}_{n}^2\right) +2 \left(m_{n}\tilde{v}_{n}   - \tilde{m}_{n} v_{\parallel,n}  \right)^2
 \pm 2 \left(m_{n}\tilde{v}_{n}   - \tilde{m}_{n} v_{\parallel,n}  \right) \sqrt{V_{n}}}{(v_{\parallel,n}^2-\tilde{v}_{n}^2)^2} ,
  \label{p0_sol}
\end{eqnarray}
where
\begin{eqnarray}
 V_{n}&=& \left(v_{\parallel,n}^2 - \tilde{v}_{n}^2 \right) \left(2  n v_{\perp,n}^2 |qB|  +m_{n}^2-\tilde{m}_{n}^2\right)+\left(m_{n}\tilde{v}_{n}   - \tilde{m}_{n} v_{\parallel,n}  \right)^2.
\end{eqnarray}

\subsection{Fourier transform of the translation invariant part of the propagator}

By performing the Fourier transform of the translation invariant part of the propagator in Eq.~(\ref{app:G-mixed-2}), we derive 
\begin{equation}
 \bar{G}(p_{\parallel},\bm{p}_{\perp}) = ie^{-p_{\perp}^2 \ell^{2}}\sum_{n=0}^{\infty}
 (-1)^nD_{n}(p_{\parallel},\bm{p}_{\perp}) \frac{1}{\mathcal{M}_{n}-2n v_{\perp,n}^2 |qB|},
 \label{GDn-new}
\end{equation}
where the $n$th Landau level contribution is determined by
\begin{eqnarray}
 D_{n}(p_{\parallel},\bm{p}_{\perp}) &=& 2\left[v_{\parallel,n} (p_\parallel \cdot \gamma_\parallel)
   - i \gamma^1\gamma^2 \tilde{v}_{n}  (p_\parallel \cdot \gamma_\parallel)
  +m_{n}  - i\gamma^1\gamma^2 \tilde{m}_{n}\right]\left[\mathcal{P}_{+}L_n\left(2 p_{\perp}^2 \ell^{2}\right)
 -\mathcal{P}_{-}L_{n-1}\left(2 p_{\perp}^2 \ell^{2}\right)\right] \nonumber\\
 &+& 4v_{\perp,n}\,(\bm{p}_{\perp}\cdot\bm{\gamma}_{\perp}) L_{n-1}^1\left(2 p_{\perp}^2 \ell^{2}\right).
 \label{Dn}
\end{eqnarray}
In derivation, we used the following table integrals:
\begin{eqnarray}
 \int_{0}^{2\pi} e^{-i (\bm{k}_\perp\cdot\bm{u}_\perp) } d\phi &=& 2\pi J_{0}(k_\perp u_\perp),
 \label{J0}
\\
 \int_{0}^{2\pi} (\bm{\gamma}_\perp\cdot\hat{\bm u}_\perp) e^{-i(\bm{k}\cdot\bm{u}_\perp)} d\phi &=&
 2i \pi (\bm{\gamma}_\perp\cdot\hat{\bm{k}}_\perp) J_{1}(k_\perp u_\perp) ,
 \label{integral-gamma}
\\
 \int_0^\infty  r^{\nu+1} e^{-\beta r^2}L^\nu_n\left(\alpha r^2\right)J_\nu(rk)dr
&=& \frac{k^\nu}{(2\beta)^{1+\nu}}\left(\frac{\beta-\alpha}{\beta}\right)^{n}
 e^{-\frac{k^2}{4\beta} }L^\nu_n\left( \frac{\alpha k^2}{4\beta(\alpha-\beta)}\right).
 \label{integral-J0}
\end{eqnarray}

\subsection{Free fermion propagator}

The free fermion propagator is obtained from the full propagator by replacing $v_{\parallel,n}$, $v_{\perp,n} \to 1$, $m_{n}\to \bar{m}_{0}$, and setting zero values to spin-splitting functions $\tilde{v}_{n}$ and $\tilde{m}_{n}$. Then, the Fourier transform of the translation invariant part of the free propagator takes the form:
\begin{equation}
 \bar{S}(p_{\parallel},\bm{p}_{\perp}) = ie^{-p_{\perp}^2 \ell^{2}}\sum_{n=0}^{\infty}
 (-1)^n \frac{D^{(0)}_{n}(p_{\parallel},\bm{p}_{\perp})}{p_\parallel^2-\bar{m}_{0}^2 -2n|qB |},
 \label{free-prop}
\end{equation}
where
\begin{eqnarray}
 D^{(0)}_{n}(p_{\parallel},\bm{p}_{\perp}) &=& 2\left[(p_\parallel \cdot \gamma_\parallel) +\bar{m}_{0}\right]\left[\mathcal{P}_{+}L_n\left(2 p_{\perp}^2 \ell^{2}\right) -\mathcal{P}_{-}L_{n-1}\left(2 p_{\perp}^2 \ell^{2}\right)\right]
 + 4 (\bm{p}_{\perp}\cdot\bm{\gamma}_{\perp}) L_{n-1}^1\left(2 p_{\perp}^2 \ell^{2}\right).
 \label{Dn0-free}  
\end{eqnarray}

\section{Inverse fermion propagator in the Landau-level representation}
\label{app:inverse-propagator}

By definition, the inverse of the full propagator is given by the following matrix element:
\begin{eqnarray}
 G^{-1}(p_{\parallel}, u_{\perp}, u_{\perp}^{\prime}) &=& 
  -i \langle u_{\perp} |  \left[ v_\parallel (p_\parallel\cdot\gamma_\parallel)
 - v_\perp (\bm{\pi}_\perp \cdot\bm{\gamma}_\perp) 
 +i\gamma^1\gamma^2 (p_\parallel\cdot\gamma_\parallel) \tilde{v}
 -m-i\gamma^1\gamma^2 \tilde{m}
 \right] |  u_{\perp}^{\prime} \rangle .
 \label{inversefull}
\end{eqnarray}
As in the derivation of the propagator in Appendix~\ref{app:propagator}, we insert the unit operator $\sum_{n,p} | n p\rangle \langle p n|$ in front of $| u^\prime_\perp \rangle$ to derive the following representation for the inverse propagator:
\begin{equation}
 G^{-1}(p_{\parallel}, u_{\perp}, u_{\perp}^{\prime}) = e^{i\Phi( u_{\perp}, u_{\perp}^{\prime})} 
 \bar{G}^{-1}(p_{\parallel}, u_{\perp}- u_{\perp}^{\prime}),
\end{equation}
where the translation invariant part of the propagator is given by a sum over Landau levels
\begin{eqnarray}
 \bar{G}^{-1}(p_{\parallel}, u_{\perp} )
 &=& -i  \frac{e^{-u_{\perp}^2/(4\ell^2)}}{2\pi \ell^2}\sum\limits_{n=0}^{\infty}
 \Bigg\{ \left[ v_{\parallel,n} (p_\parallel\cdot\gamma_\parallel)
 +i\gamma^1\gamma^2 (p_\parallel\cdot\gamma_\parallel) \tilde{v}_{n}
 -m_{n} -i\gamma^1\gamma^2\tilde{m}_{n} \right] \left[ \mathcal{P}_{+} L_n\left(\frac{u_{\perp}^2}{2\ell^2}\right)
 + \mathcal{P}_{-} L_{n-1}\left(\frac{u_{\perp}^2}{2\ell^2}\right)\right]  \nonumber\\
 &+&  
 \frac{1}{\ell^2}v_{\perp,n}  (\bm{u}_{\perp}\cdot  \bm{\gamma}_\perp) L^1_{n-1}\left(\frac{u_{\perp}^2}{2\ell^2}\right)
 \Bigg\} .
 \label{app:inversefull-1}
\end{eqnarray}
Recall that, by definition, $L_{-1}^\alpha \equiv 0$.  

The corresponding Fourier transform reads
\begin{eqnarray}
 \bar{G}^{-1}(p_{\parallel},\bm{p}_{\perp})
 &=& -2i  e^{-p_{\perp}^2 \ell^2}  \sum\limits_{n=0}^{\infty} (-1)^n
 \left[ v_{\parallel,n} (p_\parallel\cdot\gamma_\parallel)
 +i\gamma^1\gamma^2 (p_\parallel\cdot\gamma_\parallel) \tilde{v}_{n}
 -m_{n} -i\gamma^1\gamma^2 \tilde{m}_{n} \right] \left[\mathcal{P}_{+} L_n(2p_{\perp}^2 \ell^2)- \mathcal{P}_{-} L_{n-1}(2p_{\perp}^2 \ell^2) \right] \nonumber\\
 &-&  
 4 i e^{-p_{\perp}^2 \ell^2}  \sum\limits_{n=0}^{\infty} (-1)^n v_{\perp,n} (\bm{\gamma}_\perp\cdot \bm{p}_{\perp}) L^1_{n-1}(2p_{\perp}^2 \ell^2) .
 \label{app:inversefull-2}
\end{eqnarray}

\subsection{Self-energy in the Landau-level representation}
\label{app:self-energy-LL}

By making use of the inverse full and free propagators, we derive
\begin{equation}
\bar{\Sigma} (p_{\parallel}, u_{\perp} )  = i \bar{S}^{-1}(p_{\parallel}, u_{\perp}) -i  \bar{G}^{-1}(p_{\parallel}, u_{\perp}) .
   \label{app:self-energy-0}
\end{equation}
By using the Landau-level representation for the inverse propagator, we obtain
 \begin{eqnarray}
\bar{\Sigma} (p_{\parallel}, u_{\perp} )  
&=& -\frac{e^{-u_{\perp}^2/(4\ell^2)}}{2\pi \ell^2}\sum\limits_{n=0}^{\infty}
 \Bigg\{ \left[ \delta v_{\parallel,n} (p_\parallel\cdot\gamma_\parallel)
 + i\gamma^1\gamma^2 (p_\parallel\cdot\gamma_\parallel) \tilde{v}_{n}
 -\delta m_{n} - i\gamma^1\gamma^2\tilde{m}_{n}\right] \left[ \mathcal{P}_{+} L_n\left(\frac{u_{\perp}^2}{2\ell^2}\right)+ \mathcal{P}_{-} L_{n-1}\left(\frac{u_{\perp}^2}{2\ell^2}\right)\right]  \nonumber\\
 &+&\frac{\delta v_{\perp,n} }{\ell^2} (\bm{u}_{\perp}\cdot  \bm{\gamma}_\perp) L^1_{n-1}\left(\frac{u_{\perp}^2}{2\ell^2}\right)
 \Bigg\} .
  \label{app:self-energy-1}
\end{eqnarray}
where $\delta v_{\parallel,n} = v_{\parallel,n}-1$, $\delta v_{\perp,n} = v_{\perp,n}-1$, and $\delta m_{n} = m_{n}-\bar{m}_{0}$. The corresponding Fourier transform reads
\begin{eqnarray}
 \bar{\Sigma}(p_{\parallel},\bm{p}_{\perp})
 &=& - 2  e^{-p_{\perp}^2 \ell^2}  \sum\limits_{n=0}^{\infty} (-1)^n
 \left[ \delta v_{\parallel,n} (p_\parallel\cdot\gamma_\parallel)
 +i\gamma^1\gamma^2 (p_\parallel\cdot\gamma_\parallel) \tilde{v}_{n}
 -\delta m_{n} - i\gamma^1\gamma^2 \tilde{m}_{n} \right] \left[\mathcal{P}_{+} L_n(2p_{\perp}^2 \ell^2)- \mathcal{P}_{-} L_{n-1}(2p_{\perp}^2 \ell^2) \right] \nonumber\\
 &-&  
 4  e^{-p_{\perp}^2 \ell^2}  \sum\limits_{n=0}^{\infty} (-1)^n \delta v_{\perp,n} (\bm{\gamma}_\perp\cdot \bm{p}_{\perp}) L^1_{n-1}(2p_{\perp}^2 \ell^2) .
 \label{app:self-energy-2}
\end{eqnarray}

\section{Calculation of the kernels}
\label{app:kernels}

In the derivation of the Landau-level representation for the five component functions of the self-energy, see Eqs.~(\ref{Laguerre-projection-1}) through (\ref{Laguerre-projection-5}), one encounters the two different types of kernel functions defined by the following expressions:
\begin{eqnarray}
K_{n,n^\prime} &=&  (-1)^{n+n^\prime} \frac{2\ell^2}{\pi }  \int d^2 \bm{k}_{\perp}  e^{- \bm{k}_\perp^2\ell^2}
 e^{-(\bm{k}_\perp -\bm{q}_\perp )^2\ell^2} 
 L_{n^{\prime}} \left(2\bm{k}_\perp^2 \ell^2\right) 
 L_{n} \left(2(\bm{k}_\perp -\bm{q}_\perp)^2 \ell^2\right) , \\
\bar{K}_{n,n^\prime}  &=&  (-1)^{n+n^\prime} \frac{8\ell^4}{\pi }  \int d^2 \bm{k}_{\perp}  e^{- \bm{k}_\perp^2\ell^2}
 e^{-(\bm{k}_\perp -\bm{q}_\perp )^2\ell^2} 
\left(\bm{k}_\perp\cdot (\bm{k}_\perp -\bm{q}_\perp)\right)
 L^{1}_{n^{\prime}-1} \left(2 \bm{k}_\perp^2 \ell^2\right) 
 L^{1}_{n-1} \left(2(\bm{k}_\perp -\bm{q}_\perp)^2 \ell^2\right). 
\end{eqnarray}
To calculate the first kernel, it is convenient to start by noting the following Fourier transform:
\begin{eqnarray}
\frac{1}{4\pi \ell^2}\int d^2 \bm{u}_\perp e^{-u_\perp^2/(4\ell^2)} L_{n}\left(\frac{u_\perp^2}{2\ell^2}\right)
e^{ -i  \bm{p}_\perp\cdot  \bm{u}_\perp} 
&=& \frac{1}{4\pi \ell^2}\int_{0}^{\infty} u_\perp d u_\perp e^{-u_\perp^2/(4\ell^2)} L_{n}\left(\frac{u_\perp^2}{4\pi \ell^2}\right) \int_{0}^{2\pi} d\phi e^{ -i  p_\perp u_\perp \cos\phi} \nonumber\\
&=& \frac{1}{2} \int_{0}^{\infty} \bar{r} d\bar{r} e^{-\bar{r}^2/4} L_{n}\left(\frac{\bar{r}^2}{2}\right) J_0\left( p_\perp\ell \bar{r} \right)=(-1)^{n}  e^{-p_\perp^2\ell^2}L_{n}\left(2p_\perp^2\ell^2\right) ,
\end{eqnarray}
where we introduced the following dimensionless variable $\bar{r}=u_\perp/\ell$ and used table integral 7.419 1 in Ref.~\cite{Gradshteyn:1943cpj}. Similarly, in the calculation of the second kernel, it is useful to utilize another Fourier transform
\begin{eqnarray}
\frac{i}{8\pi \ell^4}\int d^2 \bm{u}_\perp  (\bm{u}_\perp\cdot\bm{a})  e^{-u_\perp^2/(4\ell^2)} L^{1}_{n}\left(\frac{u_\perp^2}{2\ell^2}\right) e^{ -i  \bm{p}_\perp\cdot  \bm{u}_\perp} 
&=& \frac{(\hat{\bm{p}}_\perp\cdot\bm{a}) }{4\ell} \int_{0}^{\infty} \bar{r}^2 d\bar{r} e^{-\bar{r}^2/4} L^{1}_{n}\left(\frac{\bar{r}^2}{2}\right) J_1 \left( p_\perp\ell \bar{r} \right) \nonumber\\
&=&  (-1)^n (\bm{p}_\perp\cdot\bm{a})  
e^{-p_\perp^2\ell^2}L^{1}_{n}\left(2p_\perp^2\ell^2\right) ,
\end{eqnarray}
where $\bm{a}$ is an arbitrary transverse 2D vector. In the derivation, we used table integral 7.419 4 in Ref.~\cite{Gradshteyn:1943cpj}. 

By making use of the first result, we derive
\begin{eqnarray}
K_{n,n^\prime} &=& \int \frac{d^2 \bm{k}_{\perp} }{8\pi^3\ell^2}
\int d^2 \bm{u}_\perp e^{-u_\perp^2/(4\ell^2)} L_{n^\prime}\left(\frac{u_\perp^2}{2\ell^2}\right)
e^{ -i  \bm{k}_\perp \cdot  \bm{u}_\perp} 
\int d^2 \bm{u}^\prime_\perp e^{-(u_\perp^\prime)^2/(4\ell^2)} L_{n}\left(\frac{(\bm{u}_\perp^\prime)^2}{2\ell^2}\right)
e^{ -i (\bm{k}_\perp -\bm{q}_\perp)\cdot  \bm{u}^\prime_\perp} \nonumber\\
&=&\int  \frac{d^2 \bm{u}_\perp}{2\pi \ell^2}
 e^{-u_\perp^2/(2\ell^2)} 
L_{n}\left(\frac{u_\perp^2}{2\ell^2}\right)
L_{n^\prime}\left(\frac{u_\perp^2}{2\ell^2}\right)
e^{ -i  \bm{q}_\perp \cdot  \bm{u}_\perp} \nonumber\\
&=&\frac{1}{\ell^2}\int_{0}^{\infty} u_\perp d u_\perp 
e^{-u_\perp^2/(2\ell^2)} 
L_{n}\left(\frac{u_\perp^2}{2\ell^2}\right)
L_{n^\prime}\left(\frac{u_\perp^2}{2\ell^2}\right)
J_0\left( q_\perp u_\perp \right) 
= \mathcal{I}_0^{n,n^\prime}\left( \frac{q_\perp^2 \ell^2}{2} \right) .
\end{eqnarray}
By making use of the second result, we derive
\begin{eqnarray}
\bar{K}_{n,n^\prime} &=&  - \int \frac{  d^2 \bm{k}_{\perp}  }{8\pi^3 \ell^4}
\int d^2 \bm{u}_\perp e^{-u_\perp^2/(4\ell^2)} 
 L^{1}_{n^{\prime}-1} \left(\frac{u_\perp^2}{2\ell^2}\right)
 e^{ -i  \bm{k}_\perp \cdot  \bm{u}_\perp} 
\int d^2 \bm{u}^\prime_\perp 
\left(\bm{u}_\perp \cdot \bm{u}^\prime_\perp \right)
e^{-(u_\perp^\prime)^2/(4\ell^2)} 
 L^{1}_{n-1} \left(\frac{(u_\perp^\prime)^2}{2\ell^2}\right) 
e^{ -i (\bm{k}_\perp -\bm{q}_\perp)\cdot  \bm{u}^\prime_\perp} \nonumber\\
&=&\int  \frac{ d^2 \bm{u}_\perp}{2\pi\ell^4} u_\perp^2
e^{-u_\perp^2/(2\ell^2)} 
 L^{1}_{n-1} \left(\frac{u_\perp^2}{2\ell^2}\right)L^{1}_{n^\prime-1} \left(\frac{u_\perp^2}{2\ell^2}\right)
 e^{ -i  \bm{q}_\perp \cdot  \bm{u}_\perp} \nonumber\\
&=& \frac{1}{\ell^4} \int_{0}^{\infty} u_\perp^3 d u_\perp 
e^{-u_\perp^2/(2\ell^2)} 
 L^{1}_{n-1} \left(\frac{u_\perp^2}{2\ell^2}\right)L^{1}_{n^\prime-1} \left(\frac{u_\perp^2}{2\ell^2}\right)
J_0\left( q_\perp u_\perp \right) = \mathcal{I}_2^{n-1,n^\prime-1}\left( \frac{q_\perp^2 \ell^2}{2} \right) . 
\end{eqnarray}
where $\mathcal{I}_{0}^{n,n^\prime}(\xi)$ and $\mathcal{I}_{2}^{n,n^\prime}(\xi)$ are the same function that were introduced in Ref.~\cite{Wang:2021ebh}, i.e.,
\begin{eqnarray}
\mathcal{I}_{0}^{n,n^{\prime}}(\xi)&=&
 \frac{(n^\prime)!}{n!} e^{-\xi}  \xi^{n-n^\prime} \left(L_{n^\prime}^{n-n^\prime}\left(\xi\right)\right)^2 
= \frac{n!}{(n^\prime)!}e^{-\xi} \xi^{n^\prime-n} \left(L_{n}^{n^\prime-n}\left(\xi\right)\right)^2 ,
\label{I0f-LL-form1}  
\\
\mathcal{I}_{2}^{n,n^{\prime}}(\xi)&=& 
2 \frac{(n^\prime+1)!}{n!}e^{-\xi}  \xi^{n-n^\prime}  
L_{n^\prime}^{n-n^\prime}\left(\xi\right) L_{n^\prime+1}^{n-n^\prime}\left(\xi\right) 
= 2 \frac{(n+1)!}{(n^\prime)!} e^{-\xi}  \xi^{n^\prime-n}  
L_{n}^{n^\prime-n}\left(\xi\right) L_{n+1}^{n^\prime-n}\left(\xi\right)  .
\label{I2f-LL-form1} 
\end{eqnarray}
Note that $\mathcal{I}_{2}^{n,n^\prime}(\xi)$ can be also expressed in terms of $\mathcal{I}_{0}^{n,n^\prime}(\xi)$ \cite{Wang:2021ebh}, i.e., 
\begin{equation}
\mathcal{I}_{2}^{n,n^{\prime}}(\xi) =
\frac{n+n^{\prime}+2}{2}\left[\mathcal{I}_{0}^{n,n^{\prime}}(\xi) +\mathcal{I}_{0}^{n+1,n^{\prime}+1}(\xi) \right]
-\frac{\xi}{2}\left[\mathcal{I}_{0}^{n+1,n^{\prime}}(\xi) +\mathcal{I}_{0}^{n,n^{\prime}+1}(\xi) \right] .
\label{I2-I0}
\end{equation}
By definition, $\mathcal{I}_{0}^{n,n^{\prime}}(\xi)$ and $\mathcal{I}_{2}^{n,n^{\prime}}(\xi)$ vanish when either of their upper indices becomes negative.

\section{Wave functions for fermions in a magnetic field}
\label{Wave-functions}

Let us consider the spinor wave function in a given Landau (labeled by index $n$) level with a positive energy:
\begin{equation}
\Psi_{n,p} (u) = e^{-i p_\parallel \cdot u_\parallel} \left[ \psi_{n,p} (u_\perp) \mathcal{P}_{+} +i \psi_{n-1,p} (u_\perp) \mathcal{P}_{-} \right] v .
\label{wave-function}
\end{equation}
Substituting it into the Dirac equation gives
\begin{eqnarray}
 \left( i\gamma^\mu D_{\mu} - \bar{m}_{0}\right)\Psi_{n,p} (u) 
 &=& \left[ \left(p_\parallel\cdot\gamma_\parallel \right)-\left(\bm{\pi}_\perp\cdot \bm{\gamma}_\perp \right) - \bar{m}_{0}\right]\Psi_{n,p} (u) \nonumber\\
 &=&e^{-i p_\parallel \cdot u_\parallel} \left[ \psi_{n,p} (u_\perp) \mathcal{P}_{+} +i \psi_{n-1,p} (u_\perp) \mathcal{P}_{-} \right] 
 \left[ \left(p_\parallel\cdot\gamma_\parallel \right) + \gamma^1\sqrt{2n|qB|}  - \bar{m}_{0} \right]  v ,
\end{eqnarray}
where we used the property in Eq.~(\ref{property3}). The spinor $v$ must satisfy the equation:
\begin{equation}
 \left[\left(p_\parallel\cdot\gamma_\parallel \right) + \gamma^1\sqrt{2n|qB|}  - \bar{m}_{0} \right] v=0 .
 \end{equation}
It has nonzero solution when 
\begin{equation}
\mbox{Det} \left[\left(p_\parallel\cdot\gamma_\parallel \right) + \gamma^1\sqrt{2n|qB|}  - \bar{m}_{0} \right] 
= \left[ p_\parallel^2- 2n|qB| - \bar{m}_{0}^2 \right] ^2
=0.
 \end{equation}
Using the following representation of Dirac matrices:
\begin{equation}
\gamma^0 =\left(\begin{array}{ll}
\mathbb{I}_2 & 0\\
0 & - \mathbb{I}_2
\end{array}\right), \qquad
\gamma^i =\left(\begin{array}{ll}
0 & \sigma_i \\
-\sigma_i & 0 
\end{array}\right),
 \end{equation}
 where $\sigma_i$ are the Pauli matrices, we derive explicit solutions for spinor $v$, i.e.,
\begin{equation}
v= \sqrt{ \bar{m}_{0}+E_{n,p_z}} 
\left(\begin{array}{c}
a_1 \\
a_2 \\
\frac{p_z a_1 -\sqrt{2n |qB|}a_2}{\bar{m}_{0}+E_{n,p_z}}\\
-\frac{\sqrt{2n |qB|}a_1+p_z a_2}{\bar{m}_{0}+E_{n,p_z}}
\end{array}
\right) .
 \end{equation}
These spinor are similar but different from those in Ref.~\cite{Bhattacharya:2002aj}. However, here we use a slightly different ansatz for the wave function (\ref{wave-function}) and a different normalization convension for the spinors, i.e.,
\begin{equation}
\bar{v} v =  2\bar{m}_{0} (a_1^2+a_2^2).
 \end{equation}
Therefore, the final spinor wave function reads
\begin{equation}
\Psi_{n,p} (u) =  e^{-i p_\parallel \cdot u_\parallel}  \left(\begin{array}{c}
\sqrt{ \bar{m}_{0}+E_{n,p_z}} \psi_{n,p} (u_\perp) a_1 \\
i \sqrt{ \bar{m}_{0}+E_{n,p_z}}\psi_{n-1,p} (u_\perp) a_2 \\
\frac{p_z a_1 -\sqrt{2n |qB|}a_2}{\sqrt{ \bar{m}_{0}+E_{n,p_z}} }\psi_{n,p} (u_\perp)\\
-i\frac{\sqrt{2n |qB|}a_1+p_z a_2}{\sqrt{ \bar{m}_{0}+E_{n,p_z}} }\psi_{n-1,p} (u_\perp)
\end{array}
\right) ,
 \end{equation}
 where we set $s_\perp=1$ for simplicity. (Note that $\psi_{n,p} (u_\perp)$ and $i\psi_{n-1,p} (u_\perp)$ switch places when $s_\perp=-1$.) Two independent states $\Psi_{n,p,s} (u)$ are obtained by setting either (i) $a_1=1$, $a_2=0$ or (ii) $a_1=0$, $a_2=1$.  Then, we check that the sum over both spin states gives
 \begin{eqnarray}
\sum_s \Psi_{n,p,s} (u)  \bar{\Psi}_{n,p,s} (u^\prime) &=&  \frac{e^{-i p_\parallel \cdot (u_\parallel-u_\parallel^\prime)}  }{2}
\Big[ \left(\psi_{n,p}(u_\perp)\psi_{n,p}^{*}(u_\perp^\prime)+\psi_{n-1,p}(u_\perp)\psi_{n-1,p}^{*}(u_\perp^\prime) \right) \left(E_{n,p_z}\gamma^0-p_z\gamma^3+\bar{m}_{0}\right)  \nonumber\\
&&+i \gamma^1 \gamma^2\left(\psi_{n,p}(u_\perp)\psi_{n,p}^{*}(u_\perp^\prime)-\psi_{n-1,p}(u_\perp)\psi_{n-1,p}^{*}(u_\perp^\prime) \right) \left(E_{n,p_z}\gamma^0-p_z\gamma^3+\bar{m}_{0}\right)  \nonumber\\
&&+\sqrt{2n|qB|} (i \gamma^1 +\gamma^2) \psi_{n-1,p}(u_\perp)\psi_{n,p}^{*}(u_\perp^\prime) 
+\sqrt{2n|qB|} (-i \gamma^1 +\gamma^2)  \psi_{n,p}(u_\perp)\psi_{n-1,p}^{*}(u_\perp^\prime) \Big].
\end{eqnarray}
Finally, when also  integrated over the quantum number $p$, we obtain
\begin{eqnarray}
	\int dp\sum_s \Psi_{n,p,s} (u)  \bar{\Psi}_{n,p,s} (u^\prime) &=&  e^{-i p_\parallel \cdot (u_\parallel-u_\parallel^\prime)} \frac{e^{-(u_{\perp}-u_{\perp}^{\prime})^2/(2\ell^2)+i\Phi( u_{\perp}, u_{\perp}^{\prime})}}{2\pi \ell^2}\nonumber\\
	&\times&\bigg[\left(E_{n,p_z}\gamma^0-p_z\gamma^3+\bar{m}_{0}\right)\big[\mathcal{P}_{+}L_n\left(\zeta\right)+\mathcal{P}_{-}L_{n-1}\left(\zeta\right)\big]	+  \frac{(\bm{u}_{\perp}\cdot  \bm{\gamma}_\perp) }{\ell^2}  L^1_{n-1}(\zeta)\bigg],
\label{SumPhi-PsiBar}
\end{eqnarray}
where we used the shorthand notation $\zeta = (u_{\perp}-u_{\perp}^{\prime})^2/(2\ell^2)$. 
By making use of this result, the expression for the fermion damping rate (\ref{damping-rate-def}) becomes
\begin{eqnarray}
\Gamma_n(p_z) &=& \frac{1}{2p_0} \int \frac{d^2 \bm{u}_\perp  e^{-\zeta}}{ 2\pi  \ell^2} 
  \mbox{Tr} \Bigg\{ \left[ \left[ (p_\parallel\cdot\gamma_\parallel)+\bar{m}_{0} \right] \left[ \mathcal{P}_{+} L_n(\zeta)
 + \mathcal{P}_{-} L_{n-1}(\zeta)\right]
 +  \frac{(\bm{u}_{\perp}\cdot  \bm{\gamma}_\perp) }{\ell^2}  L^1_{n-1}(\zeta)\right]\nonumber\\ 
 &\times& \sum\limits_{n^\prime=0}^{\infty}
 \Big[ \left[  \mbox{Im}\delta v_{\parallel,n^\prime} (p_\parallel\cdot\gamma_\parallel)
 + i\gamma^1\gamma^2 (p_\parallel\cdot\gamma_\parallel)  \mbox{Im}\tilde{v}_{n^\prime}
 - \mbox{Im}\delta m_{n^\prime} - i\gamma^1\gamma^2 \mbox{Im}\tilde{m}_{n^\prime}\right] \left[ \mathcal{P}_{+} L_{n^\prime}(\zeta)+ \mathcal{P}_{-} L_{n^\prime-1}(\zeta)\right] \nonumber\\
 &&+\frac{ \mbox{Im}\delta v_{\perp,n^\prime}}{\ell^2}  (\bm{u}_{\perp}\cdot  \bm{\gamma}_\perp) L^1_{n^\prime-1}(\zeta)
 \Big]
 \Bigg\} .
 \label{damping-rate-app}
\end{eqnarray}
After integrating over $\zeta$, this reduces to the final expression for the spin-averaged damping rate, which is given in Eq.~(\ref{damping-rate-ave}) in the main text.


\begin{thebibliography}{70}%
\makeatletter
\providecommand \@ifxundefined [1]{%
 \@ifx{#1\undefined}
}%
\providecommand \@ifnum [1]{%
 \ifnum #1\expandafter \@firstoftwo
 \else \expandafter \@secondoftwo
 \fi
}%
\providecommand \@ifx [1]{%
 \ifx #1\expandafter \@firstoftwo
 \else \expandafter \@secondoftwo
 \fi
}%
\providecommand \natexlab [1]{#1}%
\providecommand \enquote  [1]{``#1''}%
\providecommand \bibnamefont  [1]{#1}%
\providecommand \bibfnamefont [1]{#1}%
\providecommand \citenamefont [1]{#1}%
\providecommand \href@noop [0]{\@secondoftwo}%
\providecommand \href [0]{\begingroup \@sanitize@url \@href}%
\providecommand \@href[1]{\@@startlink{#1}\@@href}%
\providecommand \@@href[1]{\endgroup#1\@@endlink}%
\providecommand \@sanitize@url [0]{\catcode `\\12\catcode `\$12\catcode
  `\&12\catcode `\#12\catcode `\^12\catcode `\_12\catcode `\%12\relax}%
\providecommand \@@startlink[1]{}%
\providecommand \@@endlink[0]{}%
\providecommand \url  [0]{\begingroup\@sanitize@url \@url }%
\providecommand \@url [1]{\endgroup\@href {#1}{\urlprefix }}%
\providecommand \urlprefix  [0]{URL }%
\providecommand \Eprint [0]{\href }%
\providecommand \doibase [0]{https://doi.org/}%
\providecommand \selectlanguage [0]{\@gobble}%
\providecommand \bibinfo  [0]{\@secondoftwo}%
\providecommand \bibfield  [0]{\@secondoftwo}%
\providecommand \translation [1]{[#1]}%
\providecommand \BibitemOpen [0]{}%
\providecommand \bibitemStop [0]{}%
\providecommand \bibitemNoStop [0]{.\EOS\space}%
\providecommand \EOS [0]{\spacefactor3000\relax}%
\providecommand \BibitemShut  [1]{\csname bibitem#1\endcsname}%
\let\auto@bib@innerbib\@empty
\bibitem [{\citenamefont {Vachaspati}(1991)}]{Vachaspati:1991nm}%
  \BibitemOpen
  \bibfield  {author} {\bibinfo {author} {\bibfnamefont {T.}~\bibnamefont
  {Vachaspati}},\ }\bibfield  {title} {\bibinfo {title} {{Magnetic fields from
  cosmological phase transitions}},\ }\href
  {https://doi.org/10.1016/0370-2693(91)90051-Q} {\bibfield  {journal}
  {\bibinfo  {journal} {Phys. Lett. B}\ }\textbf {\bibinfo {volume} {265}},\
  \bibinfo {pages} {258} (\bibinfo {year} {1991})}\BibitemShut {NoStop}%
\bibitem [{\citenamefont {Widrow}(2002)}]{Widrow:2002ud}%
  \BibitemOpen
  \bibfield  {author} {\bibinfo {author} {\bibfnamefont {L.~M.}\ \bibnamefont
  {Widrow}},\ }\bibfield  {title} {\bibinfo {title} {{Origin of galactic and
  extragalactic magnetic fields}},\ }\href
  {https://doi.org/10.1103/RevModPhys.74.775} {\bibfield  {journal} {\bibinfo
  {journal} {Rev. Mod. Phys.}\ }\textbf {\bibinfo {volume} {74}},\ \bibinfo
  {pages} {775} (\bibinfo {year} {2002})},\ \Eprint
  {https://arxiv.org/abs/astro-ph/0207240} {arXiv:astro-ph/0207240}
  \BibitemShut {NoStop}%
\bibitem [{\citenamefont {Raffelt}(1996)}]{Raffelt:1996wa}%
  \BibitemOpen
  \bibfield  {author} {\bibinfo {author} {\bibfnamefont {G.~G.}\ \bibnamefont
  {Raffelt}},\ }\href@noop {} {\emph {\bibinfo {title} {{Stars as Laboratories
  for Fundamental Physics: The Astrophysics of Neutrinos, Axions, and Other
  Weakly Interacting Particles}}}}\ (\bibinfo  {publisher} {University of
  Chicago Press},\ \bibinfo {address} {Chicago, IL},\ \bibinfo {year}
  {1996})\BibitemShut {NoStop}%
\bibitem [{\citenamefont {Harding}\ and\ \citenamefont
  {Lai}(2006)}]{Harding:2006qn}%
  \BibitemOpen
  \bibfield  {author} {\bibinfo {author} {\bibfnamefont {A.~K.}\ \bibnamefont
  {Harding}}\ and\ \bibinfo {author} {\bibfnamefont {D.}~\bibnamefont {Lai}},\
  }\bibfield  {title} {\bibinfo {title} {{Physics of Strongly Magnetized
  Neutron Stars}},\ }\href {https://doi.org/10.1088/0034-4885/69/9/R03}
  {\bibfield  {journal} {\bibinfo  {journal} {Rept. Prog. Phys.}\ }\textbf
  {\bibinfo {volume} {69}},\ \bibinfo {pages} {2631} (\bibinfo {year}
  {2006})},\ \Eprint {https://arxiv.org/abs/astro-ph/0606674}
  {arXiv:astro-ph/0606674} \BibitemShut {NoStop}%
\bibitem [{\citenamefont {Liao}(2015)}]{Liao:2014ava}%
  \BibitemOpen
  \bibfield  {author} {\bibinfo {author} {\bibfnamefont {J.}~\bibnamefont
  {Liao}},\ }\bibfield  {title} {\bibinfo {title} {{Anomalous transport effects
  and possible environmental symmetry
  \textquoteleft{}violation\textquoteright{} in heavy-ion collisions}},\ }\href
  {https://doi.org/10.1007/s12043-015-0984-x} {\bibfield  {journal} {\bibinfo
  {journal} {Pramana}\ }\textbf {\bibinfo {volume} {84}},\ \bibinfo {pages}
  {901} (\bibinfo {year} {2015})},\ \Eprint {https://arxiv.org/abs/1401.2500}
  {arXiv:1401.2500 [hep-ph]} \BibitemShut {NoStop}%
\bibitem [{\citenamefont {Kharzeev}\ \emph {et~al.}(2016)\citenamefont
  {Kharzeev}, \citenamefont {Liao}, \citenamefont {Voloshin},\ and\
  \citenamefont {Wang}}]{Kharzeev:2015znc}%
  \BibitemOpen
  \bibfield  {author} {\bibinfo {author} {\bibfnamefont {D.~E.}\ \bibnamefont
  {Kharzeev}}, \bibinfo {author} {\bibfnamefont {J.}~\bibnamefont {Liao}},
  \bibinfo {author} {\bibfnamefont {S.~A.}\ \bibnamefont {Voloshin}},\ and\
  \bibinfo {author} {\bibfnamefont {G.}~\bibnamefont {Wang}},\ }\bibfield
  {title} {\bibinfo {title} {{Chiral magnetic and vortical effects in
  high-energy nuclear collisions\textemdash{}A status report}},\ }\href
  {https://doi.org/10.1016/j.ppnp.2016.01.001} {\bibfield  {journal} {\bibinfo
  {journal} {Prog. Part. Nucl. Phys.}\ }\textbf {\bibinfo {volume} {88}},\
  \bibinfo {pages} {1} (\bibinfo {year} {2016})},\ \Eprint
  {https://arxiv.org/abs/1511.04050} {arXiv:1511.04050 [hep-ph]} \BibitemShut
  {NoStop}%
\bibitem [{\citenamefont {Huang}(2021)}]{Huang:2020xyr}%
  \BibitemOpen
  \bibfield  {author} {\bibinfo {author} {\bibfnamefont {X.-G.}\ \bibnamefont
  {Huang}},\ }\bibfield  {title} {\bibinfo {title} {{Vorticity and Spin
  Polarization \textemdash{} A Theoretical Perspective}},\ }\href
  {https://doi.org/10.1016/j.nuclphysa.2020.121752} {\bibfield  {journal}
  {\bibinfo  {journal} {Nucl. Phys. A}\ }\textbf {\bibinfo {volume} {1005}},\
  \bibinfo {pages} {121752} (\bibinfo {year} {2021})},\ \Eprint
  {https://arxiv.org/abs/2002.07549} {arXiv:2002.07549 [nucl-th]} \BibitemShut
  {NoStop}%
\bibitem [{\citenamefont {Shovkovy}(2022)}]{Shovkovy:2021yyw}%
  \BibitemOpen
  \bibfield  {author} {\bibinfo {author} {\bibfnamefont {I.~A.}\ \bibnamefont
  {Shovkovy}},\ }\bibfield  {title} {\bibinfo {title} {{Anomalous plasma:
  chiral magnetic effect and all that}},\ }in\ \href
  {https://doi.org/10.1142/9789811262357\_0018} {\emph {\bibinfo {booktitle}
  {Peter Suranyi 87th Birthday Festschrift}}},\ \bibinfo {editor} {edited by\
  \bibinfo {editor} {\bibfnamefont {P.}~\bibnamefont {Argyres}}, \bibinfo
  {editor} {\bibfnamefont {G.}~\bibnamefont {Dunne}}, \bibinfo {editor}
  {\bibfnamefont {G.}~\bibnamefont {Semenoff}},\ and\ \bibinfo {editor}
  {\bibfnamefont {R.}~\bibnamefont {Wijewardhana}}}\ (\bibinfo  {publisher}
  {World Scientific},\ \bibinfo {address} {Singapore},\ \bibinfo {year}
  {2022})\ Chap.~\bibinfo {chapter} {18}, pp.\ \bibinfo {pages} {291--316},\
  \Eprint {https://arxiv.org/abs/2111.11416} {arXiv:2111.11416 [nucl-th]}
  \BibitemShut {NoStop}%
\bibitem [{\citenamefont {Kaspi}\ and\ \citenamefont
  {Beloborodov}(2017)}]{Kaspi:2017fwg}%
  \BibitemOpen
  \bibfield  {author} {\bibinfo {author} {\bibfnamefont {V.~M.}\ \bibnamefont
  {Kaspi}}\ and\ \bibinfo {author} {\bibfnamefont {A.}~\bibnamefont
  {Beloborodov}},\ }\bibfield  {title} {\bibinfo {title} {{Magnetars}},\ }\href
  {https://doi.org/10.1146/annurev-astro-081915-023329} {\bibfield  {journal}
  {\bibinfo  {journal} {Ann. Rev. Astron. Astrophys.}\ }\textbf {\bibinfo
  {volume} {55}},\ \bibinfo {pages} {261} (\bibinfo {year} {2017})},\ \Eprint
  {https://arxiv.org/abs/1703.00068} {arXiv:1703.00068 [astro-ph.HE]}
  \BibitemShut {NoStop}%
\bibitem [{\citenamefont {Hardy}\ and\ \citenamefont
  {Thoma}(2001)}]{Hardy:2000gg}%
  \BibitemOpen
  \bibfield  {author} {\bibinfo {author} {\bibfnamefont {S.~J.}\ \bibnamefont
  {Hardy}}\ and\ \bibinfo {author} {\bibfnamefont {M.~H.}\ \bibnamefont
  {Thoma}},\ }\bibfield  {title} {\bibinfo {title} {{Neutrino electron
  processes in a strongly magnetized thermal plasma}},\ }\href
  {https://doi.org/10.1103/PhysRevD.63.025014} {\bibfield  {journal} {\bibinfo
  {journal} {Phys. Rev. D}\ }\textbf {\bibinfo {volume} {63}},\ \bibinfo
  {pages} {025014} (\bibinfo {year} {2001})},\ \Eprint
  {https://arxiv.org/abs/astro-ph/0008473} {arXiv:astro-ph/0008473}
  \BibitemShut {NoStop}%
\bibitem [{\citenamefont {Granot}\ \emph {et~al.}(2015)\citenamefont {Granot},
  \citenamefont {Piran}, \citenamefont {Bromberg}, \citenamefont {Racusin},\
  and\ \citenamefont {Daigne}}]{Granot:2015xba}%
  \BibitemOpen
  \bibfield  {author} {\bibinfo {author} {\bibfnamefont {J.}~\bibnamefont
  {Granot}}, \bibinfo {author} {\bibfnamefont {T.}~\bibnamefont {Piran}},
  \bibinfo {author} {\bibfnamefont {O.}~\bibnamefont {Bromberg}}, \bibinfo
  {author} {\bibfnamefont {J.~L.}\ \bibnamefont {Racusin}},\ and\ \bibinfo
  {author} {\bibfnamefont {F.}~\bibnamefont {Daigne}},\ }\bibfield  {title}
  {\bibinfo {title} {{Gamma-Ray Bursts as Sources of Strong Magnetic Fields}},\
  }\href {https://doi.org/10.1007/s11214-015-0191-6} {\bibfield  {journal}
  {\bibinfo  {journal} {Space Sci. Rev.}\ }\textbf {\bibinfo {volume} {191}},\
  \bibinfo {pages} {471} (\bibinfo {year} {2015})},\ \Eprint
  {https://arxiv.org/abs/1507.08671} {arXiv:1507.08671 [astro-ph.HE]}
  \BibitemShut {NoStop}%
\bibitem [{\citenamefont {Skokov}\ \emph {et~al.}(2009)\citenamefont {Skokov},
  \citenamefont {Illarionov},\ and\ \citenamefont {Toneev}}]{Skokov:2009qp}%
  \BibitemOpen
  \bibfield  {author} {\bibinfo {author} {\bibfnamefont {V.}~\bibnamefont
  {Skokov}}, \bibinfo {author} {\bibfnamefont {A.~Y.}\ \bibnamefont
  {Illarionov}},\ and\ \bibinfo {author} {\bibfnamefont {V.}~\bibnamefont
  {Toneev}},\ }\bibfield  {title} {\bibinfo {title} {{Estimate of the magnetic
  field strength in heavy-ion collisions}},\ }\href
  {https://doi.org/10.1142/S0217751X09047570} {\bibfield  {journal} {\bibinfo
  {journal} {Int. J. Mod. Phys.}\ }\textbf {\bibinfo {volume} {A24}},\ \bibinfo
  {pages} {5925} (\bibinfo {year} {2009})},\ \Eprint
  {https://arxiv.org/abs/0907.1396} {arXiv:0907.1396} \BibitemShut {NoStop}%
\bibitem [{\citenamefont {Voronyuk}\ \emph {et~al.}(2011)\citenamefont
  {Voronyuk}, \citenamefont {Toneev}, \citenamefont {Cassing}, \citenamefont
  {Bratkovskaya}, \citenamefont {Konchakovski} \emph
  {et~al.}}]{Voronyuk:2011jd}%
  \BibitemOpen
  \bibfield  {author} {\bibinfo {author} {\bibfnamefont {V.}~\bibnamefont
  {Voronyuk}}, \bibinfo {author} {\bibfnamefont {V.}~\bibnamefont {Toneev}},
  \bibinfo {author} {\bibfnamefont {W.}~\bibnamefont {Cassing}}, \bibinfo
  {author} {\bibfnamefont {E.}~\bibnamefont {Bratkovskaya}}, \bibinfo {author}
  {\bibfnamefont {V.}~\bibnamefont {Konchakovski}}, \emph {et~al.},\ }\bibfield
   {title} {\bibinfo {title} {{(Electro-)magnetic field evolution in
  relativistic heavy-ion collisions}},\ }\href
  {https://doi.org/10.1103/PhysRevC.83.054911} {\bibfield  {journal} {\bibinfo
  {journal} {Phys. Rev.}\ }\textbf {\bibinfo {volume} {C83}},\ \bibinfo {pages}
  {054911} (\bibinfo {year} {2011})},\ \Eprint
  {https://arxiv.org/abs/1103.4239} {arXiv:1103.4239} \BibitemShut {NoStop}%
\bibitem [{\citenamefont {Deng}\ and\ \citenamefont
  {Huang}(2012)}]{Deng:2012pc}%
  \BibitemOpen
  \bibfield  {author} {\bibinfo {author} {\bibfnamefont {W.-T.}\ \bibnamefont
  {Deng}}\ and\ \bibinfo {author} {\bibfnamefont {X.-G.}\ \bibnamefont
  {Huang}},\ }\bibfield  {title} {\bibinfo {title} {{Event-by-event generation
  of electromagnetic fields in heavy-ion collisions}},\ }\href
  {https://doi.org/10.1103/PhysRevC.85.044907} {\bibfield  {journal} {\bibinfo
  {journal} {Phys. Rev.}\ }\textbf {\bibinfo {volume} {C85}},\ \bibinfo {pages}
  {044907} (\bibinfo {year} {2012})},\ \Eprint
  {https://arxiv.org/abs/1201.5108} {arXiv:1201.5108} \BibitemShut {NoStop}%
\bibitem [{\citenamefont {Bloczynski}\ \emph {et~al.}(2013)\citenamefont
  {Bloczynski}, \citenamefont {Huang}, \citenamefont {Zhang},\ and\
  \citenamefont {Liao}}]{Bloczynski:2012en}%
  \BibitemOpen
  \bibfield  {author} {\bibinfo {author} {\bibfnamefont {J.}~\bibnamefont
  {Bloczynski}}, \bibinfo {author} {\bibfnamefont {X.-G.}\ \bibnamefont
  {Huang}}, \bibinfo {author} {\bibfnamefont {X.}~\bibnamefont {Zhang}},\ and\
  \bibinfo {author} {\bibfnamefont {J.}~\bibnamefont {Liao}},\ }\bibfield
  {title} {\bibinfo {title} {{Azimuthally fluctuating magnetic field and its
  impacts on observables in heavy-ion collisions}},\ }\href
  {https://doi.org/10.1016/j.physletb.2012.12.030} {\bibfield  {journal}
  {\bibinfo  {journal} {Phys. Lett.}\ }\textbf {\bibinfo {volume} {B718}},\
  \bibinfo {pages} {1529} (\bibinfo {year} {2013})},\ \Eprint
  {https://arxiv.org/abs/1209.6594} {arXiv:1209.6594} \BibitemShut {NoStop}%
\bibitem [{\citenamefont {Guo}\ \emph {et~al.}(2020)\citenamefont {Guo},
  \citenamefont {Liao},\ and\ \citenamefont {Wang}}]{Guo:2019mgh}%
  \BibitemOpen
  \bibfield  {author} {\bibinfo {author} {\bibfnamefont {X.}~\bibnamefont
  {Guo}}, \bibinfo {author} {\bibfnamefont {J.}~\bibnamefont {Liao}},\ and\
  \bibinfo {author} {\bibfnamefont {E.}~\bibnamefont {Wang}},\ }\bibfield
  {title} {\bibinfo {title} {{Spin Hydrodynamic Generation in the Charged
  Subatomic Swirl}},\ }\href {https://doi.org/10.1038/s41598-020-59129-6}
  {\bibfield  {journal} {\bibinfo  {journal} {Sci. Rep.}\ }\textbf {\bibinfo
  {volume} {10}},\ \bibinfo {pages} {2196} (\bibinfo {year} {2020})},\ \Eprint
  {https://arxiv.org/abs/1904.04704} {arXiv:1904.04704 [hep-ph]} \BibitemShut
  {NoStop}%
\bibitem [{\citenamefont {McLerran}\ and\ \citenamefont
  {Skokov}(2014)}]{McLerran:2013hla}%
  \BibitemOpen
  \bibfield  {author} {\bibinfo {author} {\bibfnamefont {L.}~\bibnamefont
  {McLerran}}\ and\ \bibinfo {author} {\bibfnamefont {V.}~\bibnamefont
  {Skokov}},\ }\bibfield  {title} {\bibinfo {title} {{Comments About the
  Electromagnetic Field in Heavy-Ion Collisions}},\ }\href
  {https://doi.org/10.1016/j.nuclphysa.2014.05.008} {\bibfield  {journal}
  {\bibinfo  {journal} {Nucl. Phys. A}\ }\textbf {\bibinfo {volume} {929}},\
  \bibinfo {pages} {184} (\bibinfo {year} {2014})},\ \Eprint
  {https://arxiv.org/abs/1305.0774} {arXiv:1305.0774 [hep-ph]} \BibitemShut
  {NoStop}%
\bibitem [{\citenamefont {Tuchin}(2013{\natexlab{a}})}]{Tuchin:2013apa}%
  \BibitemOpen
  \bibfield  {author} {\bibinfo {author} {\bibfnamefont {K.}~\bibnamefont
  {Tuchin}},\ }\bibfield  {title} {\bibinfo {title} {{Time and space dependence
  of the electromagnetic field in relativistic heavy-ion collisions}},\ }\href
  {https://doi.org/10.1103/PhysRevC.88.024911} {\bibfield  {journal} {\bibinfo
  {journal} {Phys. Rev. C}\ }\textbf {\bibinfo {volume} {88}},\ \bibinfo
  {pages} {024911} (\bibinfo {year} {2013}{\natexlab{a}})},\ \Eprint
  {https://arxiv.org/abs/1305.5806} {arXiv:1305.5806 [hep-ph]} \BibitemShut
  {NoStop}%
\bibitem [{\citenamefont {Gursoy}\ \emph {et~al.}(2014)\citenamefont {Gursoy},
  \citenamefont {Kharzeev},\ and\ \citenamefont {Rajagopal}}]{Gursoy:2014aka}%
  \BibitemOpen
  \bibfield  {author} {\bibinfo {author} {\bibfnamefont {U.}~\bibnamefont
  {Gursoy}}, \bibinfo {author} {\bibfnamefont {D.}~\bibnamefont {Kharzeev}},\
  and\ \bibinfo {author} {\bibfnamefont {K.}~\bibnamefont {Rajagopal}},\
  }\bibfield  {title} {\bibinfo {title} {{Magnetohydrodynamics, charged
  currents and directed flow in heavy ion collisions}},\ }\href
  {https://doi.org/10.1103/PhysRevC.89.054905} {\bibfield  {journal} {\bibinfo
  {journal} {Phys. Rev. C}\ }\textbf {\bibinfo {volume} {89}},\ \bibinfo
  {pages} {054905} (\bibinfo {year} {2014})},\ \Eprint
  {https://arxiv.org/abs/1401.3805} {arXiv:1401.3805 [hep-ph]} \BibitemShut
  {NoStop}%
\bibitem [{\citenamefont {Zhong}\ \emph {et~al.}(2014)\citenamefont {Zhong},
  \citenamefont {Yang}, \citenamefont {Cai},\ and\ \citenamefont
  {Feng}}]{Zhong:2014cda}%
  \BibitemOpen
  \bibfield  {author} {\bibinfo {author} {\bibfnamefont {Y.}~\bibnamefont
  {Zhong}}, \bibinfo {author} {\bibfnamefont {C.-B.}\ \bibnamefont {Yang}},
  \bibinfo {author} {\bibfnamefont {X.}~\bibnamefont {Cai}},\ and\ \bibinfo
  {author} {\bibfnamefont {S.-Q.}\ \bibnamefont {Feng}},\ }\bibfield  {title}
  {\bibinfo {title} {{A systematic study of magnetic field in Relativistic
  Heavy-ion Collisions in the RHIC and LHC energy regions}},\ }\href
  {https://doi.org/10.1155/2014/193039} {\bibfield  {journal} {\bibinfo
  {journal} {Adv. High Energy Phys.}\ }\textbf {\bibinfo {volume} {2014}},\
  \bibinfo {pages} {193039} (\bibinfo {year} {2014})},\ \Eprint
  {https://arxiv.org/abs/1408.5694} {arXiv:1408.5694 [hep-ph]} \BibitemShut
  {NoStop}%
\bibitem [{\citenamefont {Tuchin}(2016)}]{Tuchin:2015oka}%
  \BibitemOpen
  \bibfield  {author} {\bibinfo {author} {\bibfnamefont {K.}~\bibnamefont
  {Tuchin}},\ }\bibfield  {title} {\bibinfo {title} {{Initial value problem for
  magnetic fields in heavy ion collisions}},\ }\href
  {https://doi.org/10.1103/PhysRevC.93.014905} {\bibfield  {journal} {\bibinfo
  {journal} {Phys. Rev. C}\ }\textbf {\bibinfo {volume} {93}},\ \bibinfo
  {pages} {014905} (\bibinfo {year} {2016})},\ \Eprint
  {https://arxiv.org/abs/1508.06925} {arXiv:1508.06925 [hep-ph]} \BibitemShut
  {NoStop}%
\bibitem [{\citenamefont {Li}\ \emph {et~al.}(2016)\citenamefont {Li},
  \citenamefont {Sheng},\ and\ \citenamefont {Wang}}]{Li:2016tel}%
  \BibitemOpen
  \bibfield  {author} {\bibinfo {author} {\bibfnamefont {H.}~\bibnamefont
  {Li}}, \bibinfo {author} {\bibfnamefont {X.-l.}\ \bibnamefont {Sheng}},\ and\
  \bibinfo {author} {\bibfnamefont {Q.}~\bibnamefont {Wang}},\ }\bibfield
  {title} {\bibinfo {title} {{Electromagnetic fields with electric and chiral
  magnetic conductivities in heavy ion collisions}},\ }\href
  {https://doi.org/10.1103/PhysRevC.94.044903} {\bibfield  {journal} {\bibinfo
  {journal} {Phys. Rev. C}\ }\textbf {\bibinfo {volume} {94}},\ \bibinfo
  {pages} {044903} (\bibinfo {year} {2016})},\ \Eprint
  {https://arxiv.org/abs/1602.02223} {arXiv:1602.02223 [nucl-th]} \BibitemShut
  {NoStop}%
\bibitem [{\citenamefont {Gorbar}\ \emph {et~al.}(2021)\citenamefont {Gorbar},
  \citenamefont {Miransky}, \citenamefont {Shovkovy},\ and\ \citenamefont
  {Sukhachov}}]{Gorbar:2021ebc}%
  \BibitemOpen
  \bibfield  {author} {\bibinfo {author} {\bibfnamefont {E.~V.}\ \bibnamefont
  {Gorbar}}, \bibinfo {author} {\bibfnamefont {V.~A.}\ \bibnamefont
  {Miransky}}, \bibinfo {author} {\bibfnamefont {I.~A.}\ \bibnamefont
  {Shovkovy}},\ and\ \bibinfo {author} {\bibfnamefont {P.~O.}\ \bibnamefont
  {Sukhachov}},\ }\href {https://doi.org/10.1142/11475} {\emph {\bibinfo
  {title} {{Electronic Properties of Dirac and Weyl Semimetals}}}}\ (\bibinfo
  {publisher} {World Scientific},\ \bibinfo {address} {Singapore},\ \bibinfo
  {year} {2021})\BibitemShut {NoStop}%
\bibitem [{\citenamefont {Heisenberg}\ and\ \citenamefont
  {Euler}(1936)}]{Heisenberg:1936nmg}%
  \BibitemOpen
  \bibfield  {author} {\bibinfo {author} {\bibfnamefont {W.}~\bibnamefont
  {Heisenberg}}\ and\ \bibinfo {author} {\bibfnamefont {H.}~\bibnamefont
  {Euler}},\ }\bibfield  {title} {\bibinfo {title} {{Consequences of Dirac's
  theory of positrons}},\ }\href {https://doi.org/10.1007/BF01343663}
  {\bibfield  {journal} {\bibinfo  {journal} {Z. Phys.}\ }\textbf {\bibinfo
  {volume} {98}},\ \bibinfo {pages} {714} (\bibinfo {year} {1936})}\BibitemShut
  {NoStop}%
\bibitem [{\citenamefont {Schwinger}(1951)}]{Schwinger:1951nm}%
  \BibitemOpen
  \bibfield  {author} {\bibinfo {author} {\bibfnamefont {J.~S.}\ \bibnamefont
  {Schwinger}},\ }\bibfield  {title} {\bibinfo {title} {{On gauge invariance
  and vacuum polarization}},\ }\href {https://doi.org/10.1103/PhysRev.82.664}
  {\bibfield  {journal} {\bibinfo  {journal} {Phys. Rev.}\ }\textbf {\bibinfo
  {volume} {82}},\ \bibinfo {pages} {664} (\bibinfo {year} {1951})}\BibitemShut
  {NoStop}%
\bibitem [{\citenamefont {Dittrich}\ and\ \citenamefont
  {Reuter}(1985)}]{Dittrich:1985yb}%
  \BibitemOpen
  \bibfield  {author} {\bibinfo {author} {\bibfnamefont {W.}~\bibnamefont
  {Dittrich}}\ and\ \bibinfo {author} {\bibfnamefont {M.}~\bibnamefont
  {Reuter}},\ }\href@noop {} {\emph {\bibinfo {title} {{Effective Lagrangians
  in Quantum Electrodynamics}, {Lecture Notes in Physics}}}},\ Vol.\ \bibinfo
  {volume} {220}\ (\bibinfo  {publisher} {Springer},\ \bibinfo {address}
  {Berlin},\ \bibinfo {year} {1985})\BibitemShut {NoStop}%
\bibitem [{\citenamefont {Andersen}\ \emph {et~al.}(2016)\citenamefont
  {Andersen}, \citenamefont {Naylor},\ and\ \citenamefont
  {Tranberg}}]{Andersen:2014xxa}%
  \BibitemOpen
  \bibfield  {author} {\bibinfo {author} {\bibfnamefont {J.~O.}\ \bibnamefont
  {Andersen}}, \bibinfo {author} {\bibfnamefont {W.~R.}\ \bibnamefont
  {Naylor}},\ and\ \bibinfo {author} {\bibfnamefont {A.}~\bibnamefont
  {Tranberg}},\ }\bibfield  {title} {\bibinfo {title} {{Phase diagram of QCD in
  a magnetic field: A review}},\ }\href
  {https://doi.org/10.1103/RevModPhys.88.025001} {\bibfield  {journal}
  {\bibinfo  {journal} {Rev. Mod. Phys.}\ }\textbf {\bibinfo {volume} {88}},\
  \bibinfo {pages} {025001} (\bibinfo {year} {2016})},\ \Eprint
  {https://arxiv.org/abs/1411.7176} {arXiv:1411.7176 [hep-ph]} \BibitemShut
  {NoStop}%
\bibitem [{\citenamefont {Miransky}\ and\ \citenamefont
  {Shovkovy}(2015)}]{Miransky:2015ava}%
  \BibitemOpen
  \bibfield  {author} {\bibinfo {author} {\bibfnamefont {V.~A.}\ \bibnamefont
  {Miransky}}\ and\ \bibinfo {author} {\bibfnamefont {I.~A.}\ \bibnamefont
  {Shovkovy}},\ }\bibfield  {title} {\bibinfo {title} {{Quantum field theory in
  a magnetic field: From quantum chromodynamics to graphene and Dirac
  semimetals}},\ }\href {https://doi.org/10.1016/j.physrep.2015.02.003}
  {\bibfield  {journal} {\bibinfo  {journal} {Phys. Rept.}\ }\textbf {\bibinfo
  {volume} {576}},\ \bibinfo {pages} {1} (\bibinfo {year} {2015})},\ \Eprint
  {https://arxiv.org/abs/1503.00732} {arXiv:1503.00732 [hep-ph]} \BibitemShut
  {NoStop}%
\bibitem [{\citenamefont {Bandyopadhyay}\ \emph {et~al.}(2016)\citenamefont
  {Bandyopadhyay}, \citenamefont {Islam},\ and\ \citenamefont
  {Mustafa}}]{Bandyopadhyay:2016fyd}%
  \BibitemOpen
  \bibfield  {author} {\bibinfo {author} {\bibfnamefont {A.}~\bibnamefont
  {Bandyopadhyay}}, \bibinfo {author} {\bibfnamefont {C.~A.}\ \bibnamefont
  {Islam}},\ and\ \bibinfo {author} {\bibfnamefont {M.~G.}\ \bibnamefont
  {Mustafa}},\ }\bibfield  {title} {\bibinfo {title} {{Electromagnetic spectral
  properties and Debye screening of a strongly magnetized hot medium}},\ }\href
  {https://doi.org/10.1103/PhysRevD.94.114034} {\bibfield  {journal} {\bibinfo
  {journal} {Phys. Rev. D}\ }\textbf {\bibinfo {volume} {94}},\ \bibinfo
  {pages} {114034} (\bibinfo {year} {2016})},\ \Eprint
  {https://arxiv.org/abs/1602.06769} {arXiv:1602.06769 [hep-ph]} \BibitemShut
  {NoStop}%
\bibitem [{\citenamefont {Das}\ \emph {et~al.}(2019)\citenamefont {Das},
  \citenamefont {Haque}, \citenamefont {Mustafa},\ and\ \citenamefont
  {Roy}}]{Das:2019nzv}%
  \BibitemOpen
  \bibfield  {author} {\bibinfo {author} {\bibfnamefont {A.}~\bibnamefont
  {Das}}, \bibinfo {author} {\bibfnamefont {N.}~\bibnamefont {Haque}}, \bibinfo
  {author} {\bibfnamefont {M.~G.}\ \bibnamefont {Mustafa}},\ and\ \bibinfo
  {author} {\bibfnamefont {P.~K.}\ \bibnamefont {Roy}},\ }\bibfield  {title}
  {\bibinfo {title} {{Hard dilepton production from a weakly magnetized hot QCD
  medium}},\ }\href {https://doi.org/10.1103/PhysRevD.99.094022} {\bibfield
  {journal} {\bibinfo  {journal} {Phys. Rev.}\ }\textbf {\bibinfo {volume}
  {D99}},\ \bibinfo {pages} {094022} (\bibinfo {year} {2019})},\ \Eprint
  {https://arxiv.org/abs/1903.03528} {arXiv:1903.03528} \BibitemShut {NoStop}%
\bibitem [{\citenamefont {Ghosh}\ \emph
  {et~al.}(2020{\natexlab{a}})\citenamefont {Ghosh}, \citenamefont {Karmakar},\
  and\ \citenamefont {Mustafa}}]{Ghosh:2019kmf}%
  \BibitemOpen
  \bibfield  {author} {\bibinfo {author} {\bibfnamefont {R.}~\bibnamefont
  {Ghosh}}, \bibinfo {author} {\bibfnamefont {B.}~\bibnamefont {Karmakar}},\
  and\ \bibinfo {author} {\bibfnamefont {M.~G.}\ \bibnamefont {Mustafa}},\
  }\bibfield  {title} {\bibinfo {title} {{Soft contribution to the damping rate
  of a hard photon in a weakly magnetized hot medium}},\ }\href
  {https://doi.org/10.1103/PhysRevD.101.056007} {\bibfield  {journal} {\bibinfo
   {journal} {Phys. Rev. D}\ }\textbf {\bibinfo {volume} {101}},\ \bibinfo
  {pages} {056007} (\bibinfo {year} {2020}{\natexlab{a}})},\ \Eprint
  {https://arxiv.org/abs/1911.00744} {arXiv:1911.00744 [hep-ph]} \BibitemShut
  {NoStop}%
\bibitem [{\citenamefont {Hattori}\ and\ \citenamefont
  {Itakura}(2013{\natexlab{a}})}]{Hattori:2012je}%
  \BibitemOpen
  \bibfield  {author} {\bibinfo {author} {\bibfnamefont {K.}~\bibnamefont
  {Hattori}}\ and\ \bibinfo {author} {\bibfnamefont {K.}~\bibnamefont
  {Itakura}},\ }\bibfield  {title} {\bibinfo {title} {{Vacuum birefringence in
  strong magnetic fields: (I) Photon polarization tensor with all the Landau
  levels}},\ }\href {https://doi.org/10.1016/j.aop.2012.11.010} {\bibfield
  {journal} {\bibinfo  {journal} {Annals Phys.}\ }\textbf {\bibinfo {volume}
  {330}},\ \bibinfo {pages} {23} (\bibinfo {year} {2013}{\natexlab{a}})},\
  \Eprint {https://arxiv.org/abs/1209.2663} {arXiv:1209.2663} \BibitemShut
  {NoStop}%
\bibitem [{\citenamefont {Hattori}\ and\ \citenamefont
  {Itakura}(2013{\natexlab{b}})}]{Hattori:2012ny}%
  \BibitemOpen
  \bibfield  {author} {\bibinfo {author} {\bibfnamefont {K.}~\bibnamefont
  {Hattori}}\ and\ \bibinfo {author} {\bibfnamefont {K.}~\bibnamefont
  {Itakura}},\ }\bibfield  {title} {\bibinfo {title} {{Vacuum birefringence in
  strong magnetic fields: (II) Complex refractive index from the lowest Landau
  level}},\ }\href {https://doi.org/10.1016/j.aop.2013.03.016} {\bibfield
  {journal} {\bibinfo  {journal} {Annals Phys.}\ }\textbf {\bibinfo {volume}
  {334}},\ \bibinfo {pages} {58} (\bibinfo {year} {2013}{\natexlab{b}})},\
  \Eprint {https://arxiv.org/abs/1212.1897} {arXiv:1212.1897} \BibitemShut
  {NoStop}%
\bibitem [{\citenamefont {Tuchin}(2013{\natexlab{b}})}]{Tuchin:2013bda}%
  \BibitemOpen
  \bibfield  {author} {\bibinfo {author} {\bibfnamefont {K.}~\bibnamefont
  {Tuchin}},\ }\bibfield  {title} {\bibinfo {title} {{Magnetic contribution to
  dilepton production in heavy-ion collisions}},\ }\href
  {https://doi.org/10.1103/PhysRevC.88.024910} {\bibfield  {journal} {\bibinfo
  {journal} {Phys. Rev.}\ }\textbf {\bibinfo {volume} {C88}},\ \bibinfo {pages}
  {024910} (\bibinfo {year} {2013}{\natexlab{b}})},\ \Eprint
  {https://arxiv.org/abs/1305.0545} {arXiv:1305.0545} \BibitemShut {NoStop}%
\bibitem [{\citenamefont {Karbstein}(2013)}]{Karbstein:2013ufa}%
  \BibitemOpen
  \bibfield  {author} {\bibinfo {author} {\bibfnamefont {F.}~\bibnamefont
  {Karbstein}},\ }\bibfield  {title} {\bibinfo {title} {{Photon polarization
  tensor in a homogeneous magnetic or electric field}},\ }\href
  {https://doi.org/10.1103/PhysRevD.88.085033} {\bibfield  {journal} {\bibinfo
  {journal} {Phys. Rev.}\ }\textbf {\bibinfo {volume} {D88}},\ \bibinfo {pages}
  {085033} (\bibinfo {year} {2013})},\ \Eprint
  {https://arxiv.org/abs/1308.6184} {arXiv:1308.6184} \BibitemShut {NoStop}%
\bibitem [{\citenamefont {Ishikawa}\ \emph {et~al.}(2013)\citenamefont
  {Ishikawa}, \citenamefont {Kimura}, \citenamefont {Shigaki},\ and\
  \citenamefont {Tsuji}}]{Ishikawa:2013fxa}%
  \BibitemOpen
  \bibfield  {author} {\bibinfo {author} {\bibfnamefont {K.-I.}\ \bibnamefont
  {Ishikawa}}, \bibinfo {author} {\bibfnamefont {D.}~\bibnamefont {Kimura}},
  \bibinfo {author} {\bibfnamefont {K.}~\bibnamefont {Shigaki}},\ and\ \bibinfo
  {author} {\bibfnamefont {A.}~\bibnamefont {Tsuji}},\ }\bibfield  {title}
  {\bibinfo {title} {{A numerical evaluation of vacuum polarization tensor in
  constant external magnetic fields}},\ }\href
  {https://doi.org/10.1142/S0217751X13501005} {\bibfield  {journal} {\bibinfo
  {journal} {Int. J. Mod. Phys.}\ }\textbf {\bibinfo {volume} {A28}},\ \bibinfo
  {pages} {1350100} (\bibinfo {year} {2013})},\ \Eprint
  {https://arxiv.org/abs/1304.3655} {arXiv:1304.3655} \BibitemShut {NoStop}%
\bibitem [{\citenamefont {Sadooghi}\ and\ \citenamefont
  {Taghinavaz}(2017)}]{Sadooghi:2016jyf}%
  \BibitemOpen
  \bibfield  {author} {\bibinfo {author} {\bibfnamefont {N.}~\bibnamefont
  {Sadooghi}}\ and\ \bibinfo {author} {\bibfnamefont {F.}~\bibnamefont
  {Taghinavaz}},\ }\bibfield  {title} {\bibinfo {title} {{Dilepton production
  rate in a hot and magnetized quark-gluon plasma}},\ }\href
  {https://doi.org/10.1016/j.aop.2016.11.008} {\bibfield  {journal} {\bibinfo
  {journal} {Annals Phys.}\ }\textbf {\bibinfo {volume} {376}},\ \bibinfo
  {pages} {218} (\bibinfo {year} {2017})},\ \Eprint
  {https://arxiv.org/abs/1601.04887} {arXiv:1601.04887} \BibitemShut {NoStop}%
\bibitem [{\citenamefont {Ayala}\ \emph {et~al.}(2020)\citenamefont {Ayala},
  \citenamefont {Casta\~no Yepes}, \citenamefont {Loewe},\ and\ \citenamefont
  {Mu\~noz}}]{Ayala:2019akk}%
  \BibitemOpen
  \bibfield  {author} {\bibinfo {author} {\bibfnamefont {A.}~\bibnamefont
  {Ayala}}, \bibinfo {author} {\bibfnamefont {J.~D.}\ \bibnamefont {Casta\~no
  Yepes}}, \bibinfo {author} {\bibfnamefont {M.}~\bibnamefont {Loewe}},\ and\
  \bibinfo {author} {\bibfnamefont {E.}~\bibnamefont {Mu\~noz}},\ }\bibfield
  {title} {\bibinfo {title} {{Gluon polarization tensor in a magnetized medium:
  Analytic approach starting from the sum over Landau levels}},\ }\href
  {https://doi.org/10.1103/PhysRevD.101.036016} {\bibfield  {journal} {\bibinfo
   {journal} {Phys. Rev. D}\ }\textbf {\bibinfo {volume} {101}},\ \bibinfo
  {pages} {036016} (\bibinfo {year} {2020})},\ \Eprint
  {https://arxiv.org/abs/1912.07136} {arXiv:1912.07136 [hep-th]} \BibitemShut
  {NoStop}%
\bibitem [{\citenamefont {Ayala}\ \emph
  {et~al.}(2021{\natexlab{a}})\citenamefont {Ayala}, \citenamefont {Casta\~no
  Yepes}, \citenamefont {Hern\'andez}, \citenamefont {Salinas
  San~Mart\'\i{}n},\ and\ \citenamefont {Zamora}}]{Ayala:2020wzl}%
  \BibitemOpen
  \bibfield  {author} {\bibinfo {author} {\bibfnamefont {A.}~\bibnamefont
  {Ayala}}, \bibinfo {author} {\bibfnamefont {J.~D.}\ \bibnamefont {Casta\~no
  Yepes}}, \bibinfo {author} {\bibfnamefont {L.~A.}\ \bibnamefont
  {Hern\'andez}}, \bibinfo {author} {\bibfnamefont {J.}~\bibnamefont {Salinas
  San~Mart\'\i{}n}},\ and\ \bibinfo {author} {\bibfnamefont {R.}~\bibnamefont
  {Zamora}},\ }\bibfield  {title} {\bibinfo {title} {{Gluon polarization tensor
  and dispersion relation in a weakly magnetized medium}},\ }\href
  {https://doi.org/10.1140/epja/s10050-021-00429-4} {\bibfield  {journal}
  {\bibinfo  {journal} {Eur. Phys. J. A}\ }\textbf {\bibinfo {volume} {57}},\
  \bibinfo {pages} {140} (\bibinfo {year} {2021}{\natexlab{a}})},\ \Eprint
  {https://arxiv.org/abs/2009.00830} {arXiv:2009.00830 [hep-ph]} \BibitemShut
  {NoStop}%
\bibitem [{\citenamefont {Hattori}\ \emph {et~al.}(2021)\citenamefont
  {Hattori}, \citenamefont {Taya},\ and\ \citenamefont
  {Yoshida}}]{Hattori:2020htm}%
  \BibitemOpen
  \bibfield  {author} {\bibinfo {author} {\bibfnamefont {K.}~\bibnamefont
  {Hattori}}, \bibinfo {author} {\bibfnamefont {H.}~\bibnamefont {Taya}},\ and\
  \bibinfo {author} {\bibfnamefont {S.}~\bibnamefont {Yoshida}},\ }\bibfield
  {title} {\bibinfo {title} {{Di-lepton production from a single photon in
  strong magnetic fields: vacuum dichroism}},\ }\href
  {https://doi.org/10.1007/JHEP01(2021)093} {\bibfield  {journal} {\bibinfo
  {journal} {JHEP}\ }\textbf {\bibinfo {volume} {01}},\ \bibinfo {pages}
  {093}},\ \Eprint {https://arxiv.org/abs/2010.13492} {arXiv:2010.13492
  [hep-ph]} \BibitemShut {NoStop}%
\bibitem [{\citenamefont {Ghosh}\ and\ \citenamefont
  {Chandra}(2018)}]{Ghosh:2018xhh}%
  \BibitemOpen
  \bibfield  {author} {\bibinfo {author} {\bibfnamefont {S.}~\bibnamefont
  {Ghosh}}\ and\ \bibinfo {author} {\bibfnamefont {V.}~\bibnamefont
  {Chandra}},\ }\bibfield  {title} {\bibinfo {title} {{Electromagnetic spectral
  function and dilepton rate in a hot magnetized QCD medium}},\ }\href
  {https://doi.org/10.1103/PhysRevD.98.076006} {\bibfield  {journal} {\bibinfo
  {journal} {Phys. Rev.}\ }\textbf {\bibinfo {volume} {D98}},\ \bibinfo {pages}
  {076006} (\bibinfo {year} {2018})},\ \Eprint
  {https://arxiv.org/abs/1808.05176} {arXiv:1808.05176} \BibitemShut {NoStop}%
\bibitem [{\citenamefont {Ghosh}\ \emph
  {et~al.}(2020{\natexlab{b}})\citenamefont {Ghosh}, \citenamefont {Chaudhuri},
  \citenamefont {Sarkar},\ and\ \citenamefont {Roy}}]{Ghosh:2020xwp}%
  \BibitemOpen
  \bibfield  {author} {\bibinfo {author} {\bibfnamefont {S.}~\bibnamefont
  {Ghosh}}, \bibinfo {author} {\bibfnamefont {N.}~\bibnamefont {Chaudhuri}},
  \bibinfo {author} {\bibfnamefont {S.}~\bibnamefont {Sarkar}},\ and\ \bibinfo
  {author} {\bibfnamefont {P.}~\bibnamefont {Roy}},\ }\bibfield  {title}
  {\bibinfo {title} {{Effects of the anomalous magnetic moment of quarks on the
  dilepton production from hot and dense magnetized quark matter using the NJL
  model}},\ }\href {https://doi.org/10.1103/PhysRevD.101.096002} {\bibfield
  {journal} {\bibinfo  {journal} {Phys. Rev. D}\ }\textbf {\bibinfo {volume}
  {101}},\ \bibinfo {pages} {096002} (\bibinfo {year} {2020}{\natexlab{b}})},\
  \Eprint {https://arxiv.org/abs/2004.09203} {arXiv:2004.09203 [nucl-th]}
  \BibitemShut {NoStop}%
\bibitem [{\citenamefont {Wang}\ \emph {et~al.}(2020)\citenamefont {Wang},
  \citenamefont {Shovkovy}, \citenamefont {Yu},\ and\ \citenamefont
  {Huang}}]{Wang:2020dsr}%
  \BibitemOpen
  \bibfield  {author} {\bibinfo {author} {\bibfnamefont {X.}~\bibnamefont
  {Wang}}, \bibinfo {author} {\bibfnamefont {I.~A.}\ \bibnamefont {Shovkovy}},
  \bibinfo {author} {\bibfnamefont {L.}~\bibnamefont {Yu}},\ and\ \bibinfo
  {author} {\bibfnamefont {M.}~\bibnamefont {Huang}},\ }\bibfield  {title}
  {\bibinfo {title} {{Ellipticity of photon emission from strongly magnetized
  hot QCD plasma}},\ }\href {https://doi.org/10.1103/PhysRevD.102.076010}
  {\bibfield  {journal} {\bibinfo  {journal} {Phys. Rev. D}\ }\textbf {\bibinfo
  {volume} {102}},\ \bibinfo {pages} {076010} (\bibinfo {year} {2020})},\
  \Eprint {https://arxiv.org/abs/2006.16254} {arXiv:2006.16254} \BibitemShut
  {NoStop}%
\bibitem [{\citenamefont {Wang}\ and\ \citenamefont
  {Shovkovy}(2021)}]{Wang:2021ebh}%
  \BibitemOpen
  \bibfield  {author} {\bibinfo {author} {\bibfnamefont {X.}~\bibnamefont
  {Wang}}\ and\ \bibinfo {author} {\bibfnamefont {I.}~\bibnamefont
  {Shovkovy}},\ }\bibfield  {title} {\bibinfo {title} {{Photon polarization
  tensor in a magnetized plasma: Absorptive part}},\ }\href
  {https://doi.org/10.1103/PhysRevD.104.056017} {\bibfield  {journal} {\bibinfo
   {journal} {Phys. Rev. D}\ }\textbf {\bibinfo {volume} {104}},\ \bibinfo
  {pages} {056017} (\bibinfo {year} {2021})},\ \Eprint
  {https://arxiv.org/abs/2103.01967} {arXiv:2103.01967 [nucl-th]} \BibitemShut
  {NoStop}%
\bibitem [{\citenamefont {Chaudhuri}\ \emph {et~al.}(2021)\citenamefont
  {Chaudhuri}, \citenamefont {Ghosh}, \citenamefont {Sarkar},\ and\
  \citenamefont {Roy}}]{Chaudhuri:2021skc}%
  \BibitemOpen
  \bibfield  {author} {\bibinfo {author} {\bibfnamefont {N.}~\bibnamefont
  {Chaudhuri}}, \bibinfo {author} {\bibfnamefont {S.}~\bibnamefont {Ghosh}},
  \bibinfo {author} {\bibfnamefont {S.}~\bibnamefont {Sarkar}},\ and\ \bibinfo
  {author} {\bibfnamefont {P.}~\bibnamefont {Roy}},\ }\bibfield  {title}
  {\bibinfo {title} {{Dilepton production from magnetized quark matter with an
  anomalous magnetic moment of the quarks using a three-flavor PNJL model}},\
  }\href {https://doi.org/10.1103/PhysRevD.103.096021} {\bibfield  {journal}
  {\bibinfo  {journal} {Phys. Rev. D}\ }\textbf {\bibinfo {volume} {103}},\
  \bibinfo {pages} {096021} (\bibinfo {year} {2021})},\ \Eprint
  {https://arxiv.org/abs/2104.11425} {arXiv:2104.11425 [hep-ph]} \BibitemShut
  {NoStop}%
\bibitem [{\citenamefont {Das}\ \emph {et~al.}(2022)\citenamefont {Das},
  \citenamefont {Bandyopadhyay},\ and\ \citenamefont {Islam}}]{Das:2021fma}%
  \BibitemOpen
  \bibfield  {author} {\bibinfo {author} {\bibfnamefont {A.}~\bibnamefont
  {Das}}, \bibinfo {author} {\bibfnamefont {A.}~\bibnamefont {Bandyopadhyay}},\
  and\ \bibinfo {author} {\bibfnamefont {C.~A.}\ \bibnamefont {Islam}},\
  }\bibfield  {title} {\bibinfo {title} {{Lepton pair production from a hot and
  dense QCD medium in the presence of an arbitrary magnetic field}},\ }\href
  {https://doi.org/10.1103/PhysRevD.106.056021} {\bibfield  {journal} {\bibinfo
   {journal} {Phys. Rev. D}\ }\textbf {\bibinfo {volume} {106}},\ \bibinfo
  {pages} {056021} (\bibinfo {year} {2022})},\ \Eprint
  {https://arxiv.org/abs/2109.00019} {arXiv:2109.00019 [hep-ph]} \BibitemShut
  {NoStop}%
\bibitem [{\citenamefont {Wang}\ and\ \citenamefont
  {Shovkovy}(2022)}]{Wang:2022jxx}%
  \BibitemOpen
  \bibfield  {author} {\bibinfo {author} {\bibfnamefont {X.}~\bibnamefont
  {Wang}}\ and\ \bibinfo {author} {\bibfnamefont {I.~A.}\ \bibnamefont
  {Shovkovy}},\ }\bibfield  {title} {\bibinfo {title} {{Rate and ellipticity of
  dilepton production in a magnetized quark-gluon plasma}},\ }\href
  {https://doi.org/10.1103/PhysRevD.106.036014} {\bibfield  {journal} {\bibinfo
   {journal} {Phys. Rev. D}\ }\textbf {\bibinfo {volume} {106}},\ \bibinfo
  {pages} {036014} (\bibinfo {year} {2022})},\ \Eprint
  {https://arxiv.org/abs/2205.00276} {arXiv:2205.00276 [nucl-th]} \BibitemShut
  {NoStop}%
\bibitem [{\citenamefont {Gorbar}\ \emph {et~al.}(2012)\citenamefont {Gorbar},
  \citenamefont {Gusynin}, \citenamefont {Miransky},\ and\ \citenamefont
  {Shovkovy}}]{Gorbar:2011kc}%
  \BibitemOpen
  \bibfield  {author} {\bibinfo {author} {\bibfnamefont {E.~V.}\ \bibnamefont
  {Gorbar}}, \bibinfo {author} {\bibfnamefont {V.~P.}\ \bibnamefont {Gusynin}},
  \bibinfo {author} {\bibfnamefont {V.~A.}\ \bibnamefont {Miransky}},\ and\
  \bibinfo {author} {\bibfnamefont {I.~A.}\ \bibnamefont {Shovkovy}},\
  }\bibfield  {title} {\bibinfo {title} {{Coulomb interaction and magnetic
  catalysis in the quantum Hall effect in graphene}},\ }\href
  {https://doi.org/10.1088/0031-8949/2012/T146/014018} {\bibfield  {journal}
  {\bibinfo  {journal} {Phys. Scripta T}\ }\textbf {\bibinfo {volume} {146}},\
  \bibinfo {pages} {014018} (\bibinfo {year} {2012})},\ \Eprint
  {https://arxiv.org/abs/1105.1360} {arXiv:1105.1360 [cond-mat.mes-hall]}
  \BibitemShut {NoStop}%
\bibitem [{\citenamefont {Shovkovy}\ and\ \citenamefont
  {Xia}(2016)}]{Shovkovy:2015kja}%
  \BibitemOpen
  \bibfield  {author} {\bibinfo {author} {\bibfnamefont {I.~A.}\ \bibnamefont
  {Shovkovy}}\ and\ \bibinfo {author} {\bibfnamefont {L.}~\bibnamefont {Xia}},\
  }\bibfield  {title} {\bibinfo {title} {{Generalized Landau level
  representation: Effect of static screening in the quantum Hall effect in
  graphene}},\ }\href {https://doi.org/10.1103/PhysRevB.93.035454} {\bibfield
  {journal} {\bibinfo  {journal} {Phys. Rev. B}\ }\textbf {\bibinfo {volume}
  {93}},\ \bibinfo {pages} {035454} (\bibinfo {year} {2016})},\ \Eprint
  {https://arxiv.org/abs/1508.04471} {arXiv:1508.04471 [cond-mat.mes-hall]}
  \BibitemShut {NoStop}%
\bibitem [{\citenamefont {Gorbar}\ \emph
  {et~al.}(2013{\natexlab{a}})\citenamefont {Gorbar}, \citenamefont {Miransky},
  \citenamefont {Shovkovy},\ and\ \citenamefont {Wang}}]{Gorbar:2013uga}%
  \BibitemOpen
  \bibfield  {author} {\bibinfo {author} {\bibfnamefont {E.~V.}\ \bibnamefont
  {Gorbar}}, \bibinfo {author} {\bibfnamefont {V.~A.}\ \bibnamefont
  {Miransky}}, \bibinfo {author} {\bibfnamefont {I.~A.}\ \bibnamefont
  {Shovkovy}},\ and\ \bibinfo {author} {\bibfnamefont {X.}~\bibnamefont
  {Wang}},\ }\bibfield  {title} {\bibinfo {title} {{Chiral asymmetry in QED
  matter in a magnetic field}},\ }\href
  {https://doi.org/10.1103/PhysRevD.88.025043} {\bibfield  {journal} {\bibinfo
  {journal} {Phys. Rev. D}\ }\textbf {\bibinfo {volume} {88}},\ \bibinfo
  {pages} {025043} (\bibinfo {year} {2013}{\natexlab{a}})},\ \Eprint
  {https://arxiv.org/abs/1306.3245} {arXiv:1306.3245 [hep-ph]} \BibitemShut
  {NoStop}%
\bibitem [{\citenamefont {Gorbar}\ \emph
  {et~al.}(2013{\natexlab{b}})\citenamefont {Gorbar}, \citenamefont {Miransky},
  \citenamefont {Shovkovy},\ and\ \citenamefont {Wang}}]{Gorbar:2013upa}%
  \BibitemOpen
  \bibfield  {author} {\bibinfo {author} {\bibfnamefont {E.~V.}\ \bibnamefont
  {Gorbar}}, \bibinfo {author} {\bibfnamefont {V.~A.}\ \bibnamefont
  {Miransky}}, \bibinfo {author} {\bibfnamefont {I.~A.}\ \bibnamefont
  {Shovkovy}},\ and\ \bibinfo {author} {\bibfnamefont {X.}~\bibnamefont
  {Wang}},\ }\bibfield  {title} {\bibinfo {title} {{Radiative corrections to
  chiral separation effect in QED}},\ }\href
  {https://doi.org/10.1103/PhysRevD.88.025025} {\bibfield  {journal} {\bibinfo
  {journal} {Phys. Rev. D}\ }\textbf {\bibinfo {volume} {88}},\ \bibinfo
  {pages} {025025} (\bibinfo {year} {2013}{\natexlab{b}})},\ \Eprint
  {https://arxiv.org/abs/1304.4606} {arXiv:1304.4606 [hep-ph]} \BibitemShut
  {NoStop}%
\bibitem [{\citenamefont {Tsai}(1974)}]{Tsai:1974df}%
  \BibitemOpen
  \bibfield  {author} {\bibinfo {author} {\bibfnamefont {W.-y.}\ \bibnamefont
  {Tsai}},\ }\bibfield  {title} {\bibinfo {title} {{Modified electron
  Propagation Function in Strong Magnetic Fields}},\ }\href
  {https://doi.org/10.1103/PhysRevD.10.1342} {\bibfield  {journal} {\bibinfo
  {journal} {Phys. Rev. D}\ }\textbf {\bibinfo {volume} {10}},\ \bibinfo
  {pages} {1342} (\bibinfo {year} {1974})}\BibitemShut {NoStop}%
\bibitem [{\citenamefont {Jancovici}(1969)}]{Jancovici:1969exc}%
  \BibitemOpen
  \bibfield  {author} {\bibinfo {author} {\bibfnamefont {B.}~\bibnamefont
  {Jancovici}},\ }\bibfield  {title} {\bibinfo {title} {{Radiative correction
  to the ground-state energy of an electron in an intense magnetic field}},\
  }\href {https://doi.org/10.1103/PhysRev.187.2275} {\bibfield  {journal}
  {\bibinfo  {journal} {Phys. Rev.}\ }\textbf {\bibinfo {volume} {187}},\
  \bibinfo {pages} {2275} (\bibinfo {year} {1969})}\BibitemShut {NoStop}%
\bibitem [{\citenamefont {Gepraegs}\ \emph {et~al.}(1994)\citenamefont
  {Gepraegs}, \citenamefont {Riffert}, \citenamefont {Herold}, \citenamefont
  {Ruder},\ and\ \citenamefont {Wunner}}]{Gepraegs:1994hy}%
  \BibitemOpen
  \bibfield  {author} {\bibinfo {author} {\bibfnamefont {R.}~\bibnamefont
  {Gepraegs}}, \bibinfo {author} {\bibfnamefont {H.}~\bibnamefont {Riffert}},
  \bibinfo {author} {\bibfnamefont {H.}~\bibnamefont {Herold}}, \bibinfo
  {author} {\bibfnamefont {H.}~\bibnamefont {Ruder}},\ and\ \bibinfo {author}
  {\bibfnamefont {G.}~\bibnamefont {Wunner}},\ }\bibfield  {title} {\bibinfo
  {title} {{Electron selfenergy in a homogeneous magnetic field}},\ }\href
  {https://doi.org/10.1103/PhysRevD.49.5582} {\bibfield  {journal} {\bibinfo
  {journal} {Phys. Rev. D}\ }\textbf {\bibinfo {volume} {49}},\ \bibinfo
  {pages} {5582} (\bibinfo {year} {1994})}\BibitemShut {NoStop}%
\bibitem [{\citenamefont {Gusynin}\ and\ \citenamefont
  {Smilga}(1999)}]{Gusynin:1998nh}%
  \BibitemOpen
  \bibfield  {author} {\bibinfo {author} {\bibfnamefont {V.~P.}\ \bibnamefont
  {Gusynin}}\ and\ \bibinfo {author} {\bibfnamefont {A.~V.}\ \bibnamefont
  {Smilga}},\ }\bibfield  {title} {\bibinfo {title} {{Electron selfenergy in
  strong magnetic field: Summation of double logarithmic terms}},\ }\href
  {https://doi.org/10.1016/S0370-2693(99)00145-8} {\bibfield  {journal}
  {\bibinfo  {journal} {Phys. Lett. B}\ }\textbf {\bibinfo {volume} {450}},\
  \bibinfo {pages} {267} (\bibinfo {year} {1999})},\ \Eprint
  {https://arxiv.org/abs/hep-ph/9807486} {arXiv:hep-ph/9807486} \BibitemShut
  {NoStop}%
\bibitem [{\citenamefont {Machet}(2016)}]{Machet:2015swa}%
  \BibitemOpen
  \bibfield  {author} {\bibinfo {author} {\bibfnamefont {B.}~\bibnamefont
  {Machet}},\ }\bibfield  {title} {\bibinfo {title} {{The 1-loop self-energy of
  an electron in a strong external magnetic field revisited}},\ }\href
  {https://doi.org/10.1142/S0217751X16500718} {\bibfield  {journal} {\bibinfo
  {journal} {Int. J. Mod. Phys.}\ }\textbf {\bibinfo {volume} {31}},\ \bibinfo
  {pages} {1650071} (\bibinfo {year} {2016})},\ \Eprint
  {https://arxiv.org/abs/1510.03244} {arXiv:1510.03244 [hep-ph]} \BibitemShut
  {NoStop}%
\bibitem [{\citenamefont {Das}\ \emph {et~al.}(2018)\citenamefont {Das},
  \citenamefont {Bandyopadhyay}, \citenamefont {Roy},\ and\ \citenamefont
  {Mustafa}}]{Das:2017vfh}%
  \BibitemOpen
  \bibfield  {author} {\bibinfo {author} {\bibfnamefont {A.}~\bibnamefont
  {Das}}, \bibinfo {author} {\bibfnamefont {A.}~\bibnamefont {Bandyopadhyay}},
  \bibinfo {author} {\bibfnamefont {P.~K.}\ \bibnamefont {Roy}},\ and\ \bibinfo
  {author} {\bibfnamefont {M.~G.}\ \bibnamefont {Mustafa}},\ }\bibfield
  {title} {\bibinfo {title} {{General structure of fermion two-point function
  and its spectral representation in a hot magnetized medium}},\ }\href
  {https://doi.org/10.1103/PhysRevD.97.034024} {\bibfield  {journal} {\bibinfo
  {journal} {Phys. Rev. D}\ }\textbf {\bibinfo {volume} {97}},\ \bibinfo
  {pages} {034024} (\bibinfo {year} {2018})},\ \Eprint
  {https://arxiv.org/abs/1709.08365} {arXiv:1709.08365 [hep-ph]} \BibitemShut
  {NoStop}%
\bibitem [{\citenamefont {Ayala}\ \emph
  {et~al.}(2021{\natexlab{b}})\citenamefont {Ayala}, \citenamefont {Casta\~no
  Yepes}, \citenamefont {Loewe},\ and\ \citenamefont
  {Mu\~noz}}]{Ayala:2021lor}%
  \BibitemOpen
  \bibfield  {author} {\bibinfo {author} {\bibfnamefont {A.}~\bibnamefont
  {Ayala}}, \bibinfo {author} {\bibfnamefont {J.~D.}\ \bibnamefont {Casta\~no
  Yepes}}, \bibinfo {author} {\bibfnamefont {M.}~\bibnamefont {Loewe}},\ and\
  \bibinfo {author} {\bibfnamefont {E.}~\bibnamefont {Mu\~noz}},\ }\bibfield
  {title} {\bibinfo {title} {{Fermion mass and width in QED in a magnetic
  field}},\ }\href {https://doi.org/10.1103/PhysRevD.104.016006} {\bibfield
  {journal} {\bibinfo  {journal} {Phys. Rev. D}\ }\textbf {\bibinfo {volume}
  {104}},\ \bibinfo {pages} {016006} (\bibinfo {year} {2021}{\natexlab{b}})},\
  \Eprint {https://arxiv.org/abs/2104.04019} {arXiv:2104.04019 [hep-ph]}
  \BibitemShut {NoStop}%
\bibitem [{\citenamefont {Chaudhuri}\ \emph {et~al.}(2023)\citenamefont
  {Chaudhuri}, \citenamefont {Ghosh}, \citenamefont {Roy},\ and\ \citenamefont
  {Sarkar}}]{Chaudhuri:2023djv}%
  \BibitemOpen
  \bibfield  {author} {\bibinfo {author} {\bibfnamefont {N.}~\bibnamefont
  {Chaudhuri}}, \bibinfo {author} {\bibfnamefont {S.}~\bibnamefont {Ghosh}},
  \bibinfo {author} {\bibfnamefont {P.}~\bibnamefont {Roy}},\ and\ \bibinfo
  {author} {\bibfnamefont {S.}~\bibnamefont {Sarkar}},\ }\bibfield  {title}
  {\bibinfo {title} {{Collective modes of a massive fermion in a magnetized
  medium with finite anomalous magnetic moment}},\ }\href
  {https://doi.org/10.1103/PhysRevD.108.116006} {\bibfield  {journal} {\bibinfo
   {journal} {Phys. Rev. D}\ }\textbf {\bibinfo {volume} {108}},\ \bibinfo
  {pages} {116006} (\bibinfo {year} {2023})},\ \Eprint
  {https://arxiv.org/abs/2310.05769} {arXiv:2310.05769 [hep-ph]} \BibitemShut
  {NoStop}%
\bibitem [{\citenamefont {Braaten}\ and\ \citenamefont
  {Pisarski}(1990)}]{Braaten:1989mz}%
  \BibitemOpen
  \bibfield  {author} {\bibinfo {author} {\bibfnamefont {E.}~\bibnamefont
  {Braaten}}\ and\ \bibinfo {author} {\bibfnamefont {R.~D.}\ \bibnamefont
  {Pisarski}},\ }\bibfield  {title} {\bibinfo {title} {{Soft Amplitudes in Hot
  Gauge Theories: A General Analysis}},\ }\href
  {https://doi.org/10.1016/0550-3213(90)90508-B} {\bibfield  {journal}
  {\bibinfo  {journal} {Nucl. Phys. B}\ }\textbf {\bibinfo {volume} {337}},\
  \bibinfo {pages} {569} (\bibinfo {year} {1990})}\BibitemShut {NoStop}%
\bibitem [{\citenamefont {Rebhan}(1992)}]{Rebhan:1992ak}%
  \BibitemOpen
  \bibfield  {author} {\bibinfo {author} {\bibfnamefont {A.}~\bibnamefont
  {Rebhan}},\ }\bibfield  {title} {\bibinfo {title} {{Comment on `high
  temperature fermion propagator: Resummation and gauge dependence of the
  damping rate'}},\ }\href {https://doi.org/10.1103/PhysRevD.46.4779}
  {\bibfield  {journal} {\bibinfo  {journal} {Phys. Rev. D}\ }\textbf {\bibinfo
  {volume} {46}},\ \bibinfo {pages} {4779} (\bibinfo {year} {1992})},\ \Eprint
  {https://arxiv.org/abs/hep-ph/9204210} {arXiv:hep-ph/9204210} \BibitemShut
  {NoStop}%
\bibitem [{\citenamefont {Nakkagawa}\ \emph {et~al.}(1992)\citenamefont
  {Nakkagawa}, \citenamefont {Niegawa},\ and\ \citenamefont
  {Pire}}]{Nakkagawa:1992ew}%
  \BibitemOpen
  \bibfield  {author} {\bibinfo {author} {\bibfnamefont {H.}~\bibnamefont
  {Nakkagawa}}, \bibinfo {author} {\bibfnamefont {A.}~\bibnamefont {Niegawa}},\
  and\ \bibinfo {author} {\bibfnamefont {B.}~\bibnamefont {Pire}},\ }\bibfield
  {title} {\bibinfo {title} {{Resolution of the gauge dependence problem of the
  fermion damping rate in hot gauge theories}},\ }\href
  {https://doi.org/10.1016/0370-2693(92)91540-P} {\bibfield  {journal}
  {\bibinfo  {journal} {Phys. Lett. B}\ }\textbf {\bibinfo {volume} {294}},\
  \bibinfo {pages} {396} (\bibinfo {year} {1992})}\BibitemShut {NoStop}%
\bibitem [{\citenamefont {Gradshteyn}\ and\ \citenamefont
  {Ryzhik}(1980)}]{Gradshteyn:1943cpj}%
  \BibitemOpen
  \bibfield  {author} {\bibinfo {author} {\bibfnamefont {I.~S.}\ \bibnamefont
  {Gradshteyn}}\ and\ \bibinfo {author} {\bibfnamefont {I.~M.}\ \bibnamefont
  {Ryzhik}},\ }\href@noop {} {\emph {\bibinfo {title} {{Table of Integrals,
  Series, and Products}}}},\ \bibinfo {edition} {5th}\ ed.\ (\bibinfo
  {publisher} {Academic Press},\ \bibinfo {address} {New York},\ \bibinfo
  {year} {1980})\BibitemShut {NoStop}%
\bibitem [{\citenamefont {Ghosh}\ and\ \citenamefont
  {Shovkovy}(2024{\natexlab{a}})}]{DataFiles:2024}%
  \BibitemOpen
  \bibfield  {author} {\bibinfo {author} {\bibfnamefont {R.}~\bibnamefont
  {Ghosh}}\ and\ \bibinfo {author} {\bibfnamefont {I.~A.}\ \bibnamefont
  {Shovkovy}},\ }\href@noop {} {\bibinfo {title} {{Data Files for the Fermion
  Self-Energy Function and Damping Rate in a Strongly Magnetized Plasma}}},\
  \bibinfo {howpublished}
  {\url{https://www.dropbox.com/scl/fo/r95a1va222rph16k2ktf2/h?rlkey=dl4c5ty7vbmy7ltgj3vhhnqhl}}
  (\bibinfo {year} {2024}{\natexlab{a}})\BibitemShut {NoStop}%
\bibitem [{\citenamefont {Weldon}(1983)}]{Weldon:1983jn}%
  \BibitemOpen
  \bibfield  {author} {\bibinfo {author} {\bibfnamefont {H.~A.}\ \bibnamefont
  {Weldon}},\ }\bibfield  {title} {\bibinfo {title} {{Simple Rules for
  Discontinuities in Finite Temperature Field Theory}},\ }\href
  {https://doi.org/10.1103/PhysRevD.28.2007} {\bibfield  {journal} {\bibinfo
  {journal} {Phys. Rev. D}\ }\textbf {\bibinfo {volume} {28}},\ \bibinfo
  {pages} {2007} (\bibinfo {year} {1983})}\BibitemShut {NoStop}%
\bibitem [{\citenamefont {Hattori}\ and\ \citenamefont
  {Satow}(2016)}]{Hattori:2016cnt}%
  \BibitemOpen
  \bibfield  {author} {\bibinfo {author} {\bibfnamefont {K.}~\bibnamefont
  {Hattori}}\ and\ \bibinfo {author} {\bibfnamefont {D.}~\bibnamefont
  {Satow}},\ }\bibfield  {title} {\bibinfo {title} {{Electrical Conductivity of
  Quark-Gluon Plasma in Strong Magnetic Fields}},\ }\href
  {https://doi.org/10.1103/PhysRevD.94.114032} {\bibfield  {journal} {\bibinfo
  {journal} {Phys. Rev. D}\ }\textbf {\bibinfo {volume} {94}},\ \bibinfo
  {pages} {114032} (\bibinfo {year} {2016})},\ \Eprint
  {https://arxiv.org/abs/1610.06818} {arXiv:1610.06818 [hep-ph]} \BibitemShut
  {NoStop}%
\bibitem [{\citenamefont {Bandyopadhyay}\ \emph {et~al.}(2022)\citenamefont
  {Bandyopadhyay}, \citenamefont {Liao},\ and\ \citenamefont
  {Xing}}]{Bandyopadhyay:2021zlm}%
  \BibitemOpen
  \bibfield  {author} {\bibinfo {author} {\bibfnamefont {A.}~\bibnamefont
  {Bandyopadhyay}}, \bibinfo {author} {\bibfnamefont {J.}~\bibnamefont
  {Liao}},\ and\ \bibinfo {author} {\bibfnamefont {H.}~\bibnamefont {Xing}},\
  }\bibfield  {title} {\bibinfo {title} {{Heavy quark dynamics in a strongly
  magnetized quark-gluon plasma}},\ }\href
  {https://doi.org/10.1103/PhysRevD.105.114049} {\bibfield  {journal} {\bibinfo
   {journal} {Phys. Rev. D}\ }\textbf {\bibinfo {volume} {105}},\ \bibinfo
  {pages} {114049} (\bibinfo {year} {2022})},\ \Eprint
  {https://arxiv.org/abs/2105.02167} {arXiv:2105.02167 [hep-ph]} \BibitemShut
  {NoStop}%
\bibitem [{\citenamefont {Bandyopadhyay}(2024)}]{Bandyopadhyay:2023hiv}%
  \BibitemOpen
  \bibfield  {author} {\bibinfo {author} {\bibfnamefont {A.}~\bibnamefont
  {Bandyopadhyay}},\ }\bibfield  {title} {\bibinfo {title} {{Heavy quark
  diffusion coefficients in magnetized quark-gluon plasma}},\ }\href
  {https://doi.org/10.1103/PhysRevD.109.034013} {\bibfield  {journal} {\bibinfo
   {journal} {Phys. Rev. D}\ }\textbf {\bibinfo {volume} {109}},\ \bibinfo
  {pages} {034013} (\bibinfo {year} {2024})},\ \Eprint
  {https://arxiv.org/abs/2307.09655} {arXiv:2307.09655 [hep-ph]} \BibitemShut
  {NoStop}%
\bibitem [{\citenamefont {Ghosh}\ and\ \citenamefont
  {Shovkovy}(2024{\natexlab{b}})}]{Ghosh:2024fkg}%
  \BibitemOpen
  \bibfield  {author} {\bibinfo {author} {\bibfnamefont {R.}~\bibnamefont
  {Ghosh}}\ and\ \bibinfo {author} {\bibfnamefont {I.~A.}\ \bibnamefont
  {Shovkovy}},\ }\bibfield  {title} {\bibinfo {title} {{Electrical conductivity
  of hot relativistic plasma in a strong magnetic field}},\ }\href@noop {} {\
  (\bibinfo {year} {2024}{\natexlab{b}})},\ \Eprint
  {https://arxiv.org/abs/2404.01388} {arXiv:2404.01388 [hep-ph]} \BibitemShut
  {NoStop}%
\bibitem [{\citenamefont {Bhattacharya}\ and\ \citenamefont
  {Pal}(2004)}]{Bhattacharya:2002aj}%
  \BibitemOpen
  \bibfield  {author} {\bibinfo {author} {\bibfnamefont {K.}~\bibnamefont
  {Bhattacharya}}\ and\ \bibinfo {author} {\bibfnamefont {P.~B.}\ \bibnamefont
  {Pal}},\ }\bibfield  {title} {\bibinfo {title} {{Neutrinos and magnetic
  fields: A Short review}},\ }\href@noop {} {\bibfield  {journal} {\bibinfo
  {journal} {Proc. Indian Natl. Sci. Acad. A}\ }\textbf {\bibinfo {volume}
  {70}},\ \bibinfo {pages} {145} (\bibinfo {year} {2004})},\ \Eprint
  {https://arxiv.org/abs/hep-ph/0212118} {arXiv:hep-ph/0212118} \BibitemShut
  {NoStop}%
\end{thebibliography}
\end{document}